\documentclass[sort&compress,3p]{elsarticle}
\journal{Journal} 
\usepackage{numcompress}
\usepackage[hidelinks]{hyperref}
\usepackage{lineno,hyperref}
\usepackage{lipsum}
\usepackage{amsfonts}
\usepackage{amsmath,amssymb}
\usepackage{amsthm}
\usepackage{fourier}
\usepackage{xcolor}
\usepackage{mathrsfs}
\usepackage{graphicx}
\usepackage{float}
\usepackage{subcaption, comment}
\usepackage{tikz}
\usepackage[labelfont=bf]{caption}
\usepackage{epstopdf}
\usepackage[ntheorem]{empheq} 
\usepackage[final]{pdfpages}
\usepackage{subcaption}
\usepackage[justification=centering]{caption}
\usepackage{amsopn}
\usepackage{array}
\usepackage{empheq}
\usepackage{upgreek}
\usepackage{graphicx}
\usepackage{float}
\usepackage{subcaption}
\usepackage[makeroom]{cancel}
\usepackage{algorithm}
\usepackage{algpseudocode}
\usepackage[normalem]{ulem}

\newcommand{\vs}{\vspace{\baselineskip}}
\newcommand{\R}{\mathbb{R}}
\newcommand{\stress}{\boldsymbol{\upsigma}}
\newcommand{\strain}{\boldsymbol{\upvarepsilon}}
\newcommand{\C}{\mathbb{C}}
\newcommand{\I}{\mathbb{I}}
\newcommand{\e}{\textrm{\textbf{e}}}
\newcommand{\1}{\mathrm{\boldsymbol{1}}}
\newcommand{\s}{\textrm{\textbf{s}}}
\newcommand{\n}{\textrm{\textbf{n}}}

\newtheorem{lem}{Lemma}

\newtheorem{rmk}{Remark}
\newtheorem{assmp}{Assumption}
\newtheorem{axm}{Axiom}
\newtheorem{dfn}{Definition}
\newtheorem{prpsn}{Proposition}

\begin{document}

\begin{frontmatter}
\title{\textbf{Shear-enhanced compaction analysis of the Vaca Muerta formation.}}

\author[one]{Jose G. Hasbani}\corref{mycorrespondingauthor}
\cortext[mycorrespondingauthor]{Corresponding author}
\ead{jose.hasbani@vistaenergy.com}

\author[three]{Evan M. C. Kias}
\author[three]{Roberto Suarez-Rivera}
\author[two]{Victor M. Calo}

\address[one]{Vista Energy, Buenos Aires, Argentina}
\address[three]{W.D. Von Gonten Engineering, Houston, Texas, United States}
\address[two]{School of Electrical Engineering,  Computing and Mathematical Sciences,  Curtin University,  Bentley,  Australia}

\begin{abstract}
Laboratory measurements on Vaca Muerta formation samples show stress-dependent elastic behavior and compaction at representative in-situ conditions. Experimental results show that the analyzed samples exhibit elastoplastic deformation and shear-enhanced compaction as the main plasticity mechanism. These experimental observations conflict with the anticipated linear-elastic response prior to the brittle failure reported in several works on the geomechanical characterization of the Vaca Muerta formation. Therefore, we present a complete laboratory analysis of samples from the Vaca Muerta formation showing experimental evidence of nonlinear elastic and unrecoverable shear-enhanced compaction. We also calibrate an elastoplastic constitutive model using these experimental observations; the resulting model reproduces the observed phenomena adequately.
\end{abstract}


\end{frontmatter}

\tableofcontents
\section{Introduction}
Reservoir rocks' nonlinear and heterogeneous nature is typically simplified when analyzing deformation and failure during oil-and-gas well operations such as drilling, hydraulic fracturing, and production~\cite{ EWY1990387}. Consequently, geomechanical engineers routinely assume that the material response is linear-elastic until reaching brittle failure; this assumption allows them to use straightforward analytical approximations to model stress distribution around a wellbore~\cite{ Kirsch1898}. These oversimplifications are widely used to model wellbore stability problems and hydraulic fracture~\cite{ Lecampion2017}. Typically, the linear elasticity assumption in unconventional reservoirs has been justified based on their brittleness~\cite{ Kias2015}, implying that reservoir rocks favorable to hydraulic fracturing treatments could be accurately modeled using linear elastic fracture mechanics~\cite{ Griffith1921}. This practice generally produces acceptable results when the reservoir rock exhibits a strong linear elastic response.

The Vaca Muerta formation is Argentina's main unconventional reservoir in the Neuqu\'en Basin. This formation is composed of sedimentary rock rich in organic mudstones, limestones, and marls; it was deposited in a distal ramp~\cite{ Sagasti2014}. The mineralogy of this type of reservoir rock is dominated by calcite, quartz, mica, pyrite, and clays. Typically, Vaca Muerta mudrock is characterized as a linear elastic material~\cite{ Varela2017} in rate-independent mechanical problems or as a visco-elastic material~\cite{ Hasbani2018} in geological time-dependent basin modeling problems. However, during routine triaxial testing to characterize the elastic properties of a set of Vaca Muerta samples, we observe shear-induced compaction; this compaction as a plasticity-driven mechanism may explain specific field observations such as inefficient fracture initiation, unexpected wellbore production underperformance due to unexpected hydraulic fracture geometry, or proppant placement~\cite{ wang1994, papanastasiou1997}.

The compaction of porous rocks is often explained as the closure of porosity due to increasing effective stress, assuming that solid constituents have negligible compressibility. This mechanism is a factor in the life-cycle of conventional reservoirs in the form of permeability reduction~\cite{ Carman1997} and subsidence~\cite{ Geertsma1973}. In addition, the regional in-situ stress state impacts the rock's failure mode as demonstrated in numerous studies found in the literature addressing the brittle-ductile transition~\cite{ Byerlee1968, Wong1999, Wong1990, Wong1997, Nygård2006}. During laboratory testing, as confining pressure increases, the rock failure evolves from void volume creation through the formation and opening of micro-cracks that finally coalesce into a macroscopic material discontinuity known as brittle faulting~\cite{ Horii1986}. This failure mode involves the accumulation of irreversible plastic volumetric strain from grains rearrangement, the collapse of the pore volume or grain fragmentation and it is known as shear enhanced compaction~\cite{ Curran1979}.  Plastic deformation in mudrocks is typically attributed to the abundance of organic matter and clay inside the reservoir rock matrix ~\cite{ Wang2019}. However, the laboratory experiments conducted in this work in the Vaca Muerta mudrock samples suggest that shear-enhanced compaction, possibly controlled by grain displacement, is also a predominant mechanism of unrecoverable volumetric plastic deformation. Here, we characterize the compaction mechanism experimentally, conducting a series of triaxial tests over a set of samples from the Vaca Muerta formation.

We adopt a phenomenological nonlinear elastoplastic constitutive model that adequately captures the observed material response; we calibrate its constitutive parameters for use in engineering applications. From all available phenomenological constitutive models, we select the Modified Cam-Clay (MCC) model~\cite{ Roscoe1963, Roscoe1968} as it captures the main experimental observations: nonlinear elasticity and isotropic hardening evolution of the yield surface, which translates into unrecoverable volumetric deformation. Although, MCC was originally developed for characterizing the soils' critical state, it was extended to other types of cohesive-frictional materials~\cite{ Cier2022, Diarra2017}. In addition, we implement an implicit integration algorithm~\cite{ Borja1990, Borja1991} and simulate a triaxial test response at a Gauss point. We structure the manuscript as follows: a brief introduction in this section; Section 2 gives notation and preliminary definitions that are used throughout this work; Section 3 briefly introduces the plasticity theory in the context of thermodynamics of deformable solids; Section 4 addresses the continuous elastoplastic formulation, Section 5 discusses experimental evidence of elastoplastic response on Vaca Muerta mudstone, Section 6 details the numerical implementation, and, finally, Section 7 discusses our conclusions. 
\section{Notation and preliminary definitions}

We denote second-order tensors (i.e., matrix arrays that satisfy certain change of basis rules) using bold Greek symbols or letters, fourth-order tensors with upper case blackboard-bold letters (e.g., $\mathbb{C})$), vectors (i.e., first-order tensors) with lower case italic, bold letters, and scalars with lower case Greek symbols or letters. In addition, we adopt Einstein's summation convention (given $i=1,\,2,\,3; \ a_i\,b_i =  a_1\,b_1 + a_2\,b_2 + a_3\,b_3$) to perform component-wise tensor operations such as contractions (vector inner products) or double contractions (second-order-tensor inner products). The symbol $\otimes$ denotes the tensor product (outer product). We assume that all the operations are performed in Euclidean space (zero curvature space). Thus, we make no distinction between covariant and contravariant basis vectors. Unit basis vectors in the Euclidean Space $\R^3$ are denoted as $\e_i, i = 1,\,2,\,3$ with the property $\e_i \cdot \e_j = \delta_{ij}$ and $\e_i \cdot \e_i = \delta_{ii}=3$ for $i,\,j=1,\,2,\,3$, where $\delta_{ij} \in \R^{3\times3}$ is the \textit{Kronecker Delta, where $\delta_{ii}=1$ and $\delta_{ij}=0$ if $i\neq j$}. Therefore, we write first-order, second-order, and fourth-order tensors as follows,
\begin{align*}
\boldsymbol{a} = a_i \, \e_i, &&
\stress = \upsigma_{ij}\, \e_i \otimes \e_j, &&
\mathbb{C} = \textrm{C}_{ijkl}\, \e_i\otimes\e_j\otimes\e_k\otimes\e_l,&&
\text{with} \quad i,\,j,\,k,\,l = 1,\,2,\,3
\end{align*}
\begin{rmk}[Index expansion]
In what follows, we work in a three-dimensional Euclidean Space. Thus, the range for indexes will be omitted in the rest of this manuscript.
\end{rmk}
The \textit{transpose} of a second-order tensor $\stress \in \R^{3 \times 3}$ using index notation is defined as,
$$
\stress^T = \upsigma_{ji} \, \e_i \otimes \e_j = \upsigma_{ij} \,\e_j \otimes \e_i,
$$
and the \textit{trace} of a second-order tensor $\stress$ is given by $\text{tr}(\stress) = \stress\,:\boldsymbol{1}$, where $\boldsymbol{1}= \delta_{ij}\, \e_i \otimes \e_j$.
Given two vectors $\boldsymbol{a},\, \boldsymbol{b} \in \R^3$ and two second-order tensors $\stress,\, \strain \in \R^{3 \times 3}$, the contraction and the double contraction are,
\begin{equation*}
    \begin{aligned}
\boldsymbol{a} \cdot \boldsymbol{b} &= a_i \,\e_i \cdot b_j \, \e_j
    && \textit{(Contraction)}\\
                                    &= a_i\, b_j \, \e_i \cdot \e_j\\
                                    &= a_i \, b_j \, \delta_{ij}\\
                                    &= a_i \, b_i\\
                                    \\
\end{aligned}
\qquad\qquad
\qquad
\begin{aligned}
\stress : \strain 
    &= \upsigma_{ij} \, \e_i \otimes \e_j \, : \, \upvarepsilon_{kl} \, \e_k \otimes \e_l 
    &&\textit{(Double Contraction)}\\
	&= \upsigma_{ij}\,\upvarepsilon_{kl}\, \e_i \otimes \e_j\,:\, \e_k\,\otimes\e_l\\
	&= \upsigma_{ij}\,\upvarepsilon_{kl}\, \left(\e_i\cdot\e_k\right)\left(\e_j\cdot\e_l\right)\\
	&= \upsigma_{ij}\,\upvarepsilon_{kl}\,\delta_{ik}\,\delta_{jl}\\
	&= \upsigma_{ij}\,\upvarepsilon_{ij}
\end{aligned}
\end{equation*}
%
%
 We often make use of the symmetric fourth-order unit tensor defined as,
\begin{equation}\label{eq:fourth_order_unit_tensor}
\I = \mathrm{I}_{ijkl}\, \e_i \otimes \e_j \otimes \e_k  \otimes \e_l := \dfrac{1}{2} \big(\delta_{ik}\,\delta_{jl} + \delta_{il}\,\delta_{jk} \big)\, \e_i \otimes \e_j \otimes \e_k \otimes \e_l.
\end{equation}
The symmetric fourth-order unit tensor admits the following decomposition:
$$
\I = \I_d + \I_v,
$$
where $\I_v$ and $\I_d$ are the volumetric and deviatoric contributions given by,
\begin{align*}
\I_v &= \dfrac{1}{3} \1 \otimes \1,&&
\I_d = \I - \I_v. 
\end{align*}
The operator $\|\ast\|: \ast \mapsto \mathbb{R}$ denotes the Frobenius norm for either vectors or second-order tensors. For vectors, $\| \boldsymbol{a} \| = \sqrt{\boldsymbol{a} \cdot \boldsymbol{a}} = \sqrt{a_i \, a_i}$, whereas for second-order tensors $\| \stress \| = \sqrt{\stress : \stress} = \sqrt{\upsigma_{ij} \, \upsigma_{ij}}$.
We usually work with variables that  evolve during a pseudo-time increment $\Delta t = [t_{n+1},\, t_n],\, \forall n \in \mathbb{N}^+$. The derivative of a tensor or scalar field with respect to the pseudo-time $t$ is denoted as:
\begin{align*}
\dfrac{\partial \stress(t)}{\partial t} &:= \dot{\stress}(t) = \dot{\upsigma}_{ij}(t), \qquad \forall \stress(t) \in \R^{3\times 3} \times [0, \,T]&&
\dfrac{\partial \phi(t)}{\partial t} := \dot{\phi}(t), \qquad \forall \phi: \R^n \times [0, T] \mapsto \R.
\end{align*}
We adopt the rock mechanics convention in which compressive stresses are positive and tensile stresses negative. In addition, we denote the \textit{Effective Cauchy's Stress Tensor} by $\stress$ and the \textit{Infinitesimal Strain Tensor} by $\strain$. The \textit{Hydrostatic Stress} is $p = \frac{1}{3} \text{tr}(\stress) = \stress\,:\,\boldsymbol{1}$ and the \textit{Deviatoric Effective Stress} $\s = \stress - p\,\1$ with the property that $\text{tr}(\s) = 0$. Additionally, for a symmetric second-order tensor $\stress$, we define the following tensor invariants,
\begin{align}
\label{eq:stress_invariants}
I_1(\stress) &= \stress : \1 = \upsigma_{ii},\\
I_2(\stress) &= \dfrac{1}{2} \left(I(\stress)^2_1 - \stress\,:\, \stress \right),\\
I_3(\stress) &= \dfrac{1}{6} \left[2\, \left(\stress \cdot \stress\right) : \stress - 3\,I_1(\stress)\left(\stress : \stress \right) + I^3_1 \right) = \det(\stress).
\end{align}
 The stress invariants for the deviatoric symmetric second-order tensor are,
\begin{align}
J_1(\s) &= \mathrm{tr}(\s) = 0,\\
J_2(\s) &=\dfrac{1}{2} \s : \s,\\
J_3(\s) &= \dfrac{1}{3} (\s : \s) : \s.
\end{align}
 Finally, the deviatoric stress invariants relate to the stress invariants as follows,
\begin{align}
J_2(\s) &= \dfrac{1}{3} \left(I^2_1(\stress) - 3\,I_2(\stress)\right),\\
J_3(\s) &= \dfrac{1}{27} \left(2\,I^3_1(\stress) - 9\,I_1(\stress)\,I_2(\stress) + 27\,I_3(\stress)\right),\\
I_2(\stress) &= \dfrac{1}{3} \left(I^2_1(\stress) - 3\,J_2(\s)\right),\\
I_3(\stress) &= \dfrac{1}{27}\left(I^3_1(\stress) - 9 \, I_1(\stress)\,J_2(\s) + 27\, J_3(\s)\right). 
\end{align}
\vs
\section{Introduction to Plasticity Theory}
\subsection{Thermodynamics of deformable continua}
This section introduces the mathematical theory of plasticity in the context of the thermomechanics of deformable bodies (see~\cite{ Lubiner1985}). Let $\mathcal{B}$ be a continuum body with volume $V$ and boundary $\partial \mathcal{B}$. Let us define on $\mathcal{B}$ the following scalar fields: $\rho$, $\theta$, $u$, $s$, and $R$ representing the mass density, temperature, internal energy, entropy, and heat density, respectively. We assume that small deformations properly describe the system's kinematics. Thus, the \textit{first (energy conservation) and second (entropy production imbalance in the form of the Clasius-Duhem inequality) thermodynamic laws} in their local form adopt the following differential expressions,
\begin{itemize}
\item[1.] \textit{Energy conservation:}
\begin{equation}\label{eq:first_principle_thermo}
\rho\,\dot{u} = \stress\, : \, \dot{\strain} - \nabla \cdot \boldsymbol{q} + \rho\,R,
\end{equation}
where $\stress, \dot{\strain} \in \R^{3 \times 3}$ are Cauchy's stress and strain rate tensors, respectively, $\nabla = \dfrac{\partial}{\partial x_i} \e_i$ is the gradient operator, and $\boldsymbol{q}$ is the heat flux vector acting on $\partial \mathcal{B}$.

\item[2.] \textit{Entropy production imbalance:}
\begin{equation}\label{eq:second_principle_thermo}
\rho\, \dot{s} + \nabla \cdot \left(\dfrac{\boldsymbol{q}}{\theta} \right) - \dfrac{\rho\,R}{\theta} \geq 0.
\end{equation}
\end{itemize}
Substituting~\eqref{eq:first_principle_thermo} into~\eqref{eq:second_principle_thermo} we get the \textit{Clausius-Duhem Inequality},
\begin{equation}\label{eq:clausius_duhem_ineq}
\rho\,\dot{s} + \nabla \cdot \left(\dfrac{\boldsymbol{q}}{\theta} \right) - \dfrac{1}{\theta} \left( \rho \, \dot{u} - \stress : \dot{\strain} + \nabla \cdot \boldsymbol{q} \right) \geq 0.
\end{equation}
We define the \textit{Helmholtz free energy} per unit of mass as,
\begin{equation}\label{eq:helmholtz_free_energy}
\Psi := u - \theta\,s.
\end{equation}
Recalling that (by distributive property of the $\nabla$ operator over vector and scalar fields),
\begin{equation*}
\nabla \cdot \left(\dfrac{\boldsymbol{q}}{\theta} \right) = \dfrac{1}{\theta} \, \nabla \cdot \boldsymbol{q} - \dfrac{1}{\theta^2}\, \boldsymbol{q} \cdot \nabla \theta,
\end{equation*}
and the change of variables $\textrm{\textbf{g}} = \nabla \theta$, we rewrite~\eqref{eq:clausius_duhem_ineq} as,
\begin{equation}\label{eq:clausius_duhem_ineq_2}
\stress : \dot{\strain} - \rho \left(\dot{\Psi} + \dot{\theta}\, s\right) - \dfrac{1}{\theta} \boldsymbol{q} \cdot \textrm{\textbf{g}} \geq 0.
\end{equation}

\begin{assmp}[Thermodynamic evolution]
 We assume that the system's evolution is isothermal, leading to a purely mechanical formulation for~\eqref{eq:clausius_duhem_ineq_2},
\begin{equation}\label{eq:clausius_duhem_mech}
 \stress : \dot{\strain} - \rho \dot{\Psi} \geq 0.
 \end{equation}
\end{assmp}

\begin{rmk}
Every mechanical system must satisfy the Clausius-Duhem inequality of~\eqref{eq:clausius_duhem_mech} where the system's energy in thermodynamic equilibrium is characterized by a series set of \textit{state variables}.
\end{rmk}

\begin{axm}[Local-state postulate]
Given an arbitrary system evolution, the energetic state of a continuum medium is characterized by the same state variables that are fixed at the equilibrium state, and it is rate independent. Therefore, the Helmholtz free energy~\eqref{eq:helmholtz_free_energy} is given by,

\begin{equation}\label{eq:local_state_postulate}
\Psi = \tilde{\Psi}\left(\strain, \boldsymbol{\eta} \right),
\end{equation}
where $\boldsymbol{\eta} = {\eta_k}, \forall k \in \mathbb{N^+}$ is an \textit{internal variable} set describing the material's dissipation and its free energy rate is,
\begin{equation}
\dot{\Psi} = \dfrac{\partial\Psi}{\partial \strain} : \dot{\strain} + \dfrac{\partial \Psi}{\partial \boldsymbol{\eta}} \ast \dot{\boldsymbol{\eta}}.
\end{equation}
The symbol $"\ast"$ denotes the contraction compatible with $\dfrac{\partial \Psi}{\partial \boldsymbol{\eta}} $ and $\dot{\boldsymbol{\eta}}$.
\end{axm}

\subsection{Reversible and irreversible thermodynamic processes: elastoplasticity}

We characterize the deformation evolution of a material as reversible (non-dissipative) and irreversible (dissipative) by formalizing these concepts using the following definitions:
\begin{dfn}[Reversibility: elasticity]
A deformation process is \textit{non-dissipative}, \textit{reversible} or \textit{elastic} if and only if the internal variables remain constant throughout the deformation process (i.e., $\dot{\boldsymbol{\eta}} = 0$). Thus,~\eqref{eq:clausius_duhem_mech} reduces to
\begin{equation}\label{eq:clausius_duhem_elastic}
\stress \,:\, \dot{\strain} - \dfrac{\partial\Psi}{\partial\strain} \, :\, \dot{\strain} = 0.
\end{equation}
Letting $\Psi=\mathcal{W}$, where $\mathcal{W}$ is the strain energy density, we express the \textit{hyperelastic constitutive models} as
\begin{equation}\label{eq:hyperelastic_const_models}
\stress = \dfrac{\partial\mathcal{W}}{\partial\strain}.
\end{equation}
\end{dfn}
We define dissipative thermodynamic processes by adopting an additive decomposition of the strain tensor,
\begin{assmp}[Strain-tensor additive decomposition]\label{assmp: strain_tensor_decomp}
Let $\strain \in \R^{3 \times 3}$ be the strain tensor at a material point $\boldsymbol{x} \in \mathcal{B}$, which admits the following decomposition,
\begin{equation}\label{eq:strain_additive_decomp}
\strain = \strain^e + \strain^p,
\end{equation}
where $\strain^e$ and $\strain^p$ are the elastic and plastic components of the strain tensor, respectively. 
\end{assmp}

\begin{dfn}[Dissipative Processes]
A thermodynamic deformation process is dissipative, irreversible, or plastic if the free energy $u$ can be decomposed in a strain energy density $\mathcal{W}$ and a latent energy density $\mathcal{V}$. Thus, defining $\Psi = \tilde{\Psi}\left(\strain,\, \strain^p,\, \boldsymbol{\eta}\right)$ and considering Assumption~\ref{assmp: strain_tensor_decomp},
$$
\tilde{\Psi}\left(\strain,\,\strain^p,\,\boldsymbol{\eta}\right) = \mathcal{W}(\strain - \strain^p) + \mathcal{V}(\boldsymbol{\eta}) = \mathcal{W}(\strain^e) + \mathcal{V}(\boldsymbol{\eta}).
$$
Therefore, the Clausius-Duhem inequality~\eqref{eq:clausius_duhem_mech} becomes,
$$
\stress\,:\,\dot{\strain} - \dfrac{\partial\mathcal{W}}{\partial\strain^e}\,:\,\dot{\strain}_e - \dfrac{\partial\mathcal{V}}{\partial{\boldsymbol{\eta}}} \ast \dot{\boldsymbol{\eta}} \geq 0
$$
where $\dot{\strain} = \dot{\strain}^e + \dot{\strain}^p$ and $\stress = \dfrac{\partial\mathcal{W}}{\partial\strain^e}$, thus
\begin{equation}
\stress \,:\,\dot{\strain}^p - \dfrac{\partial\mathcal{V}}{\partial\boldsymbol{\eta}} \ast \dot{\boldsymbol{\eta}} \geq 0.
\end{equation}
\end{dfn}
\begin{rmk}[Reversible and irreversible processes]
A system's evolution is \textit{reversible} if and only if there exists an isomorphism between the initial and the final state; otherwise, the evolution of a system is \textit{irreversible}.
\end{rmk}
For irreversible processes, the Clausius-Duhem inequality is satisfied for more than one set of state variables. Thus, we enforce uniqueness using the \textit{Maximum Plastic Dissipation Condition}.

\begin{prpsn}[Maximum Plastic-Dissipation Condition]\label{prpsn:maximum_plastic_dis_cond}
Let $~\mathcal{D}^p: \R^{3 \times 3} \times \R^n \times \R^{3 \times 3} \mapsto \mathbb{R}$ be the plastic dissipation:
\begin{equation}\label{eq:max_diss_func}
\mathcal{D}^p\left(\stress,\,\dot{\boldsymbol{\eta}},\,\dot{\strain}^p\right) := \stress\,:\,\dot{\strain}^p - \dfrac{\partial\mathcal{V}}{\partial\boldsymbol{\eta}} \ast \dot{\boldsymbol{\eta}}.
\end{equation}
Let $~\mathbb{E}_{\upsigma}$ be a closed and convex set defining the admissible state variables $\mathbb{E}_{\upsigma}$ given by
\begin{equation}\label{eq:admissible_states_set}
\mathbb{E}_{\upsigma} := \left\{(\boldsymbol{\uptau},\,\dot{\boldsymbol{\upkappa}}) \in \R^{3 \times 3} \times \R^n, F_f(\boldsymbol{\uptau},\,\dot{\boldsymbol{\upkappa}}) \leq 0 \right\},
\end{equation}
where $F_f: \R^{3 \times 3} \times \R^n \mapsto \R$ bounds the admissible stress states (yield function). This assumption defines a unique state variable set  $\left(\stress,\,\dot{\boldsymbol{\eta}},\,\dot{\strain}^p\right)$ such that,
\begin{equation}
\mathcal{D}^p\left(\stress,\,\dot{\boldsymbol{\eta}},\,\dot{\strain}^p\right) = \underset{(\boldsymbol{\uptau},\,\dot{\boldsymbol{\upkappa}})\in \mathbb{E}_{\upsigma}}{\arg \max} \, \mathcal{D}^p\left(\boldsymbol{\uptau},\,\dot{\boldsymbol{\upkappa}},\,\dot{\strain}_p\right).
\end{equation}
\end{prpsn}

\begin{rmk}[Elastic Domain and Flow Surface]
The set of admissible states $\mathbb{E}_{\upsigma}$ admits a partition $\mathbb{E}_{\upsigma} = \mathrm{int}(\mathbb{E}_{\upsigma}) \bigcup \partial \mathbb{E}_{\upsigma}$, where $\mathrm{int}(\mathbb{E}_{\upsigma})$ is the elastic domain defined by,
$$
\mathrm{int}(\mathbb{E}_{\upsigma}) := \left\{(\stress,\,\dot{\boldsymbol{\eta}}) \in \R^{3 \times 3} \times \R^n,\, F_f(\stress,\,\dot{\boldsymbol{\eta}}) < 0 \right\},
$$
and $\partial \mathbb{E}_{\upsigma}$ is the flow surface defined as
$$
\partial \mathbb{E}_{\upsigma} :=\left\{(\stress,\,\dot{\boldsymbol{\eta}}) \in \R^{3 \times 3} \times \R^n,\, F_f(\stress,\,\dot{\boldsymbol{\eta}}) = 0 \right\}.
$$
\end{rmk}

 Proposition~\ref{prpsn:maximum_plastic_dis_cond} finds the \textit{state variables} that maximize $\mathcal{D}^p$ subject to the constraint $F_f(\stress) = 0$ \textit{(Consistency Condition)} and the complementary Kuhn-Tucker conditions~\cite{ Sundaram1996}. We solve this optimization problem by introducing a Lagrangian (cost function) that transforms the maximization into a minimization problem (i.e., let $-\mathcal{D}_p (\boldsymbol{\uptau},\, \dot{\boldsymbol{\upkappa}}, \dot{\strain}^p)$)
$$
\mathcal{L}(\boldsymbol{\uptau},\,\dot{\boldsymbol{\kappa}}, \lambda) := \upgamma\,F_f(\boldsymbol{\uptau},\,\dot{\boldsymbol{\kappa}}) -\mathcal{D}^p (\boldsymbol{\uptau},\, \dot{\boldsymbol{\upkappa}}, \dot{\strain}^p),
$$
where $
\mathcal{L}: \R^{3\times 3} \times \R^n \times \mathbb{R}^+ \mapsto \mathbb{R}
$ is the Lagrangian and $\upgamma \in \mathbb{R}^+$ is the \textit{Lagrange multiplier}. Henceforth, satisfying the \textit{Maximum Plastic Dissipation Condition} is equivalent to solving the following minimization problem:
\begin{equation}\label{eq:plasticity_opt_problem}
\begin{cases}
\textrm{\textit{Find}}\, \stress,\,\dot{\boldsymbol{\eta}} \in \mathbb{E}_{\upsigma}\,\textrm{\textit{and}}\, \upgamma \in \mathbb{R}^+,\, \textrm{\textit{such that}}\\
\qquad(\stress,\,\dot{\boldsymbol{\eta}}, \upgamma) := \underset{(\boldsymbol{\uptau},\,\dot{\boldsymbol{\upkappa}})\in \mathbb{E}_{\upsigma}}{\arg \min} \, \upgamma \,F_f(\boldsymbol{\uptau},\,\dot{\boldsymbol{\kappa}}) -\mathcal{D}^p (\boldsymbol{\uptau},\, \dot{\boldsymbol{\upkappa}}, \dot{\strain}^p).
\end{cases}
\end{equation}
The necessary optimal condition is $\nabla \mathcal{L}\vert_{(\stress,\,\dot{\boldsymbol{\eta}})} = 0$ and the Lagrangian is convex (concave); thus, we can deduce,
\begin{align}
\dfrac{\partial \mathcal{L}}{\partial \stress} &= - \dot{\strain}^p + \upgamma \dfrac{\partial F_f}{\partial \stress} = 0 \label{eq:lagrangian_der_sigma},\\
\dfrac{\partial \mathcal{L}}{\partial \boldsymbol{\eta}} &= \boldsymbol{\mathcal{H}} \ast \dot{\boldsymbol{\eta}} + \upgamma \,\dfrac{\partial F_f}{\partial \boldsymbol{\eta}} = 0 \label{eq:lagrangian_der_eta},\\
\dfrac{\partial \mathcal{L}}{\partial\upgamma} &= F_f(\stress, \, \dot{\boldsymbol{\eta}}) = 0 \label{eq:lagrangian_der_gamma}.
\end{align}
where $\boldsymbol{\mathcal{H}}$ is the  \textit{Hardening Modulus} given by,
$$
\boldsymbol{\mathcal{H}} := \dfrac{\partial^2\mathcal{V}}{\partial\boldsymbol{\eta} \ast \partial{\boldsymbol{\eta}}}.
$$

 From~\eqref{eq:lagrangian_der_sigma} and~\eqref{eq:lagrangian_der_eta} and considering the following change of variable $\upgamma = \dot{\lambda}$, the \textit{Generalized Associative Flow Rule} and the \textit{Generalized Hardening Law} are~\cite{ Neto2019},\\
\begin{itemize}
    \item \textit{Generalized Associative Flow Rule: }\\
    \begin{equation}\label{eq:generalized_flow_rule}
    \dot{\strain}^p = \dot{\lambda}\,\dfrac{\partial F_f}{\partial \stress}.
    \end{equation}
    \item \textit{Generalized Hardening Law:}\\
    \begin{equation}\label{eq:generalized_hardening_law}
    \dot{\boldsymbol{\eta}} = -\dot{\lambda} \,\boldsymbol{\mathcal{H}}^{-1} \dfrac{\partial F_f}{\partial \boldsymbol{\eta}}.
    \end{equation}
\end{itemize}

\begin{rmk}[Non-Associative Flow Rules]
Proposition~\ref{prpsn:maximum_plastic_dis_cond} is a sufficient condition that is too restrictive in general, but it inherently induces a generalized hardening law and a generalized associative flow rule. In practice, the associative plastic flow rule appropriately characterizes materials with crystalline micro-structural composition. Although associative flow rules are used to characterize the plastic flow of granular materials, non-associative flow rules could be more suitable. Therefore, the Clausius-Duhem Inequality in its local form should be verified independently when using non-associated plastic flow rules. Typically, a non-associated flow rule is defined as
\begin{equation}\label{eq:non_associated_flow_rule}
\dot{\strain}^p := \dot{\lambda}\, \boldsymbol{\mathcal{G}}(
\stress,\, \boldsymbol{\eta})\quad\textrm{where}\quad \boldsymbol{\mathcal{G}}(
\stress,\, \boldsymbol{\eta})\neq\dfrac{\partial\,F_f}{\partial\,\stress}.
\end{equation}
These heuristic definitions seek to reconcile experimental observations with simulations by relaxing the overly restrictive Proposition~\ref{prpsn:maximum_plastic_dis_cond}, while their major weakness is their ad-hoc nature.
\end{rmk}

\section{Continuous elastoplastic constitutive model}\label{sec:continuous_formulation}

Elastoplastic constitutive models for materials should capture the following experimental observations~\cite{ Neto2019}:
\begin{enumerate}
\item For loads below a threshold that defines a \textit{flow criterion}, the material response is reversible \textit{(elastic)}.
\item Once the material reaches the limit condition, the deformation becomes partly \textit{irrecoverable} \textit{(plastic)}.
\item Plastic deformations evolve the failure state as described by a \textit{hardening law}.
\item During unloading, the material response is \textit{elastic}.
\item The material response during the whole deformation process is \textit{quasi-stationary}.
\item The material is thermodynamically \textit{stable}.
\end{enumerate}
We describe the material's elastic response by adopting a linear Hookean model given by,
\begin{equation}\label{eq:linear_elasticity}
\stress = \C^e \, : \, \strain^e = \C^e \, : \, \left(\strain - \strain^p\right),
\end{equation}
where $\C^e$ is the fourth-order elasticity tensor defined by the bulk modulus $K$, and the shear modulus $G$ as,
\begin{equation}\label{eq:elasticity_tensor}
\C^e := K\,\1\otimes\1 + 2\,G\,\left(\I-\dfrac{1}{3}\1\otimes\1\right),
\end{equation}
with $K$ and $G$ adopt the following expressions~\cite{ Borja1990},
\begin{equation}\label{eq:bulk_modulus}
K = \dfrac{p}{\kappa(1-\phi)},
\end{equation}
\begin{equation}\label{eq:shear_modulus}
G = \dfrac{3\,K(1 - 2\nu)}{2(1 + \nu)},
\end{equation}
where $\kappa$ is the volumetric deformation recovery, $\phi$ is the rock's total porosity, and $\nu$ is Poisson's ratio.
\begin{rmk}[Pressure dependency of $K$ and $G$]
During laboratory testing conducted in Vaca Muerta samples, we observed pressure dependence on the bulk $K$  and shear $G$ moduli. We capture that  phenomena including isotropic part of the effective stress tensor $p$ in~\eqref{eq:bulk_modulus} and  \eqref{eq:shear_modulus}. However, some porous rocks in specific stress state ranges could be modelled as linear and non-pressure dependent.
\end{rmk}
\begin{rmk}[Coupling $K$ and $G$]
The coupling of elastic shear and volumetric moduli may lead to energy dissipation under cyclic loading. However, non-conservation is not an issue for monotonic loading. In addition, the definitions for $K$ and $G$ in~\eqref{eq:bulk_modulus} and~\eqref{eq:shear_modulus} are widely used in practice; thus, we adopt these expressions. 
\end{rmk}

We also assume that the porosity evolution during loading satisfies the following state equation in rate form,
\begin{equation} \label{eq:porosity_state_equation}
\dot{\phi} = -\psi\, \dot{\upvarepsilon}_v,
\end{equation}
where $\psi$ represents the porosity degradation rate during volumetric deformation (see~\ref{fig:porosity_def_profile}).

The mathematical description of an elastoplastic constitutive model involves three main components~\cite{ Simo1998}:
\begin{enumerate}
\item \textit{Yield Function:} describes the location of points where the material develops irrecoverable deformation.
\item \textit{Flow Rule:} characterizes the evolution of irrecoverable deformations
\item \textit{Hardening Law:} typifies the evolution of the yield function throughout the plastic evolution.
\end{enumerate}

\begin{figure}[h!] 
\centering
\begin{subfigure}[b]{0.4\textwidth}
\centering
\includegraphics[scale=0.3]{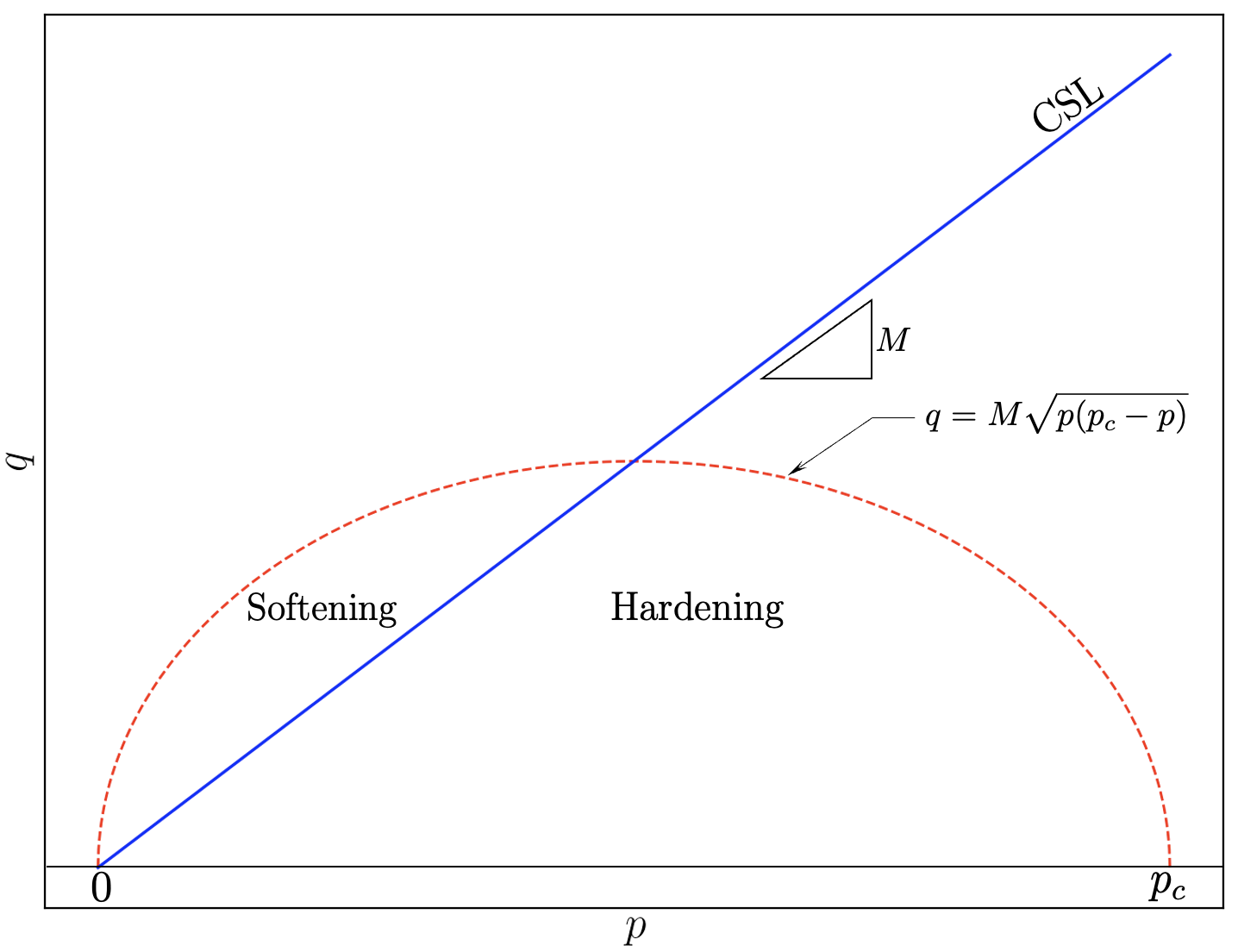}
\caption{$p-q$ projection representation.}
\end{subfigure}
\hspace{1.5cm}
\begin{subfigure}[b]{0.4\textwidth} 	 
\centering
\includegraphics[scale=0.4]{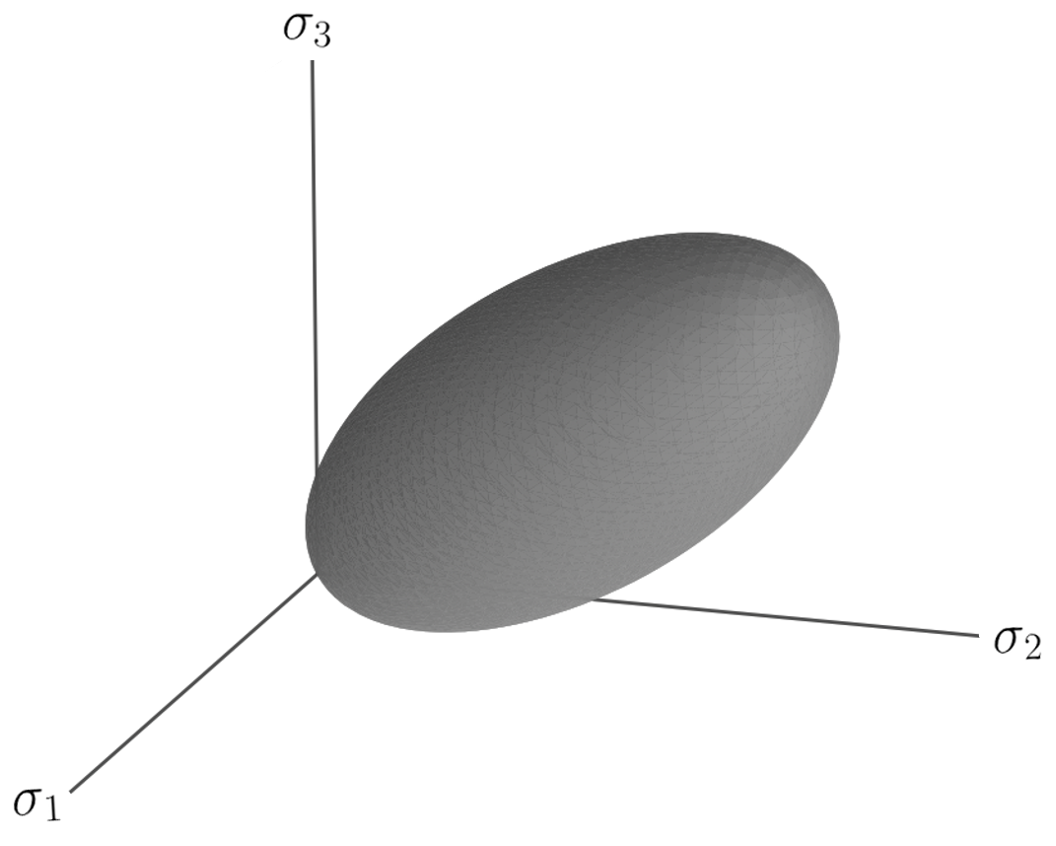}
\caption{$\upsigma_1,\, \upsigma_2,\, \upsigma_3$ representation.}
\end{subfigure}
\caption{Cam-Clay Yield Surface represented in different spaces.}
\label{fig1:cam_clay_representation}
\end{figure}

\subsection{Yield Function: Modified Cam-Clay Model}

Cam-Clay models~\cite{ Roscoe1963, Roscoe1968} are widely used for plasticity characterizations of the stress-strain response of cohesive-frictional materials subject to three-dimensional stress states~\cite{ Diarra2017, Cier2022}. These simple models can realistically represent the compaction and dilation responses of porous materials~\cite{ Borja1990, Borja1991}. Cam-Clay models capture the typical pressure sensitivity and hardening  of cohesive-frictional materials, requiring few parameters that standard laboratory testing procedures can characterize (see~\cite{ Roscoe1963, Roscoe1968}).
Recently, experimental data on limestone rocks that exhibit compaction and dilation during laboratory triaxial testing was reported in~\cite{ SuarezRivera2023}. The same material response was observed in Vaca Muerta mudstone rock during triaxial testing reported in this work. Thus, we select a constitutive model that captures this nonlinear response for practical geomechanical applications; see Figure~\ref{fig1:cam_clay_representation} for a schematic representation of the MCC yield criterion.

We adopt a \textit{Modified Cam-Clay Model} defined in the $p-q$ plane as,
\begin{align}
    \label{eq:cam_clay_Ff}
    F_f(\stress) : \, \R^{3 \times 3} \mapsto \R \quad \text{such that} \quad
    F_f (\stress) = \dfrac{q^2}{M^2} + p(p-p_c),
\end{align}
where $p$ and $q$ represent the hydrostatic pressure and the deviatoric part of the effective stress tensor given as,
\begin{align}
p &:= \dfrac{1}{3} \textrm{tr}(\stress) = \dfrac{1}{3} \stress : \1,
\label{eq:p}
\\
q &:= \sqrt{\dfrac{3}{2}} \|\s\|,
\label{eq:q}
\end{align}
and $M,\, p_c$ define the Critical State Line (CSL) slope and the hardening parameter, respectively. Alternatively, we express~\eqref{eq:cam_clay_Ff} in terms of the stress invariants as,
\begin{equation}\label{eq:cam_clay_Ff_inv}
F_f(\stress) = \dfrac{3}{2M}\, I_2(\stress) + \eta\,I^2_1(\stress) - \dfrac{1}{3} \, p_c\,I_1(\stress),
\end{equation}
where $$\eta = \dfrac{2M-9}{18M}.$$ Figures~\ref{fig1:cam_clay_representation}(a) and~(b) represent~\eqref{eq:cam_clay_Ff} and~\eqref{eq:cam_clay_Ff_inv}, respectively.

\begin{rmk}[Hardening parameter] 
The state variable $p_c$ is a hardening parameter that describes the yield surface's evolution as the hydrostatic pressure grows.
\end{rmk}
\vs
\subsection{Flow Rule}
We adopt an \textit{associative flow rule}, which expresses the rate of plastic deformation as,
\begin{equation}\label{eq:cam_clay_associative_flow_rule}
\dot{\strain}^p = \dot{\lambda}\dfrac{\partial F}{\partial \stress},
\end{equation}
where the flow direction is given by (see \ref{sec:modified_cam_clay_derivatives}),
\begin{equation}
\dfrac{\partial F_f (\stress)}{\partial \stress} = \dfrac{1}{3}(2p-p_c)\1+\sqrt{\dfrac{3}{2}}\dfrac{2q}{M^2} \n,
\end{equation}
in which $\n = \dfrac{\s}{\|\s\|}$ and $\dot{\lambda} > 0$ is the plastic multiplier.

\begin{rmk}[Non-associativity]
Although associative flow rules are used to characterize the plastic flow of granular materials, non-associative flow rules on the volumetric component of $~\dot{\strain}^p$ could be more suitable for certain type of rocks. Nevertheless, assuming associative flow to capture the plastic flow in Vaca Muerta reproduce the experimental observation without adding any additional heuristics.
\end{rmk}
\subsection{Hardening Law}
We define the hardening law in rate form as,
\begin{equation}\label{eq:hardening_law}
\dot{p}_c = \chi\, p_c \,\dot{\strain}^p_v,
\end{equation}
where $\dot{\strain}^p_v = \dot{\strain}^p : \1$ is the volumetric component of the plastic strain rate tensor and $\chi$ is given by
\begin{equation}\label{eq:multiplier_to_hard_law}
\chi = \left[(1-\phi)(\gamma - \kappa)\right]^{-1},
\end{equation}
In~\eqref{eq:multiplier_to_hard_law} $\gamma$, $\kappa$ are material parameters typically obtained in  hydrostatic cycling tests.

\begin{rmk}[On the $\gamma$ and $\kappa$ definitions]
$\gamma$ and $\kappa$ are material parameters representing compaction and bulk volumetric recovery measured in the hydrostatic cycling test. The required number of hydrostatic cycles applied during testing induces a linear trend from loading and unloading cycles. At this point, the sample is compacted at a certain level where artifacts from sample extraction for the core are mitigated. 
\end{rmk}
%

\begin{figure}[h!]
\centering
\includegraphics[scale=0.4]{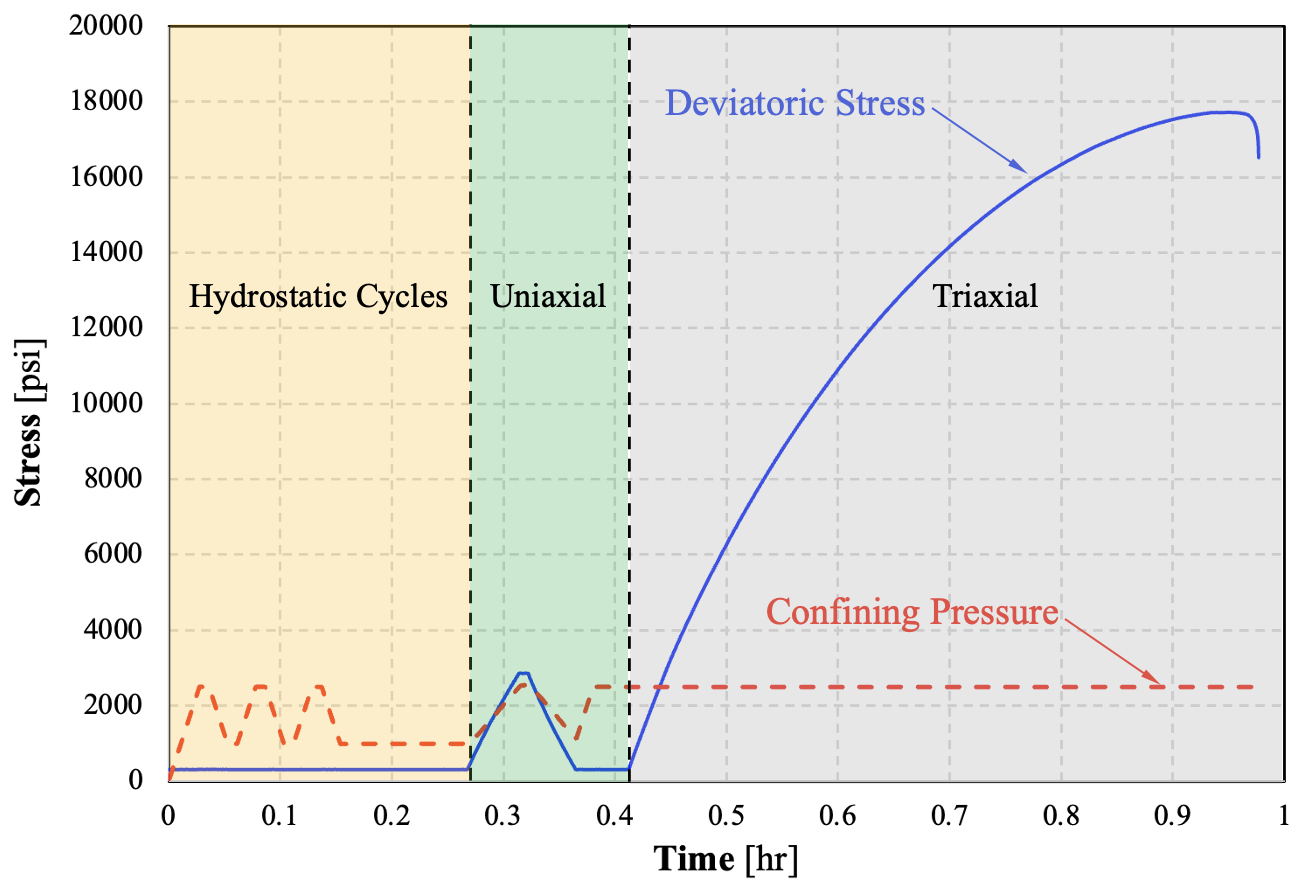}
\caption{General laboratory test program for mudstones and limestones.}
\label{fig:lab_procedure}
\end{figure}

\section{Material parameters estimation from laboratory tests}

\begin{table}[h!]
\centering
\begin{tabular}{| c | c | c | c | c |}
 \hline
 \textbf{Properties\textbackslash ID} & \textbf{2-44v} & \textbf{3-14-2v} & \textbf{4-31-1v} & \textbf{4-41v}\\
 \hline
 Diameter [in]   & 0.999 & 1.004 & 0.999 & 1.000\\
 Length [in] & 2.007 & 2.019 & 1.989 & 1.994 \\
 Volume [$\text{in}^3$] & 1.573 & 1.598 & 1.559 & 1.566\\
 Mass [gr] & 62.242 & 62.190 & 62.758 & 59.673\\
 Porosity [\%] & 12.295 & 11.035 & 9.841 & 12.300\\
 Pore Volume [$\text{in}^3$] & 0.193 & 0.176 & 0.153 & 0.192\\
 Solid Volume [$\text{in}^3$] & 1.379 & 1.422 & 1.405 & 1.373\\
 \hline
\end{tabular}
 \caption{Vaca Muerta sample properties as received.}
 \label{tab:vm_sample_properties}
\end{table}

Triaxial compression tests for vertically oriented samples were conducted at W. D. Von Gonten Engineering laboratory facilities. Each of the samples was tested in a modern servo-controlled triaxial loading system in accordance with ASTM testing standards. Laboratory tests were performed under room temperature and drained conditions using Teflon isolating jacket, cantilever radial displacement gauge, and four equally spaced axial displacement transducers (LVDT). Before conducting the triaxial test, three hydrostatic cycles and uniaxial strain compression cycles were performed before the triaxial test (see Figure~\ref{fig:lab_procedure}). The initial cycles of hydrostatic compression consolidate each sample and minimize the effect of coring-induced microcracks and other artifacts. The uniaxial-strain test seeks to evaluate the evolution of elastic properties as a function of the confining pressure. Before the failure, a triaxial compression test was used to investigate the elastic properties and the post-elastic stress-strain behavior, including dilatant and compactive plasticity. We test four Vaca Muerta mudstone samples, and Table~\ref{tab:vm_sample_properties} summarizes their properties.

\begin{figure}[h!]
\centering
\includegraphics[scale=0.25]{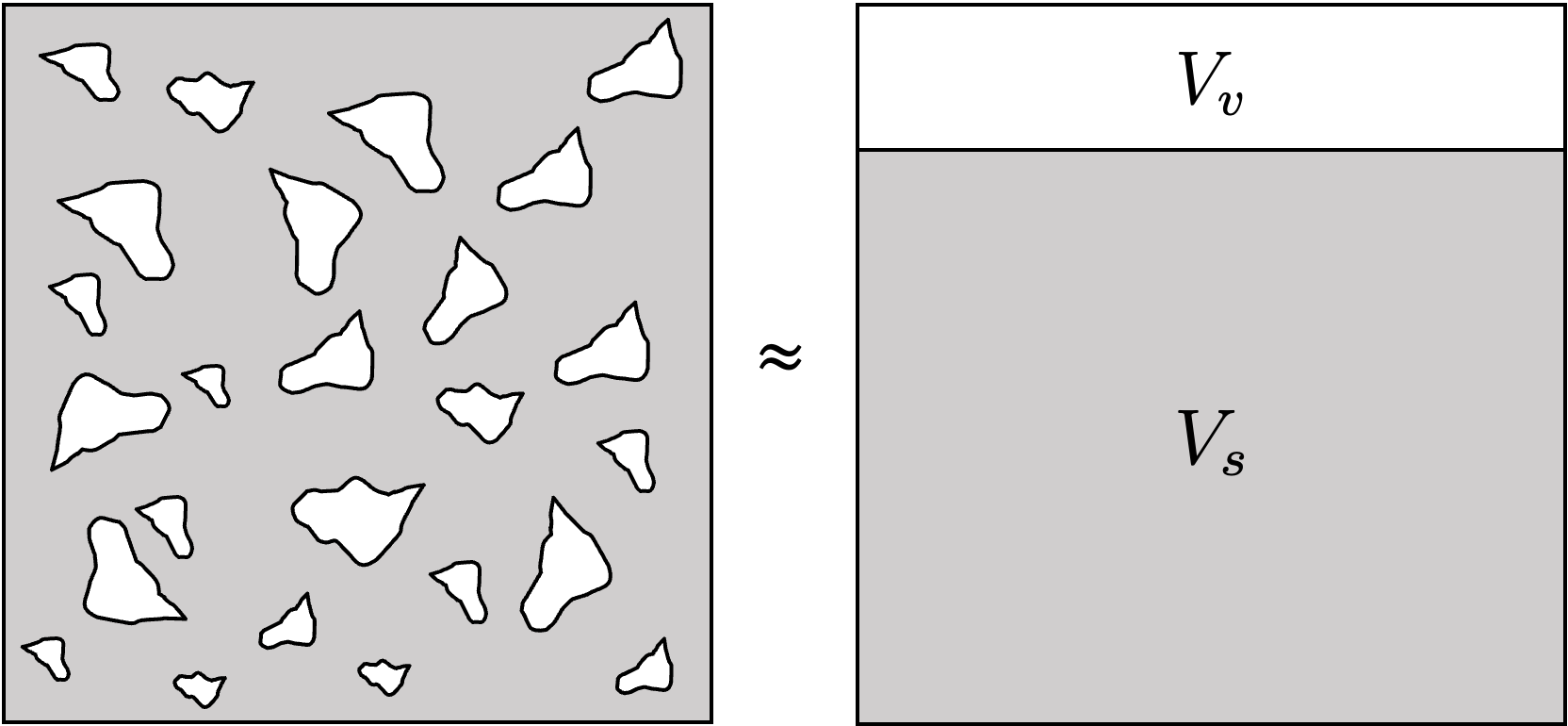}
\caption{Additive decomposition of a porous medium volume.}
\label{fig:porous_medium_decomp}
\end{figure}

We assume that the total volume of the porous medium admits an additive decomposition  $V = V_v + V_s$ (see~\cite{ Coussy2004}) where $V_v$ and $V_s$ are the pore and solid volumes, respectively (see Figure~\ref{fig:porous_medium_decomp}); therefore, $V_v$ and $V_s$ are,
\begin{equation}\label{eq:pore_volume}
V_v = \phi V,
\end{equation}
\begin{equation}\label{eq:solid_volume}
V_s = (1-\phi) V.
\end{equation}
\begin{assmp}[Matrix volume conservation]\label{assmp:matrix_incompr}
At any load step, the matrix volume remains constant; thus,
$$
\dot{V}_s = 0, \Leftrightarrow V_s = \text{constant}.
$$
This assumption is relevant as we are not destroying solid volume at laboratory conditions.
\end{assmp}

\begin{figure}[h!] 
\centering
\begin{subfigure}[b]{0.45\textwidth}
\centering
\includegraphics[scale=0.23]{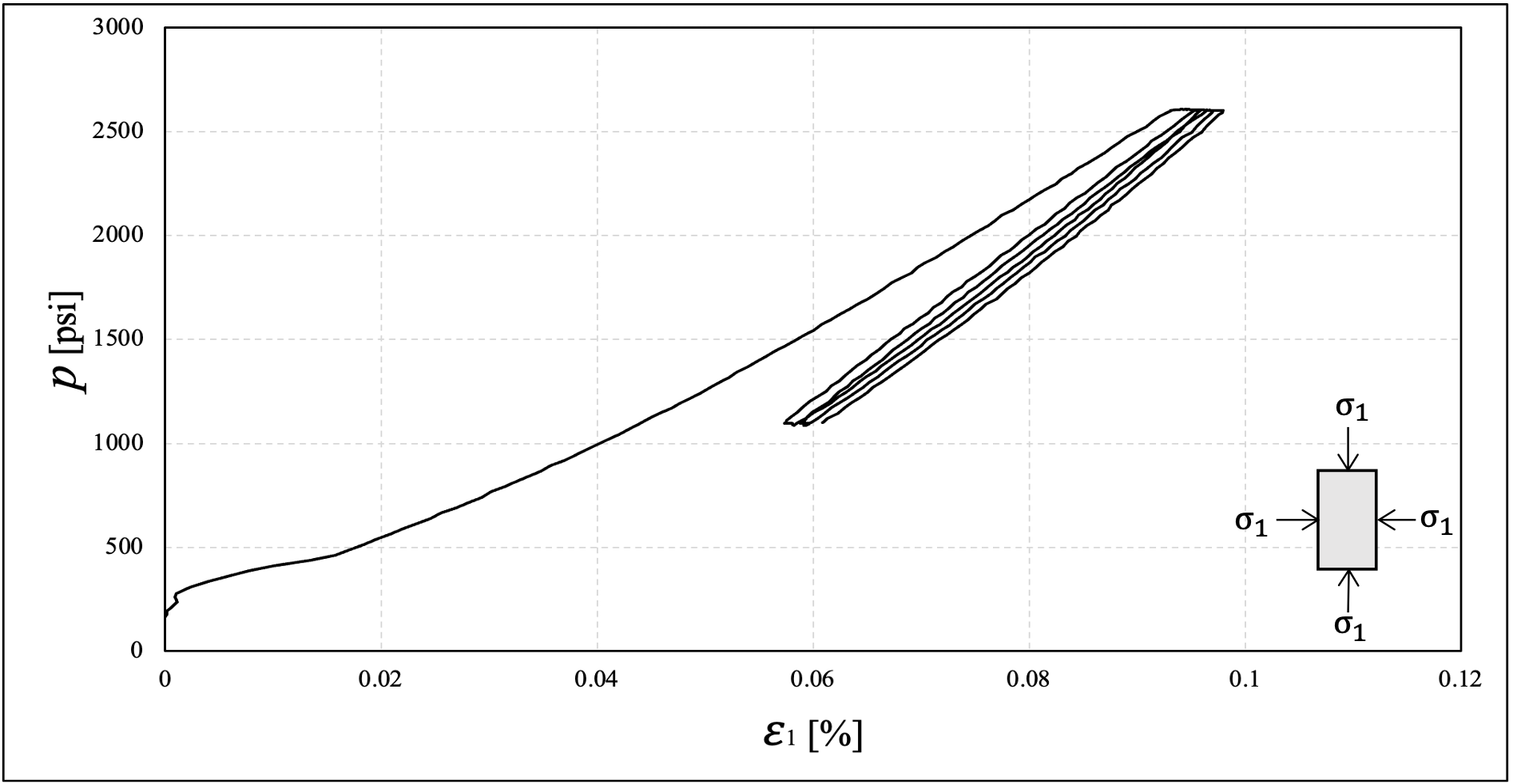}
\caption{$p$ vs $\upvarepsilon_1$.}
\end{subfigure}
\hspace{1.2cm}
\begin{subfigure}[b]{0.45\textwidth} 	 
\centering
\includegraphics[scale=0.23]{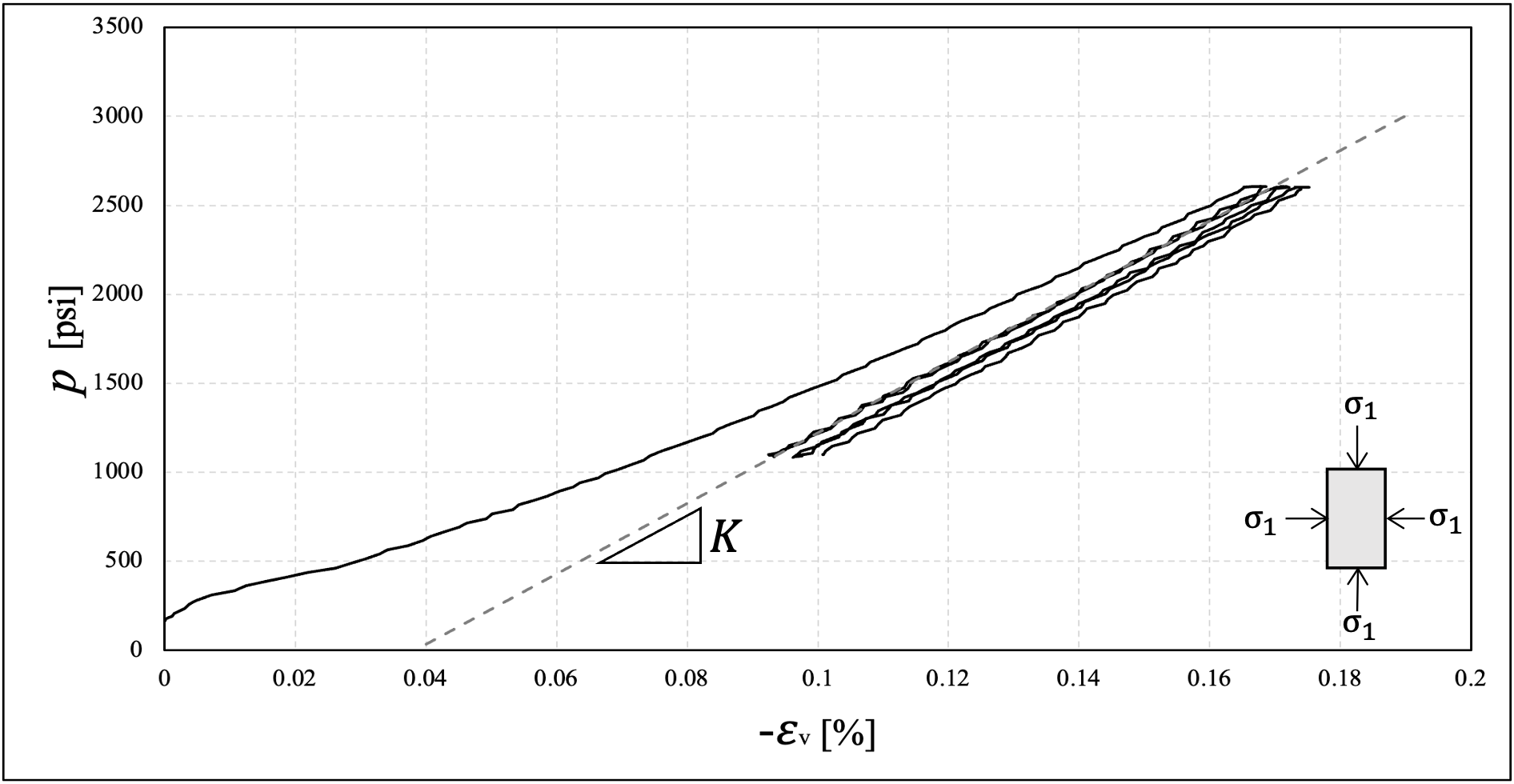}
\caption{$p$ vs $\upvarepsilon_v$.}
\end{subfigure}
\begin{subfigure}[b]{0.45\textwidth} 	 
\bigskip
\centering
\includegraphics[scale=0.23]{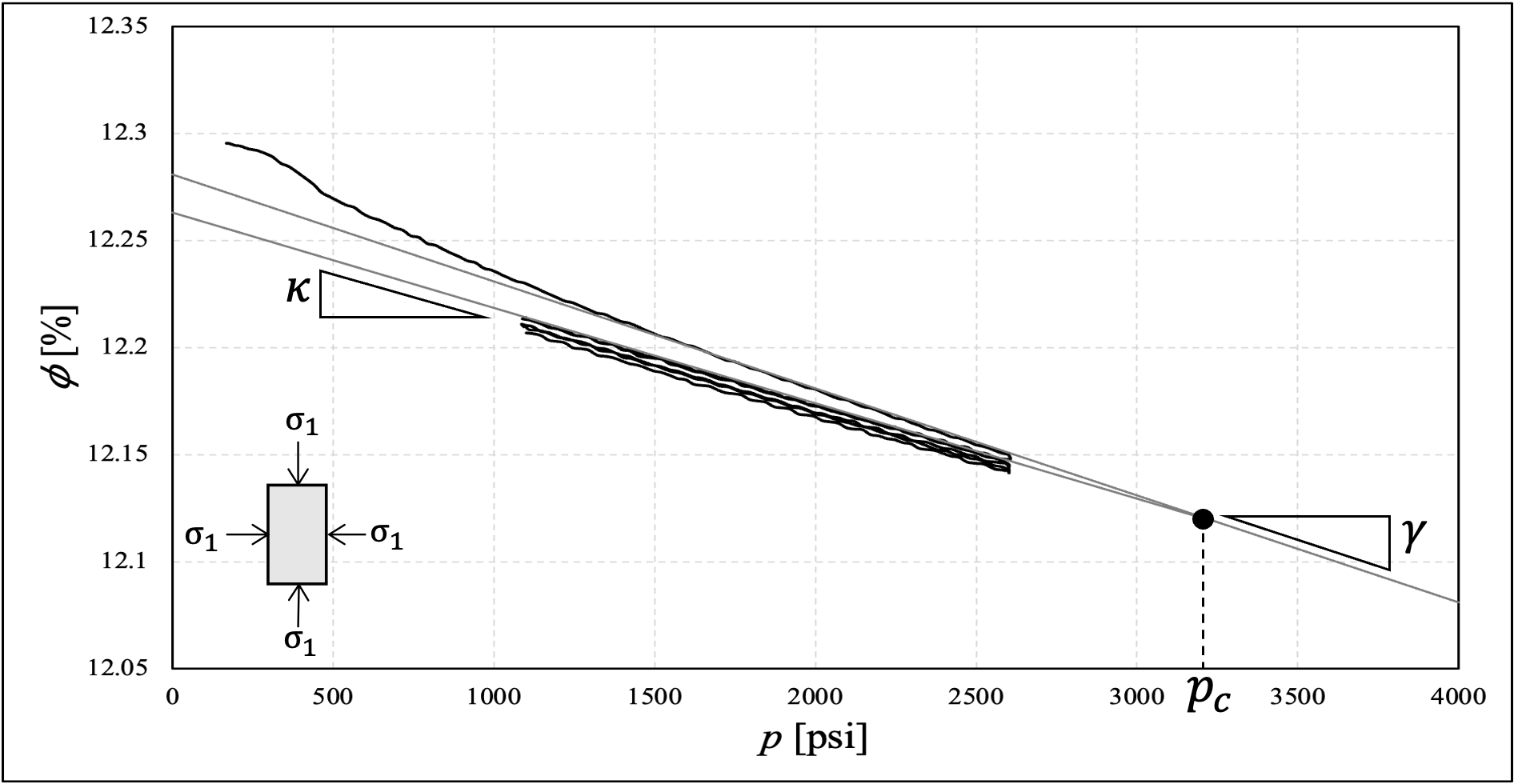}
\caption{$\phi$ vs $p$.}
\end{subfigure}
\caption{2-44v Hydrostatic Test Results.}
\label{fig:2_44v_hydrostat_results}
\end{figure}

\subsection{Material response during hydrostatic cycling}
During the hydrostatic cycling stage, deviatoric stress $q = \upsigma_1 - \upsigma_3$ remains constant and hydrostatic pressure $p = \frac{1}{3}(\upsigma_1 + 2\, \upsigma_3)$ is monotonically increased from 200 psi to 2600 psi. From this test, we determine the initial volumetric modulus during unloading $K$ considering the slope of the unloading curve from $p$ against the $\upvarepsilon_v$ chart. Additionally, $\kappa$ and $\gamma$ material parameters are determined by applying Assumption~\ref{assmp:matrix_incompr} to calculate $\phi$ and analyzing its variation as a function of $p$ and identifying the slopes of the loading and unloading curves. The initial hardening parameter $p^0_c$ is determined from the intersection of the loading and unloading curves in the $\phi-p$ plot.

\begin{figure}[h!] 
\centering
\begin{subfigure}[b]{0.45\textwidth}
\centering
\includegraphics[scale=0.23]{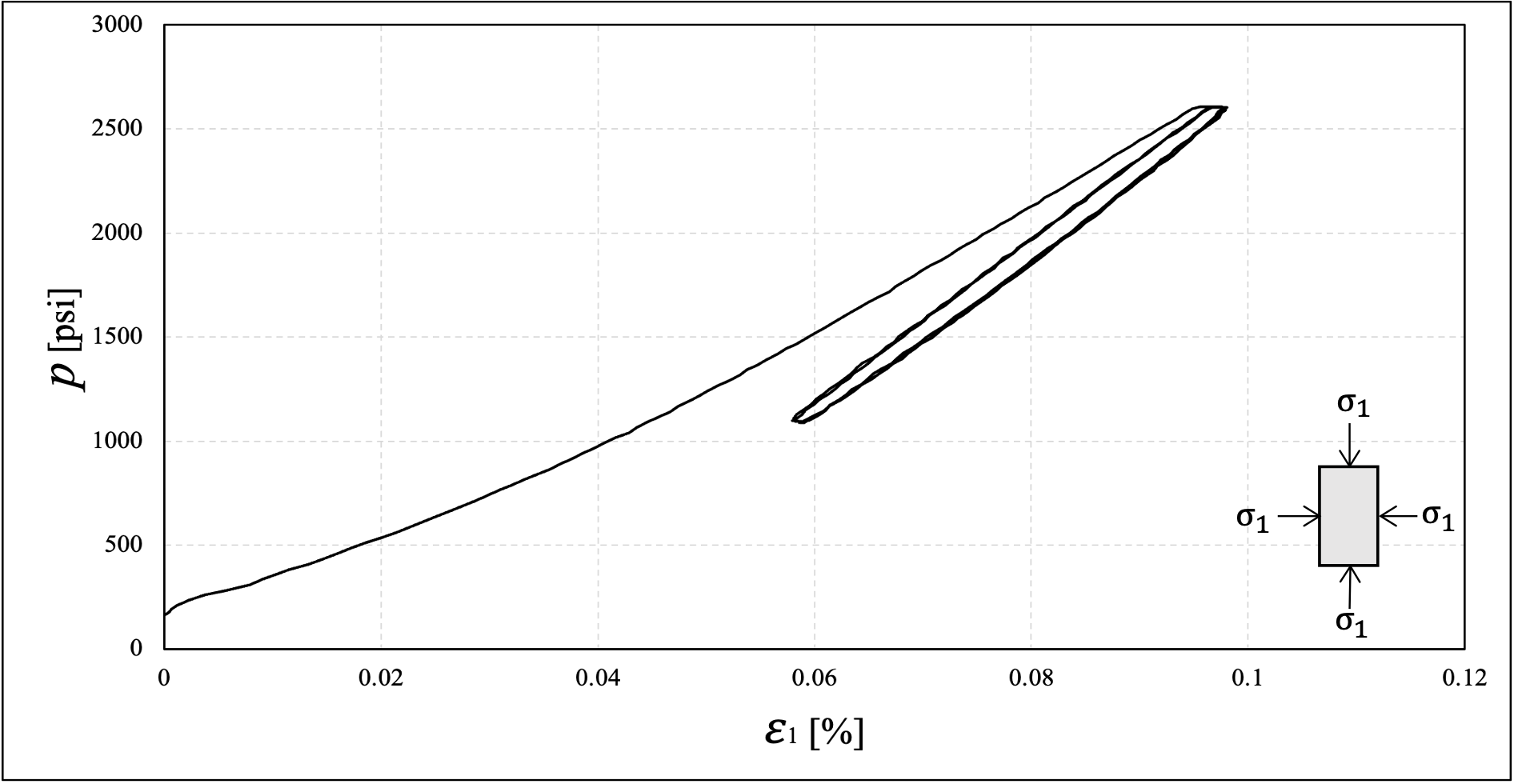}
\caption{$p$ vs $\upvarepsilon_1$.}
\end{subfigure}
\hspace{1.2cm}
\begin{subfigure}[b]{0.45\textwidth} 	 
\centering
\includegraphics[scale=0.23]{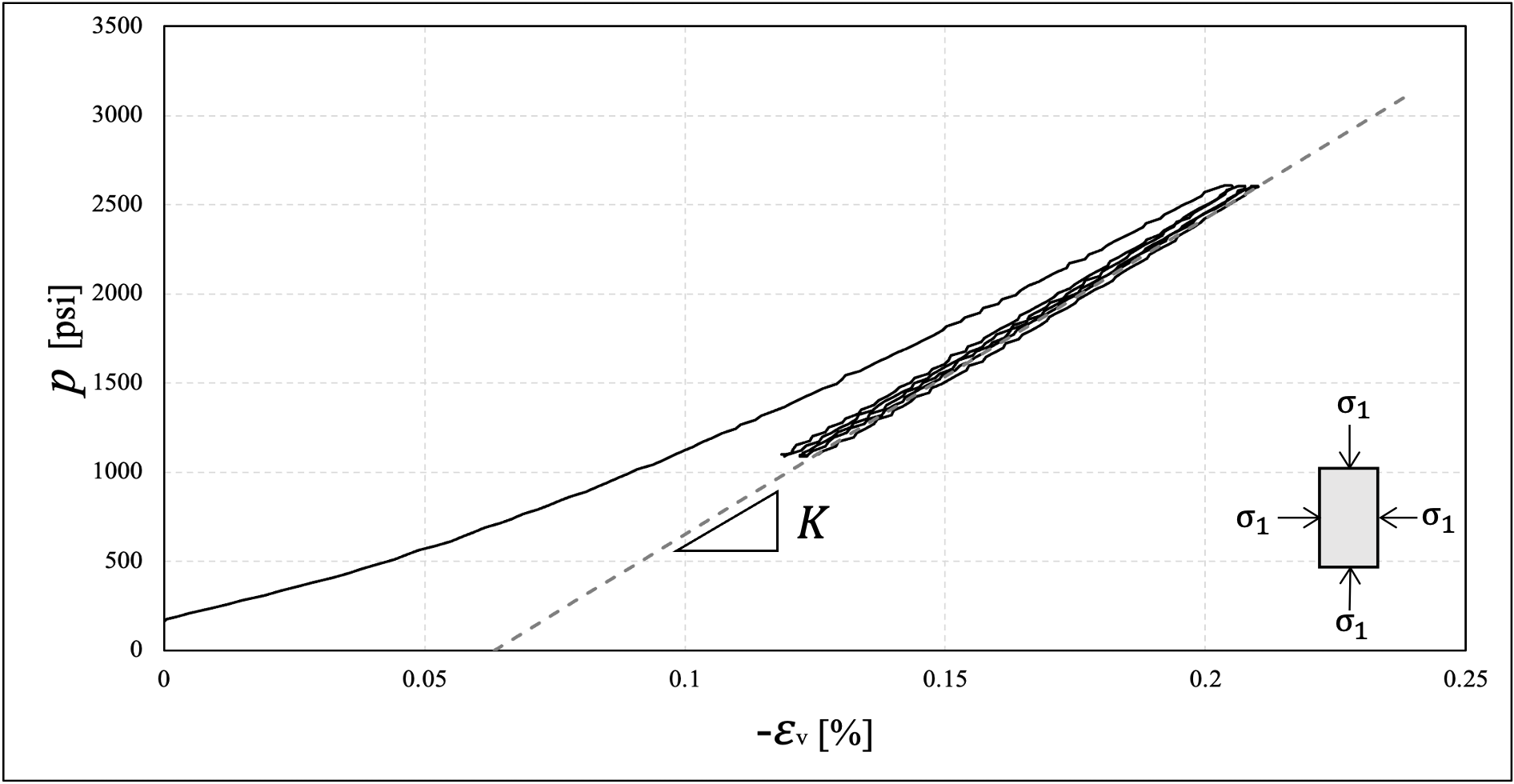}
\caption{$p$ vs $\upvarepsilon_v$.}
\end{subfigure}
\begin{subfigure}[b]{0.45\textwidth} 	 
\bigskip
\centering
\includegraphics[scale=0.23]{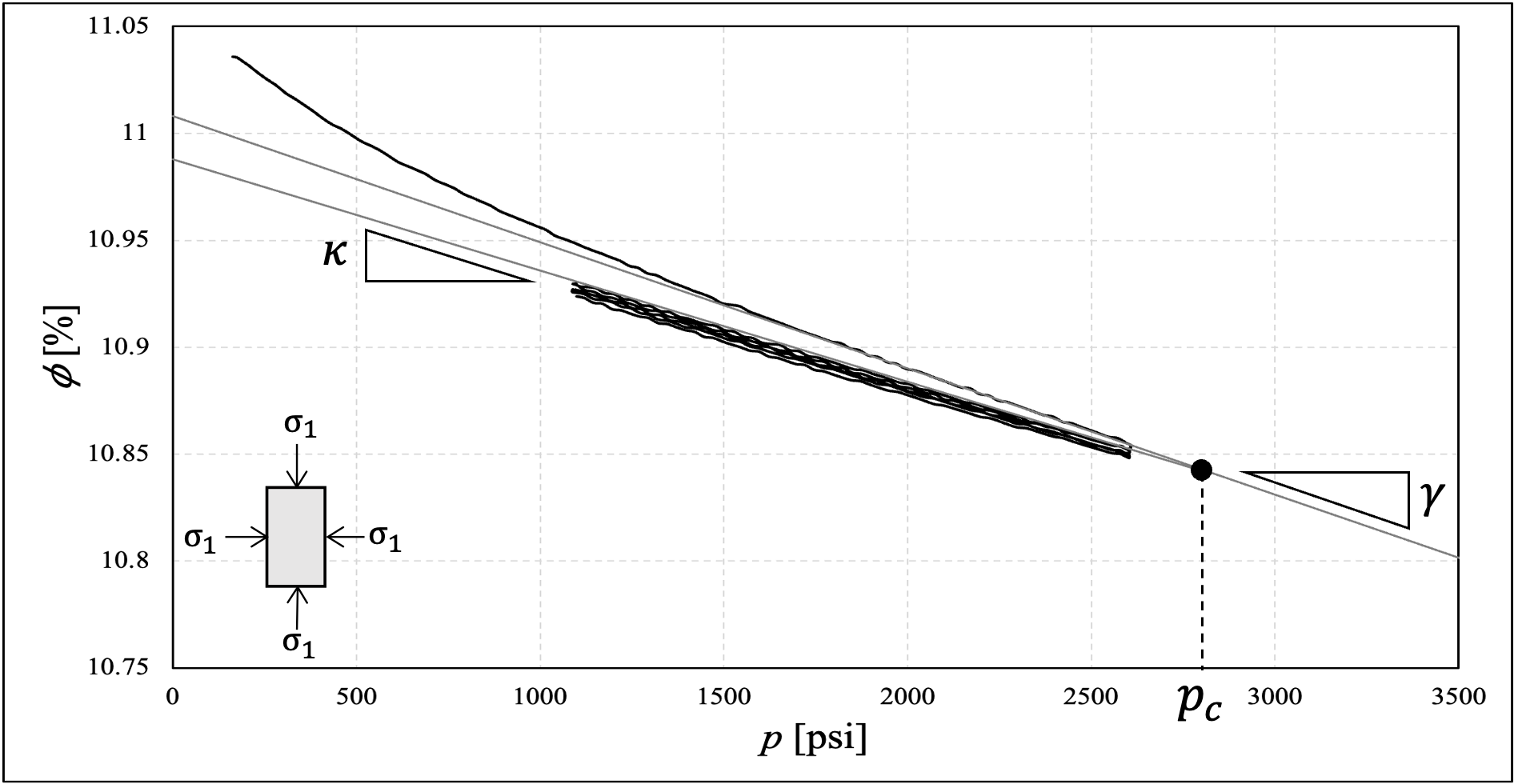}
\caption{$\phi$ vs $p$.}
\end{subfigure}
\caption{3-14-2v Hydrostatic Test Results.}
\label{fig:3_14_2v_hydrostat_results}
\end{figure}
\begin{figure}[h!] 
\centering
\begin{subfigure}[b]{0.45\textwidth}
\centering
\includegraphics[scale=0.23]{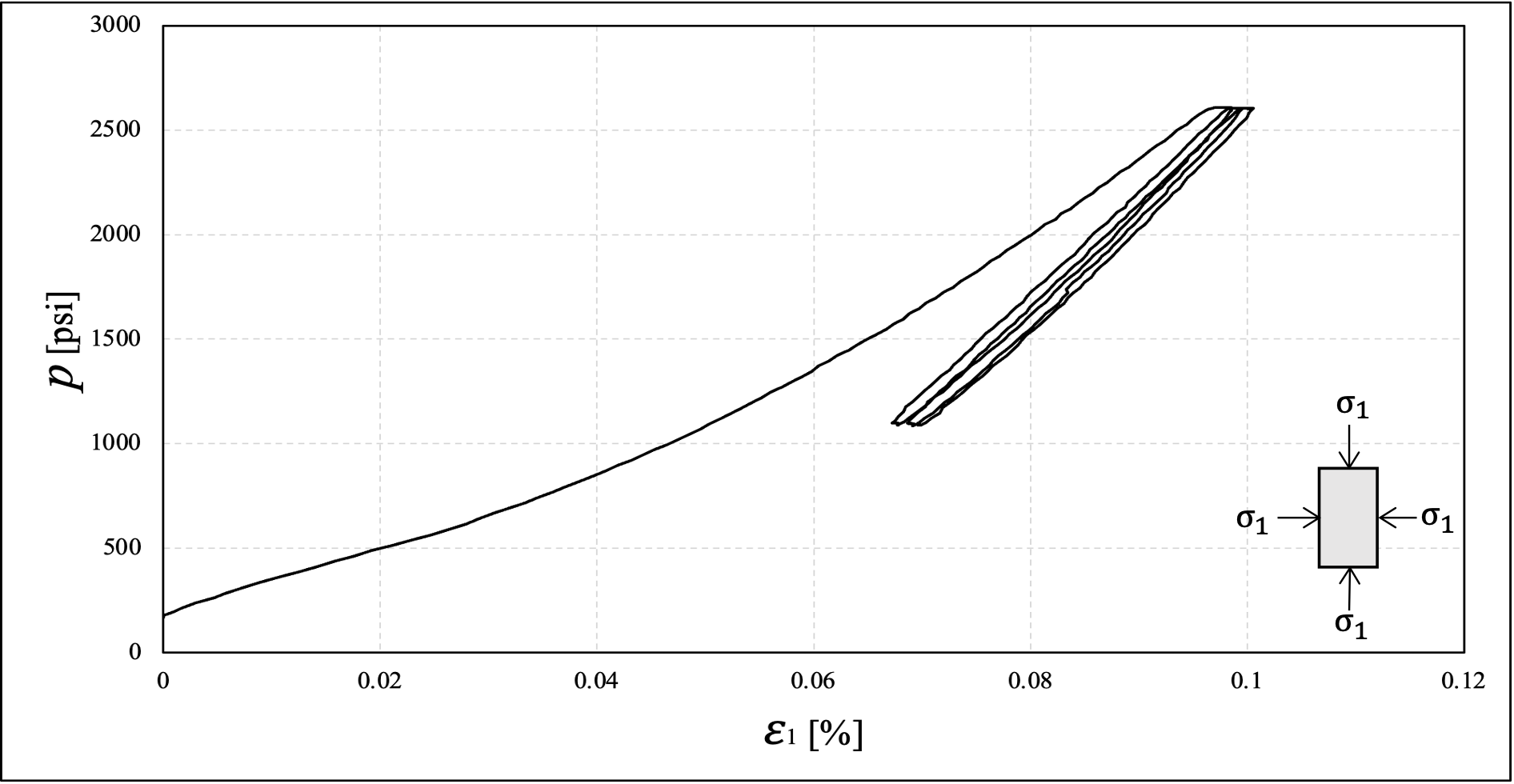}
\caption{$p$ vs $\upvarepsilon_1$.}
\end{subfigure}
\hspace{1.2cm}
\begin{subfigure}[b]{0.45\textwidth} 	 
\centering
\includegraphics[scale=0.23]{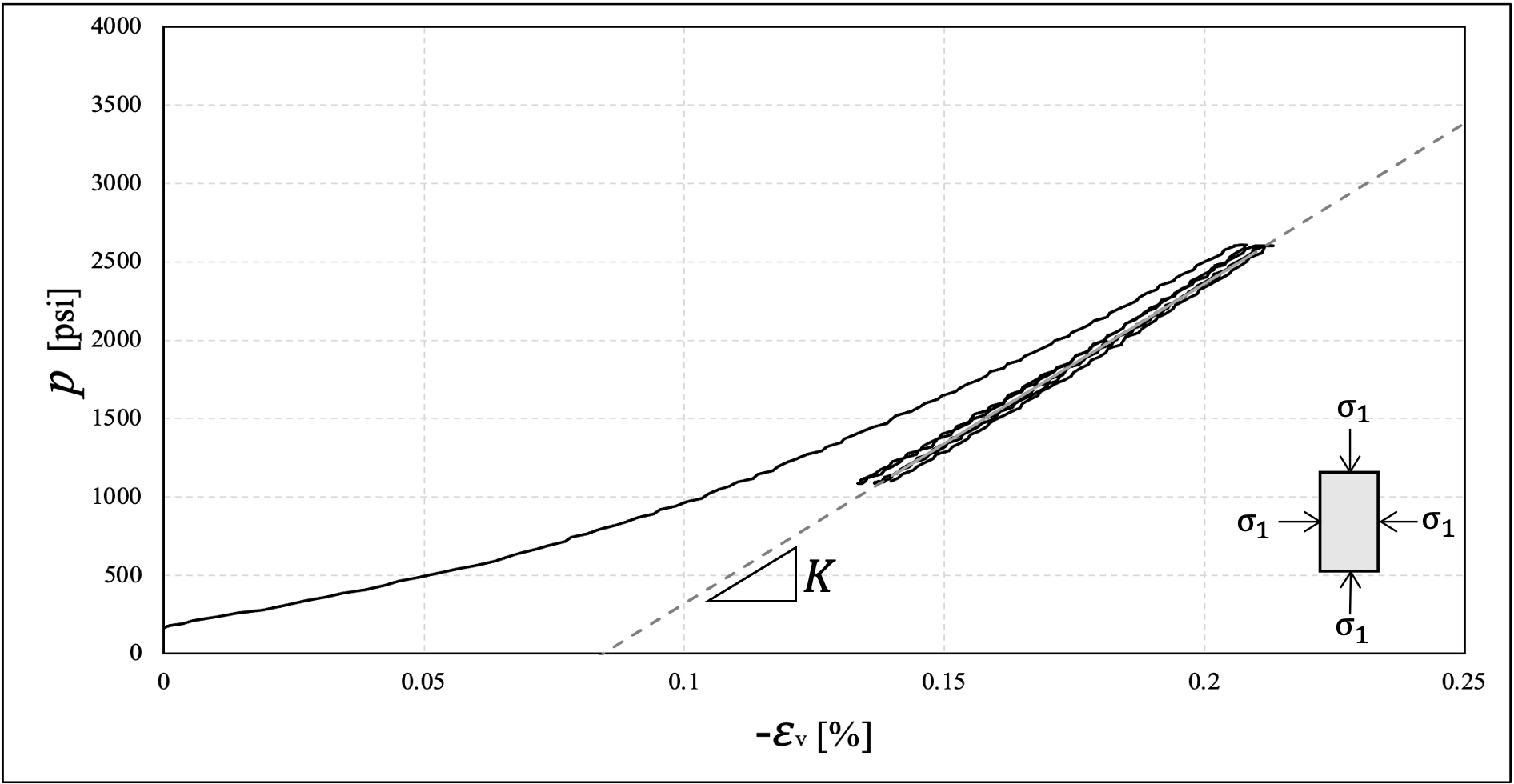}
\caption{$p$ vs $\upvarepsilon_v$.}
\end{subfigure}
\begin{subfigure}[b]{0.45\textwidth} 	 
\bigskip
\centering
\includegraphics[scale=0.23]{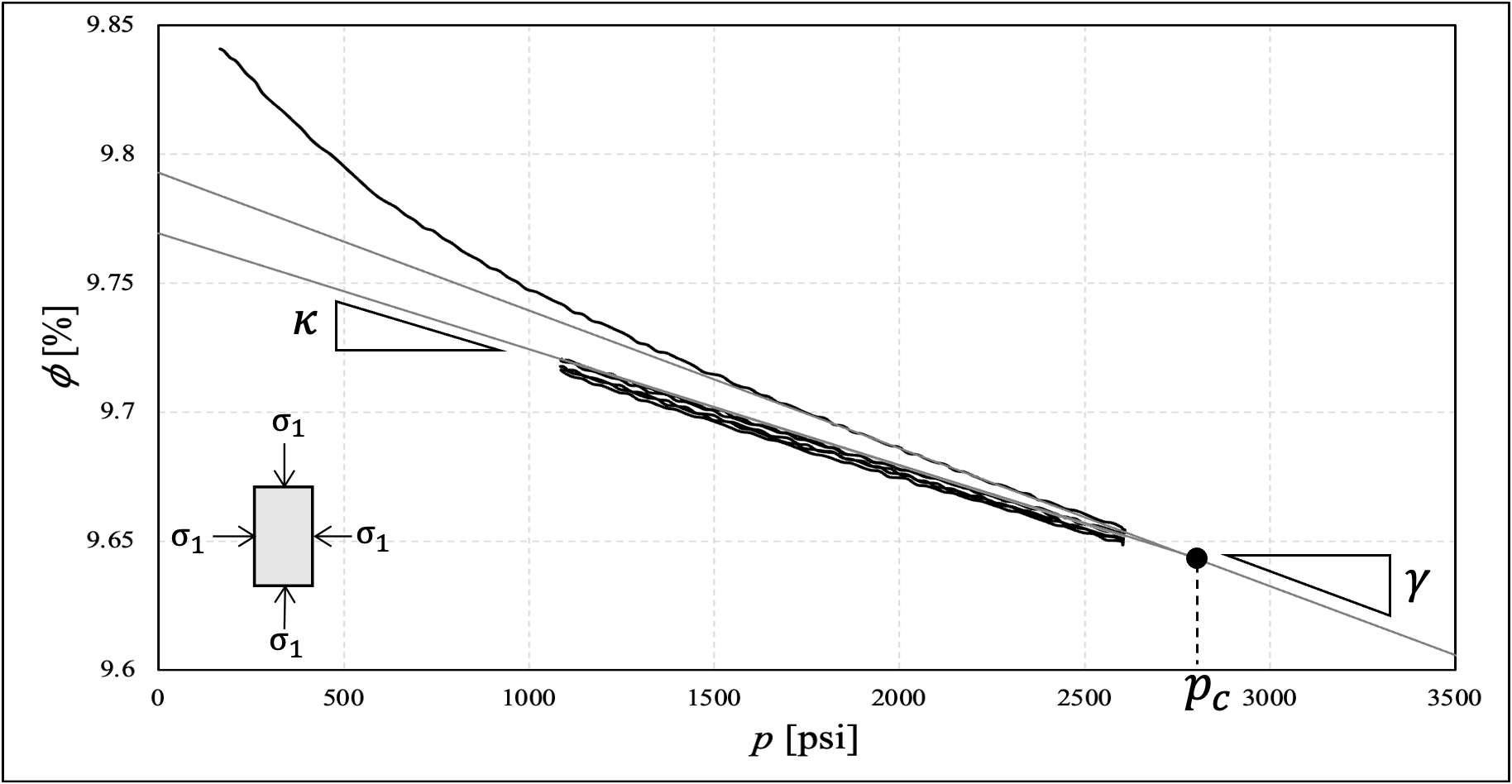}
\caption{$\phi$ vs $p$.}
\end{subfigure}
\caption{4-31-1v Hydrostatic Test Results.}
\label{fig:4_31_1v_hydrostat_results}
\end{figure}
\begin{figure}[h!] 
\centering
\begin{subfigure}[b]{0.45\textwidth}
\centering
\includegraphics[scale=0.23]{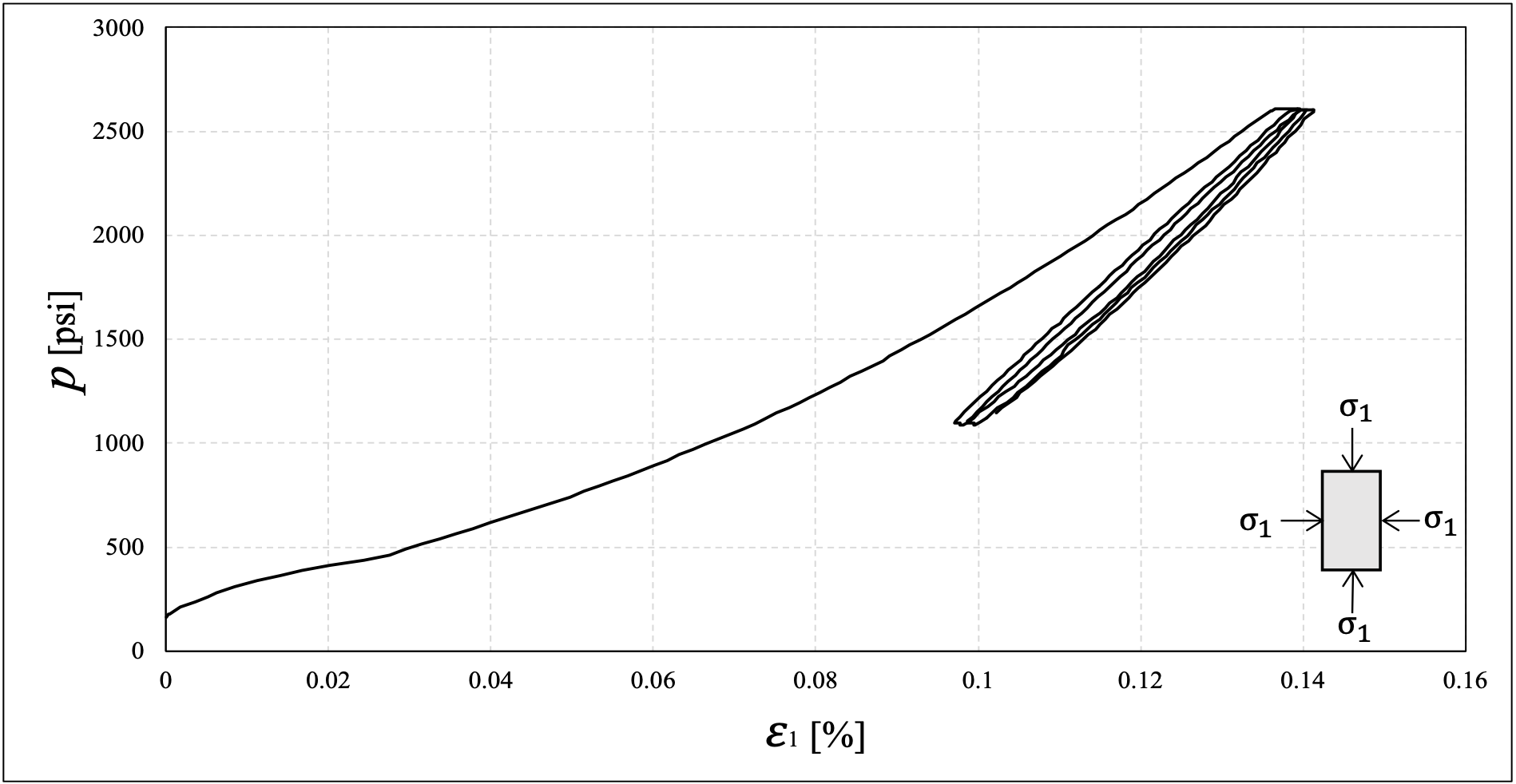}
\caption{$p$ vs $\upvarepsilon_1$.}
\end{subfigure}
\hspace{1.2cm}
\begin{subfigure}[b]{0.45\textwidth} 	 
\centering
\includegraphics[scale=0.23]{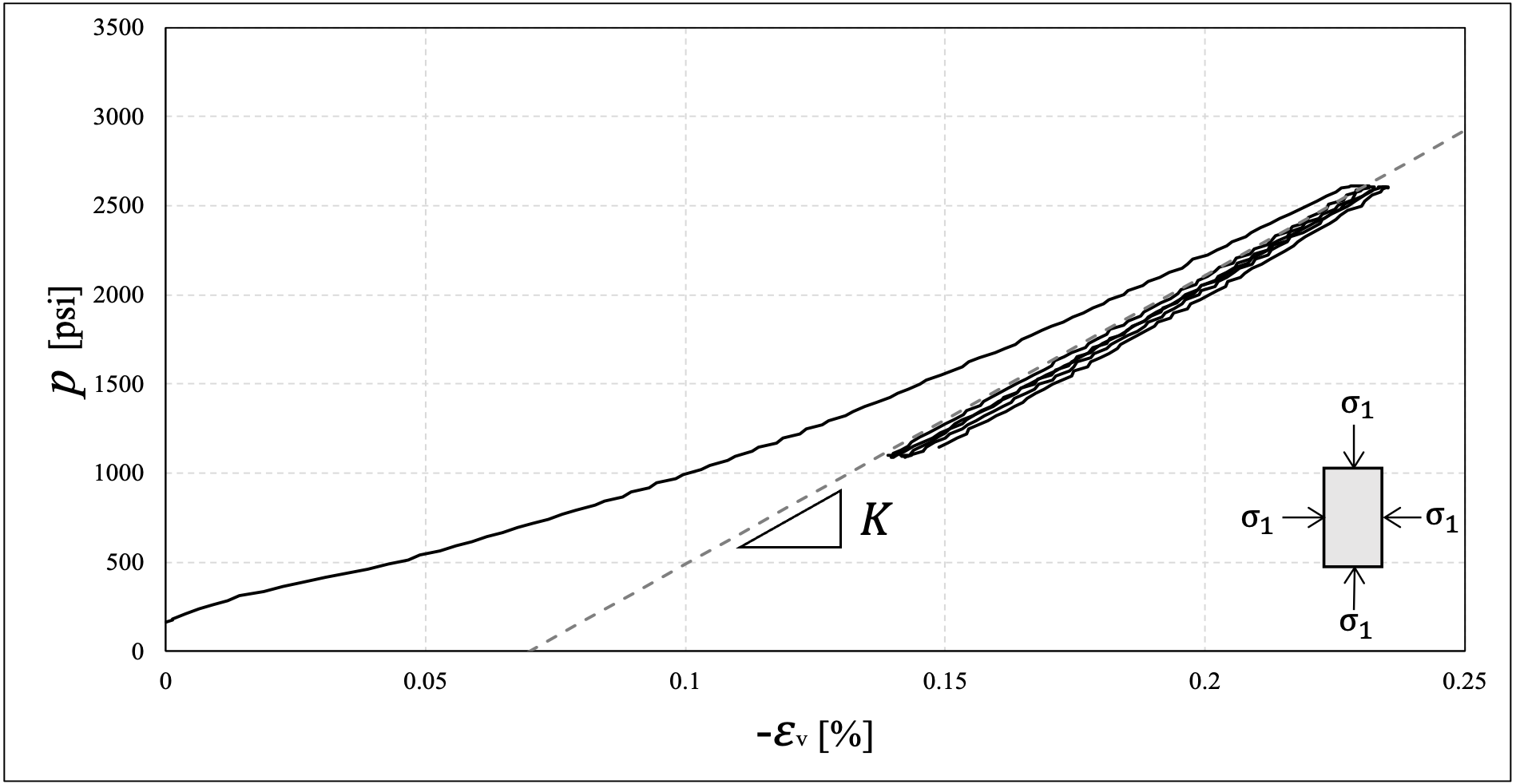}
\caption{$p$ vs $\upvarepsilon_v$.}
\end{subfigure}
\begin{subfigure}[b]{0.45\textwidth} 	 
\bigskip
\centering
\includegraphics[scale=0.23]{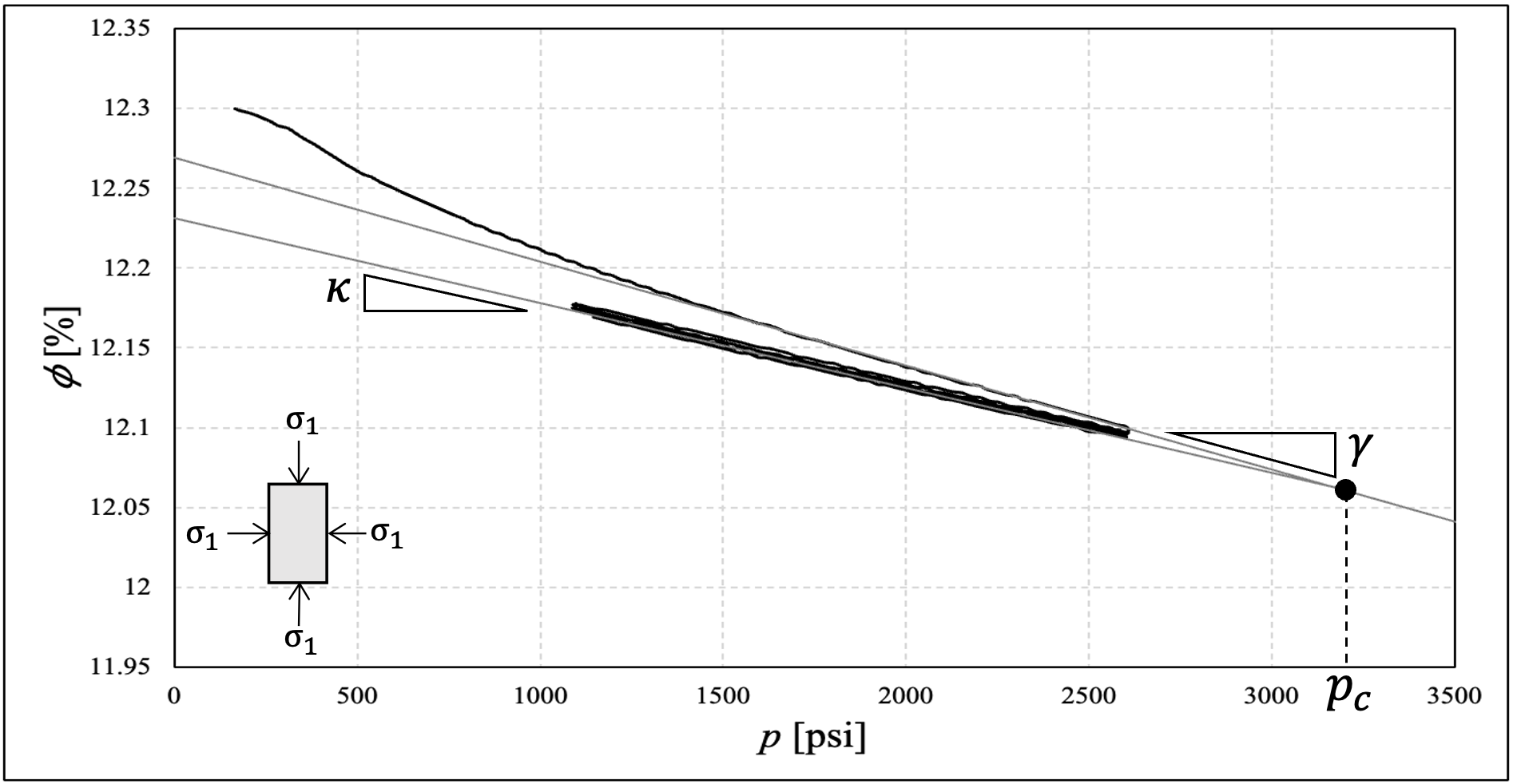}
\caption{$\phi$ vs $p$.}
\end{subfigure}
\caption{4-41v Hydrostatic Test Results.}
\label{fig:4_41v_hydrostat_results}
\end{figure}

\begin{table}[h!]
\centering
\begin{tabular}{| c | c | c | c | c |}
 \hline
 \textbf{Parameter} & \textbf{2-44v} & \textbf{3-14-2v} & \textbf{4-31-1v} & \textbf{4-41v}\\
 \hline
 $K [\text{psi}]$  & 1981000 & 1773400 & 2044400 & 1624100\\
 $p^0_c [\text{psi}]$ & 3200 & 2800 & 2800 & 3200\\
 $\gamma$   & 2.30E-03 & 2.70E-03 & 2.20E-03 & 2.50E-03\\
 $\kappa$ & 1.23E-03 & 1.45E-03 & 1.56E-03 & 1.67E-03 \\
 \hline
\end{tabular}
 \caption{Material parameters for each Vaca Muerta sample.}
 \label{tab:vm_material_parameters}
\end{table}

Figures~\ref{fig:2_44v_hydrostat_results} to~\ref{fig:4_41v_hydrostat_results} show the material response during hydrostatic cycling for Vaca Muerta samples and the resulting parameters $K,\, \kappa,\, \gamma$ and $p^0_c$. Table~\ref{tab:vm_material_parameters} summarizes the interpreted material parameters from the hydrostatic test.

\begin{rmk}[Hysteresis during unloading]
During unloading/reloading cycles, an insignificant amount of hysteresis develops. We disregard this behavior for parameter interpretation and consider only the first unloading response to define each material parameter and the initial hardening $p^0_c$.
\end{rmk}

\begin{rmk}[Maximum hydrostatic pressure]
The maximum hydrostatic pressure during cycling should be the confinement pressure at triaxial conditions. We subject each sample to 2500 psi of peak hydrostatic pressure for the data we present. Otherwise, the triaxial stress path may induce unwanted isotropic compression with a certain amount of plastic compaction. Therefore, the sample might show a different response at failure.
\end{rmk}
\subsection{Material response during triaxial test}

\begin{figure}[h!] 
\centering
\begin{subfigure}[b]{0.45\textwidth}
\centering
\includegraphics[scale=0.23]{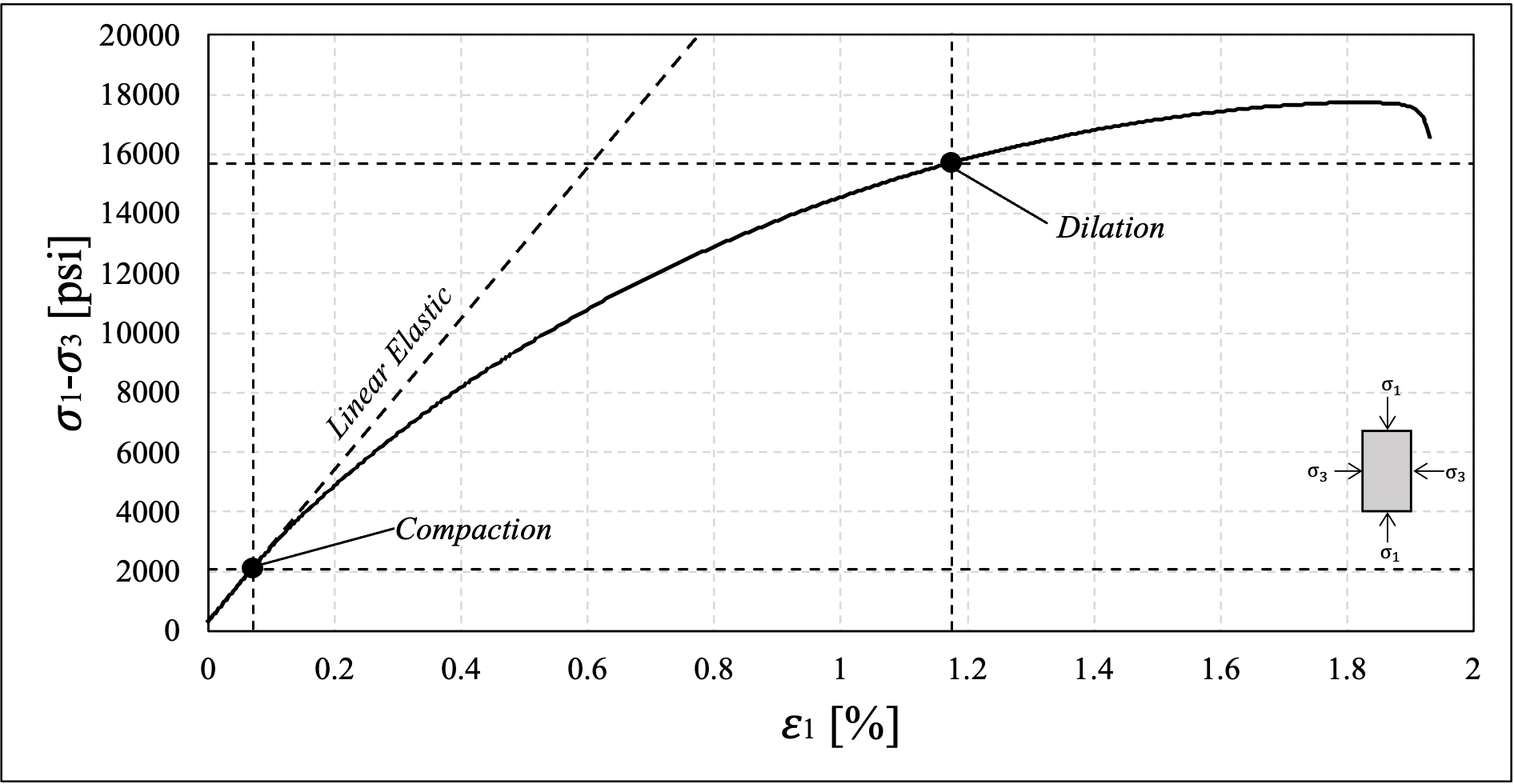}
\caption{$\upsigma_1 - \upsigma_3$ vs $\upvarepsilon_1$.}
\end{subfigure}
\hspace{1.2cm}
\begin{subfigure}[b]{0.45\textwidth} 	 
\centering
\includegraphics[scale=0.23]{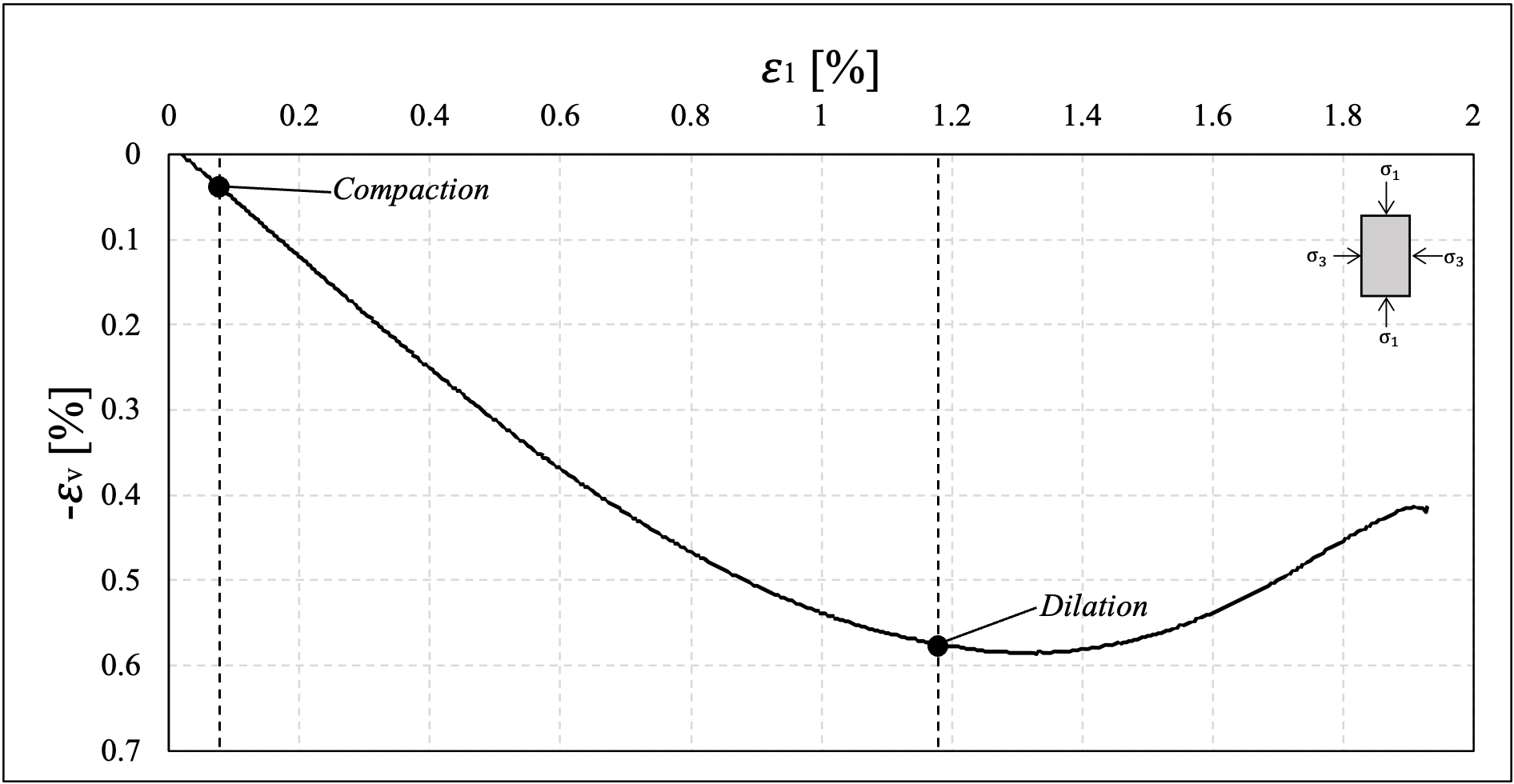}
\caption{$\upvarepsilon_v$ vs $\upvarepsilon_v$.}
\end{subfigure}
\begin{subfigure}[b]{0.45\textwidth} 	 
\bigskip
\centering
\includegraphics[scale=0.23]{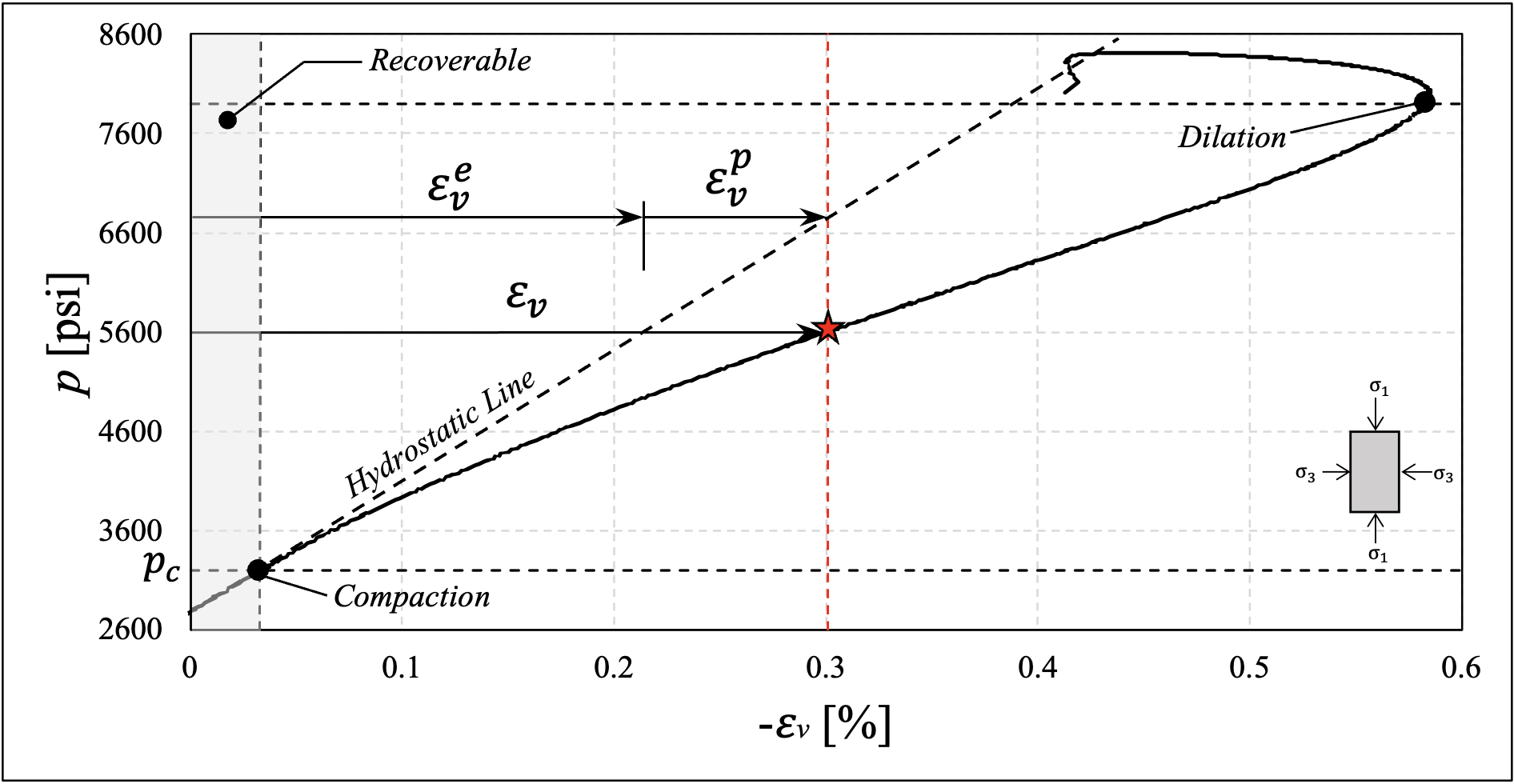}
\caption{$p$ vs $\upvarepsilon_v$.}
\end{subfigure}
\hspace{1.2cm}
\begin{subfigure}[b]{0.45\textwidth} 	 
\bigskip
\centering
\includegraphics[scale=0.23]{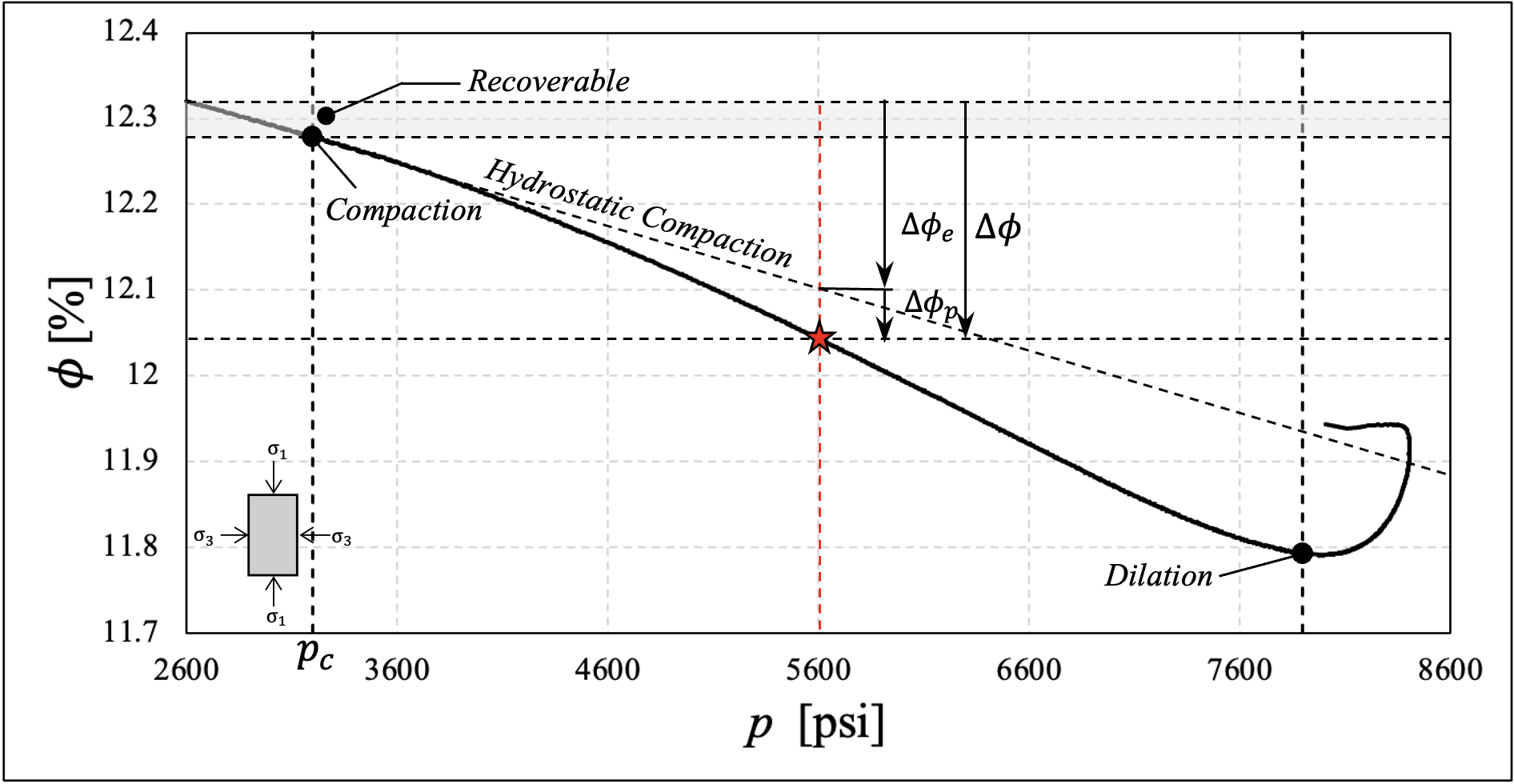}
\caption{$\phi$ vs $p$.}
\end{subfigure}
\caption{2-44v Triaxial Test Results.}
\label{fig:2_44v_triaxial_results}
\end{figure}

\begin{figure}[h!] 
\centering
\begin{subfigure}[b]{0.45\textwidth}
\centering
\includegraphics[scale=0.23]{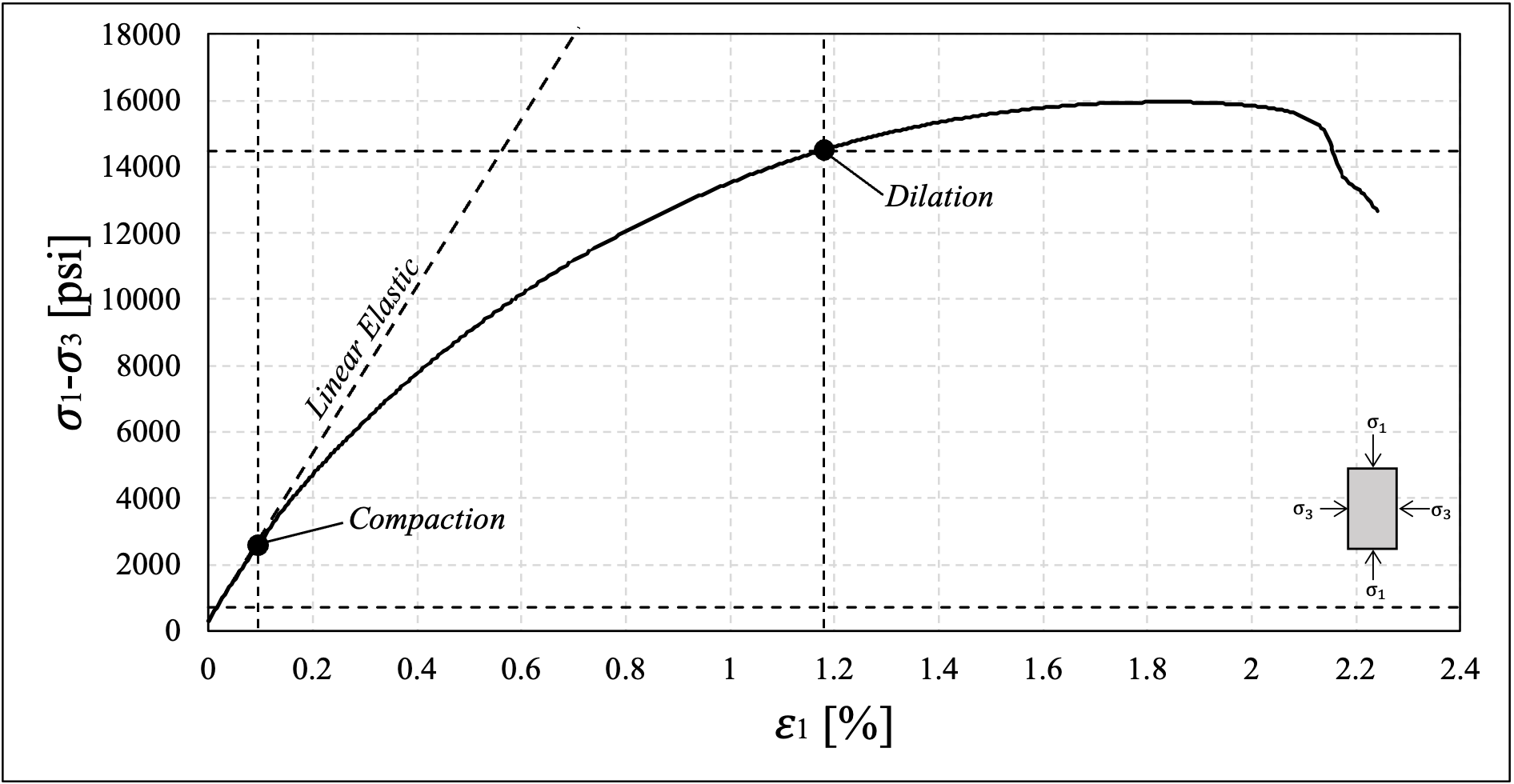}
\caption{$\upsigma_1 - \upsigma_3$ vs $\upvarepsilon_1$.}
\end{subfigure}
\hspace{1.2cm}
\begin{subfigure}[b]{0.45\textwidth} 	
\centering
\includegraphics[scale=0.23]{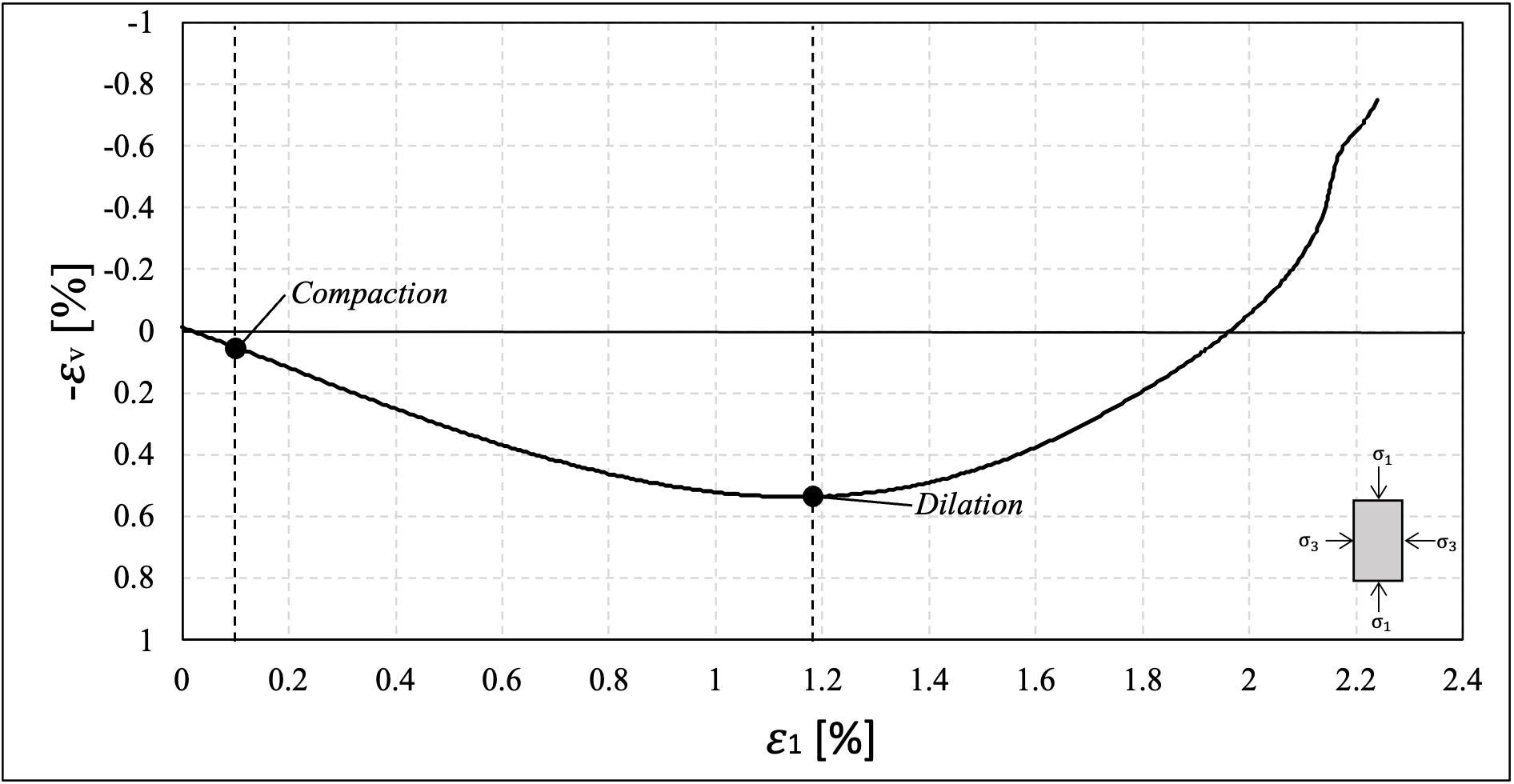}
\caption{$\upvarepsilon_v$ vs $\upvarepsilon_v$.}
\end{subfigure}
\begin{subfigure}[b]{0.45\textwidth} 	 
\bigskip
\centering
\includegraphics[scale=0.23]{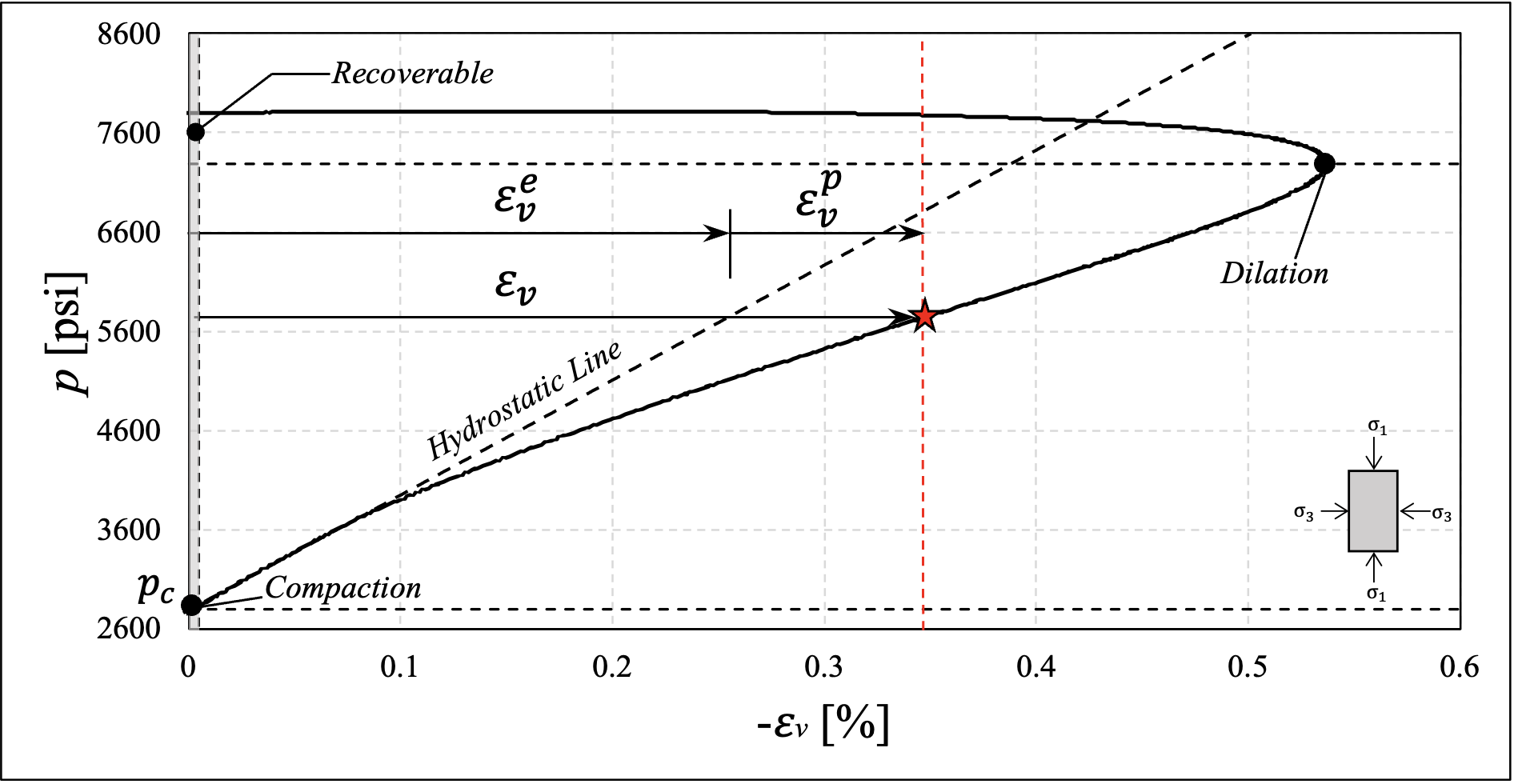}
\caption{$p$ vs $\upvarepsilon_v$.}
\end{subfigure}
\hspace{1.2cm}
\begin{subfigure}[b]{0.45\textwidth} 	 
\bigskip
\centering
\includegraphics[scale=0.23]{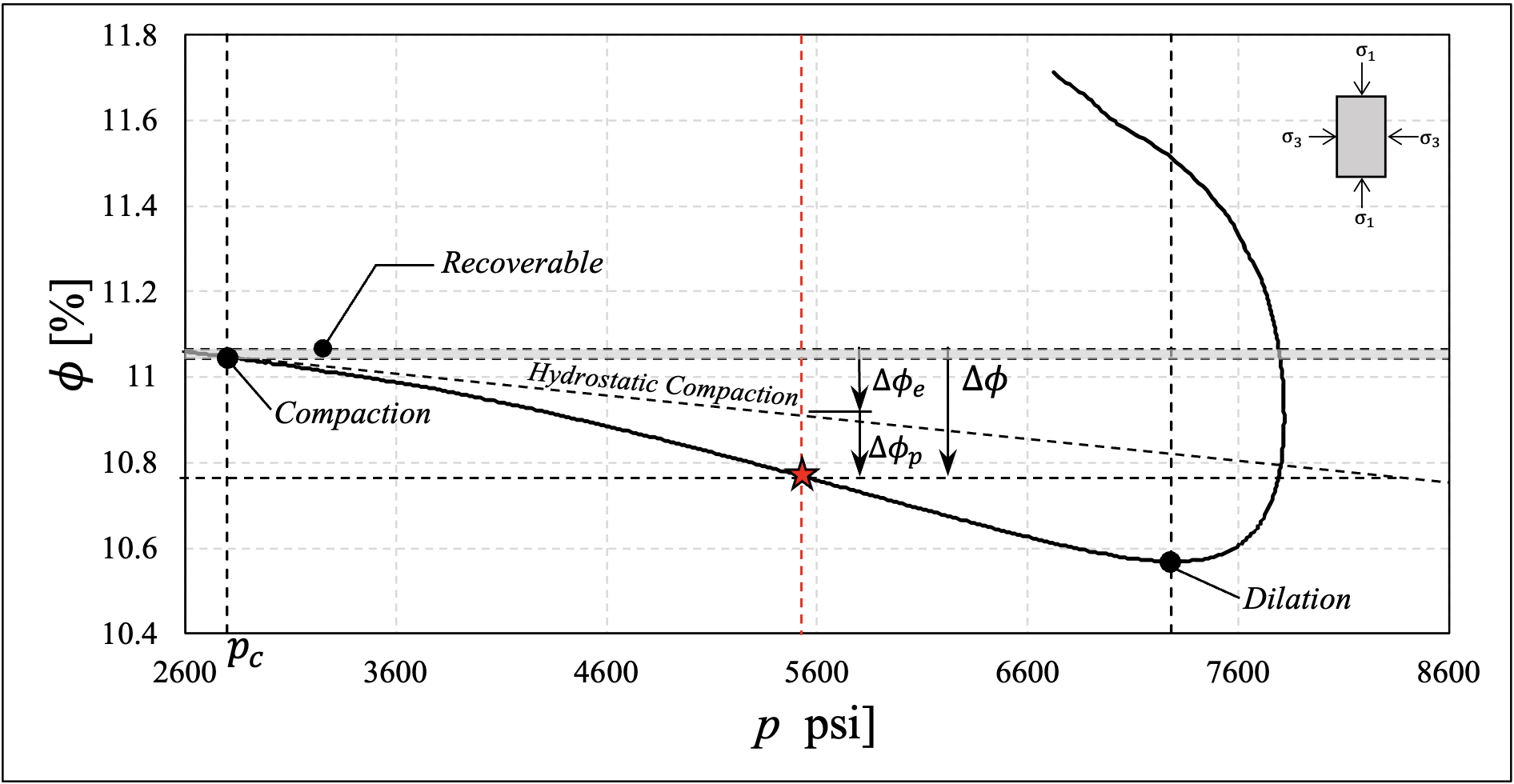}
\caption{$\phi$ vs $p$.}
\end{subfigure}
\caption{3-14-2v Triaxial Test Results.}
\label{fig:3_14_2v_triaxial_results}
\end{figure}

\begin{figure}[h!] 
\centering
\begin{subfigure}[b]{0.45\textwidth}
\centering
\includegraphics[scale=0.23]{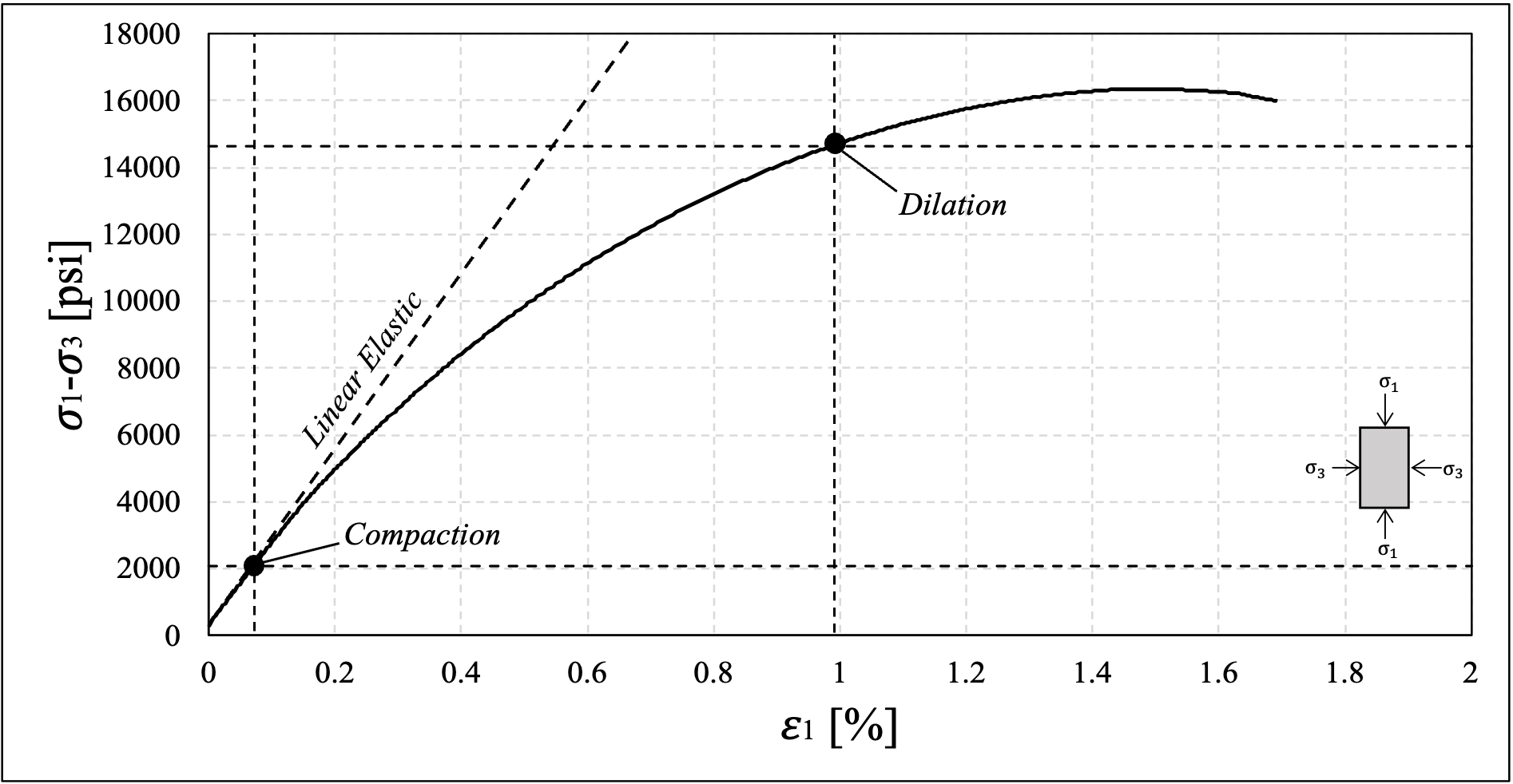}
\caption{$\upsigma_1 - \upsigma_3$ vs $\upvarepsilon_1$.}
\end{subfigure}
\hspace{1.2cm}
\begin{subfigure}[b]{0.45\textwidth} 	 
\centering
\includegraphics[scale=0.23]{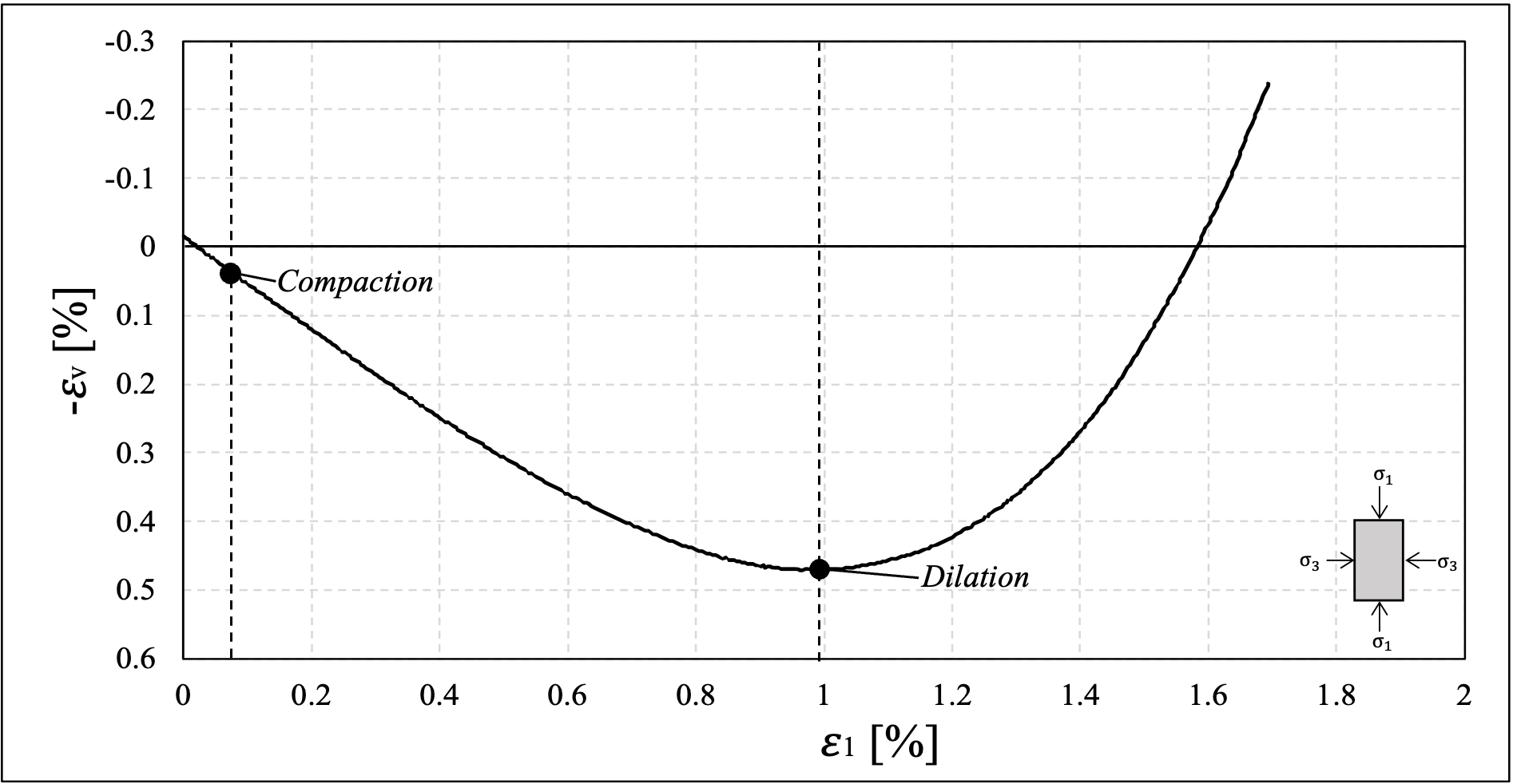}
\caption{$\upvarepsilon_v$ vs $\upvarepsilon_v$.}
\end{subfigure}
\begin{subfigure}[b]{0.45\textwidth} 	 
\bigskip
\centering
\includegraphics[scale=0.23]{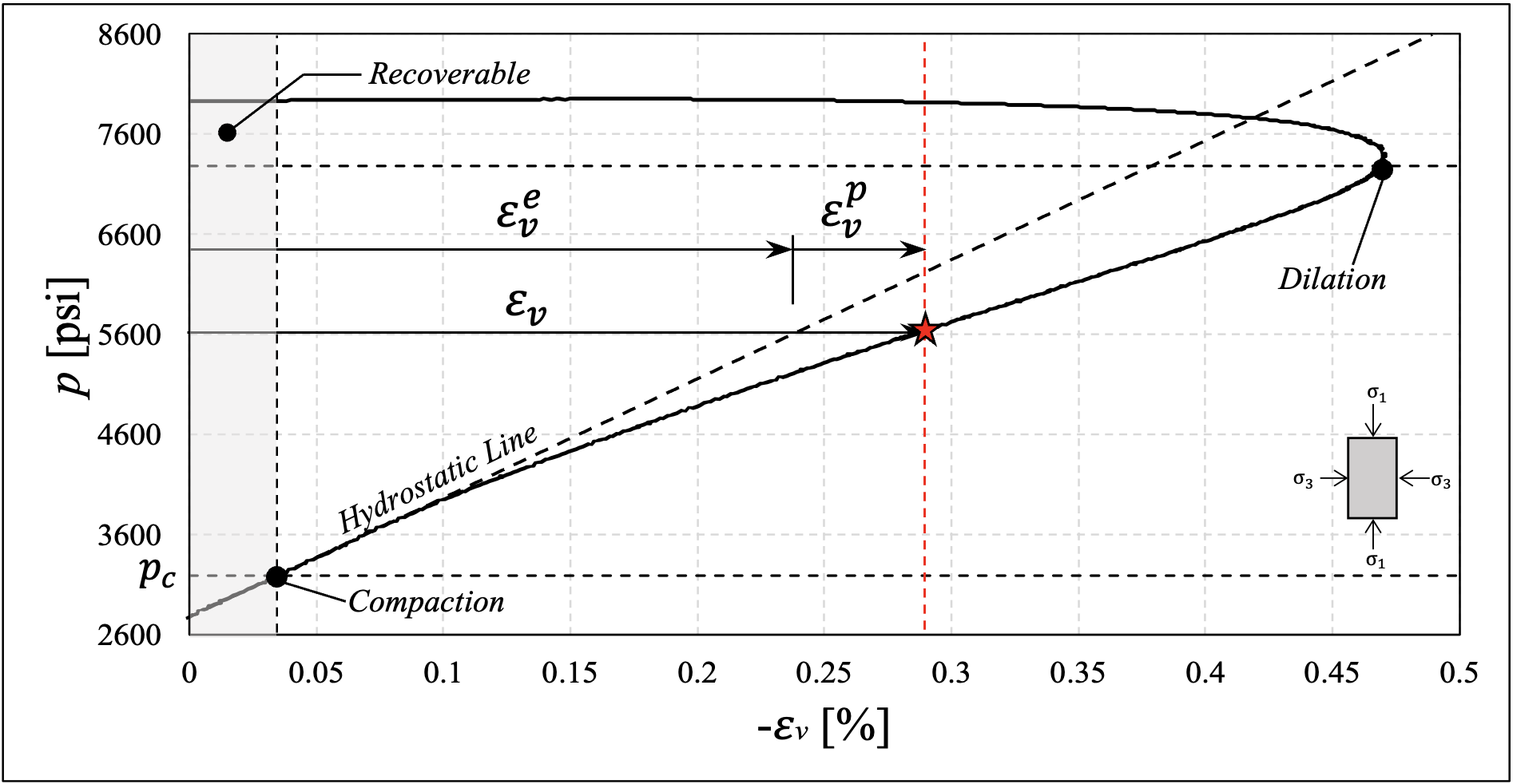}
\caption{$p$ vs $\upvarepsilon_v$.}
\end{subfigure}
\hspace{1.2cm}
\begin{subfigure}[b]{0.45\textwidth} 	 
\bigskip
\centering
\includegraphics[scale=0.23]{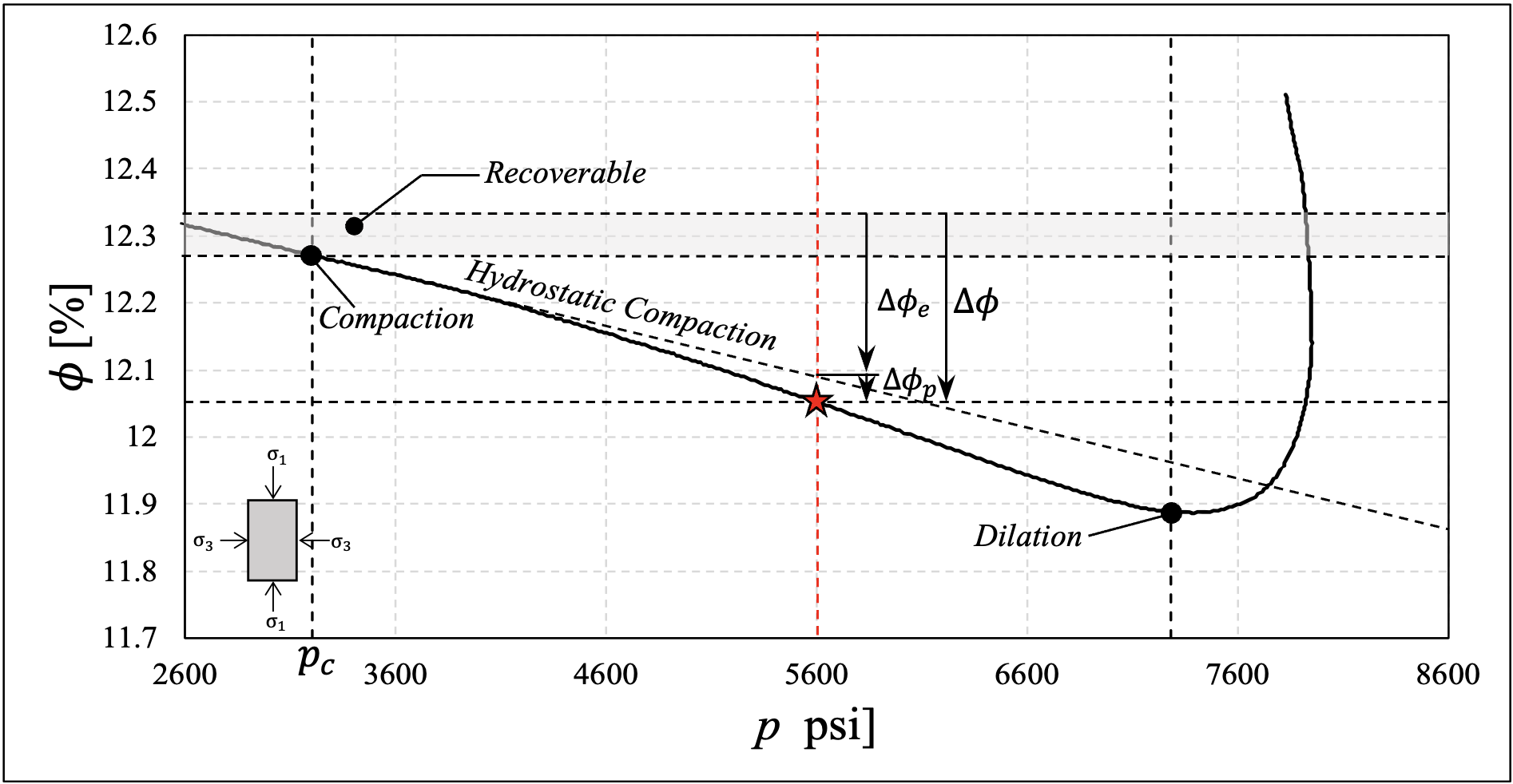}
\caption{$\phi$ vs $p$.}
\end{subfigure}
\caption{4-31-1v Triaxial Test Results.}
\label{fig:4_31_1v_triaxial_results}
\end{figure}

\begin{figure}[h!] 
\centering
\begin{subfigure}[b]{0.45\textwidth}
\centering
\includegraphics[scale=0.23]{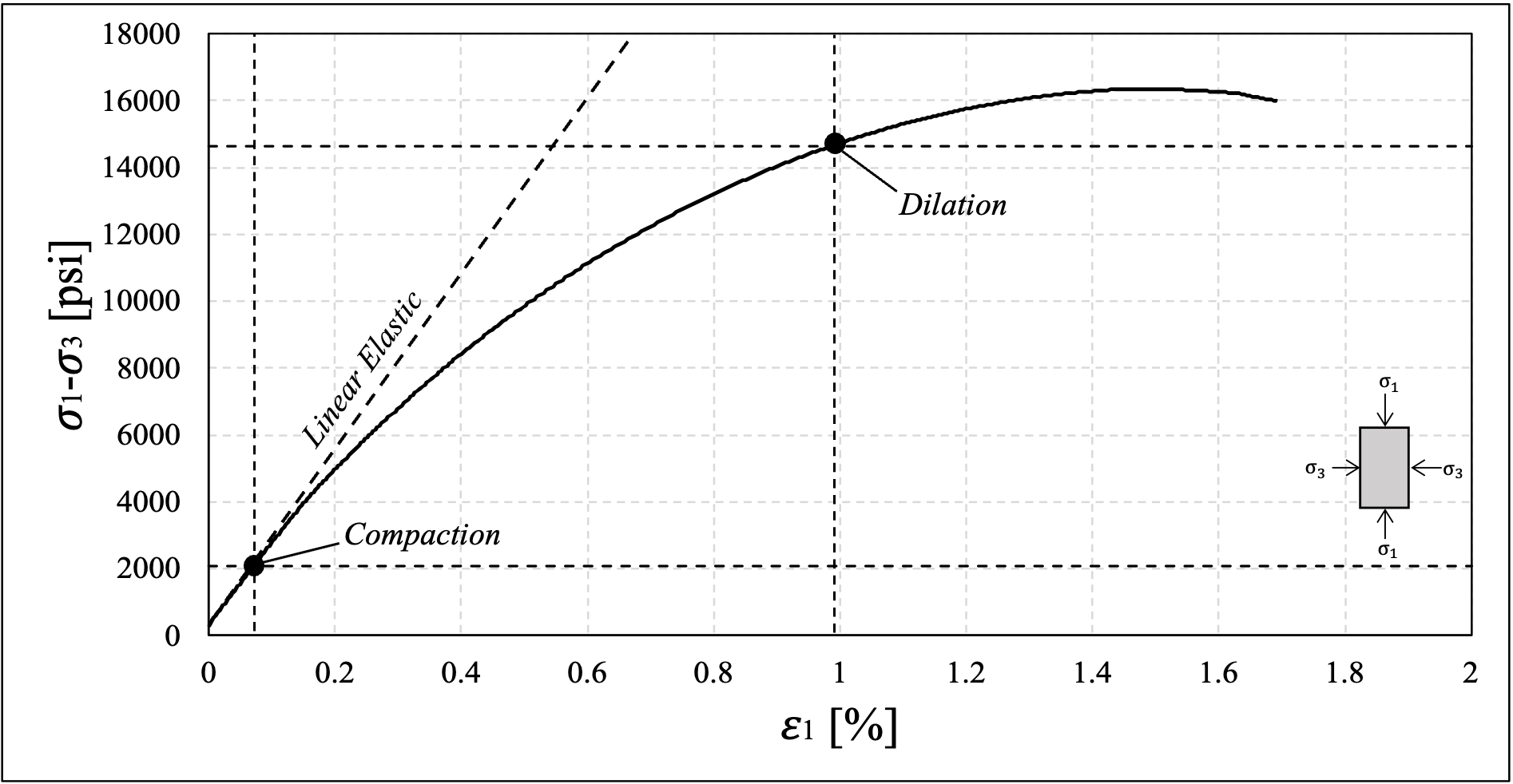}
\caption{$\upsigma_1 - \upsigma_3$ vs $\upvarepsilon_1$.}
\end{subfigure}
\hspace{1.2cm}
\begin{subfigure}[b]{0.45\textwidth} 	 
\centering
\includegraphics[scale=0.23]{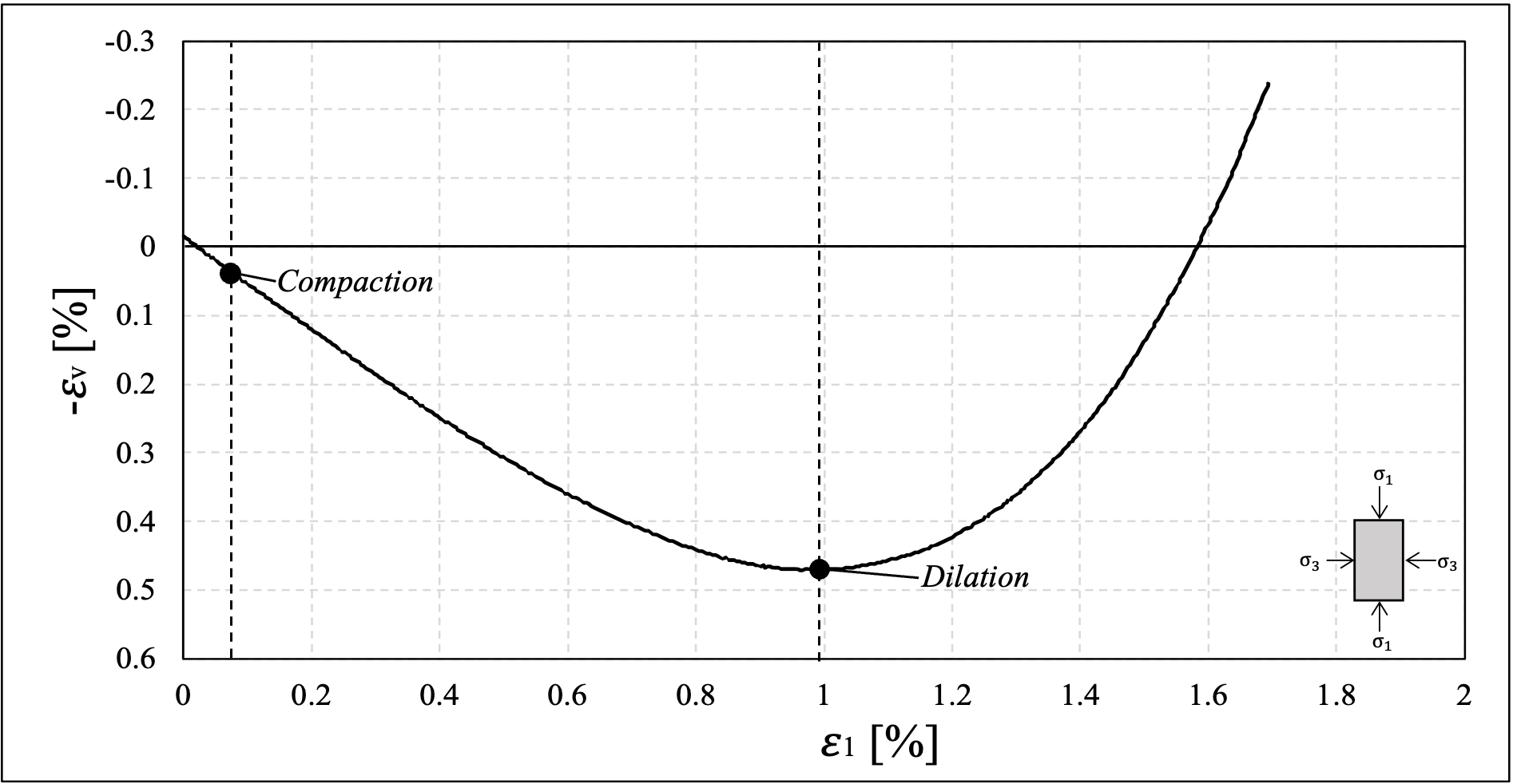}
\caption{$\upvarepsilon_v$ vs $\upvarepsilon_v$.}
\end{subfigure}
\begin{subfigure}[b]{0.45\textwidth} 	 
\bigskip
\centering
\includegraphics[scale=0.23]{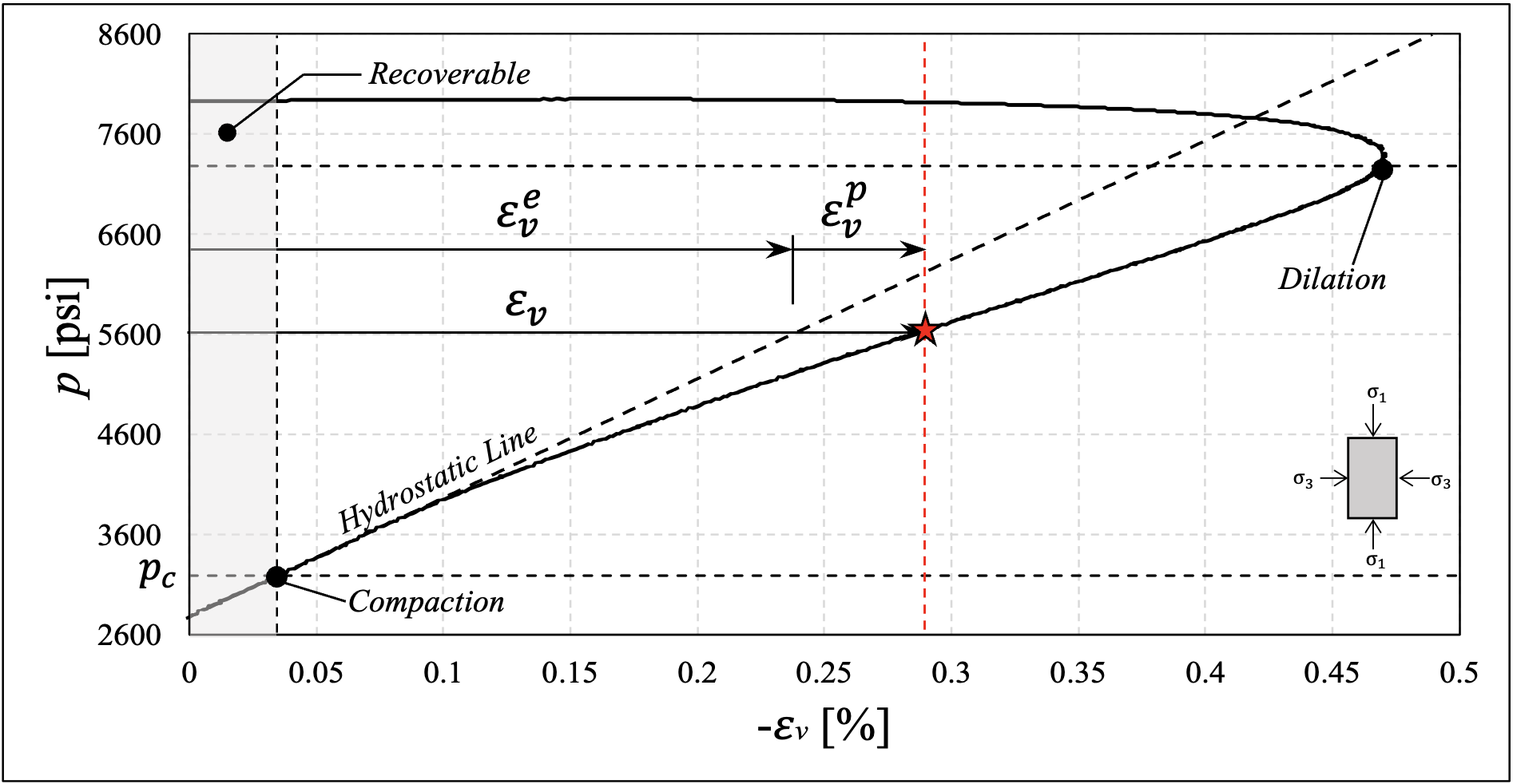}
\caption{$p$ vs $\upvarepsilon_v$.}
\end{subfigure}
\hspace{1.2cm}
\begin{subfigure}[b]{0.45\textwidth} 	 
\bigskip
\centering
\includegraphics[scale=0.23]{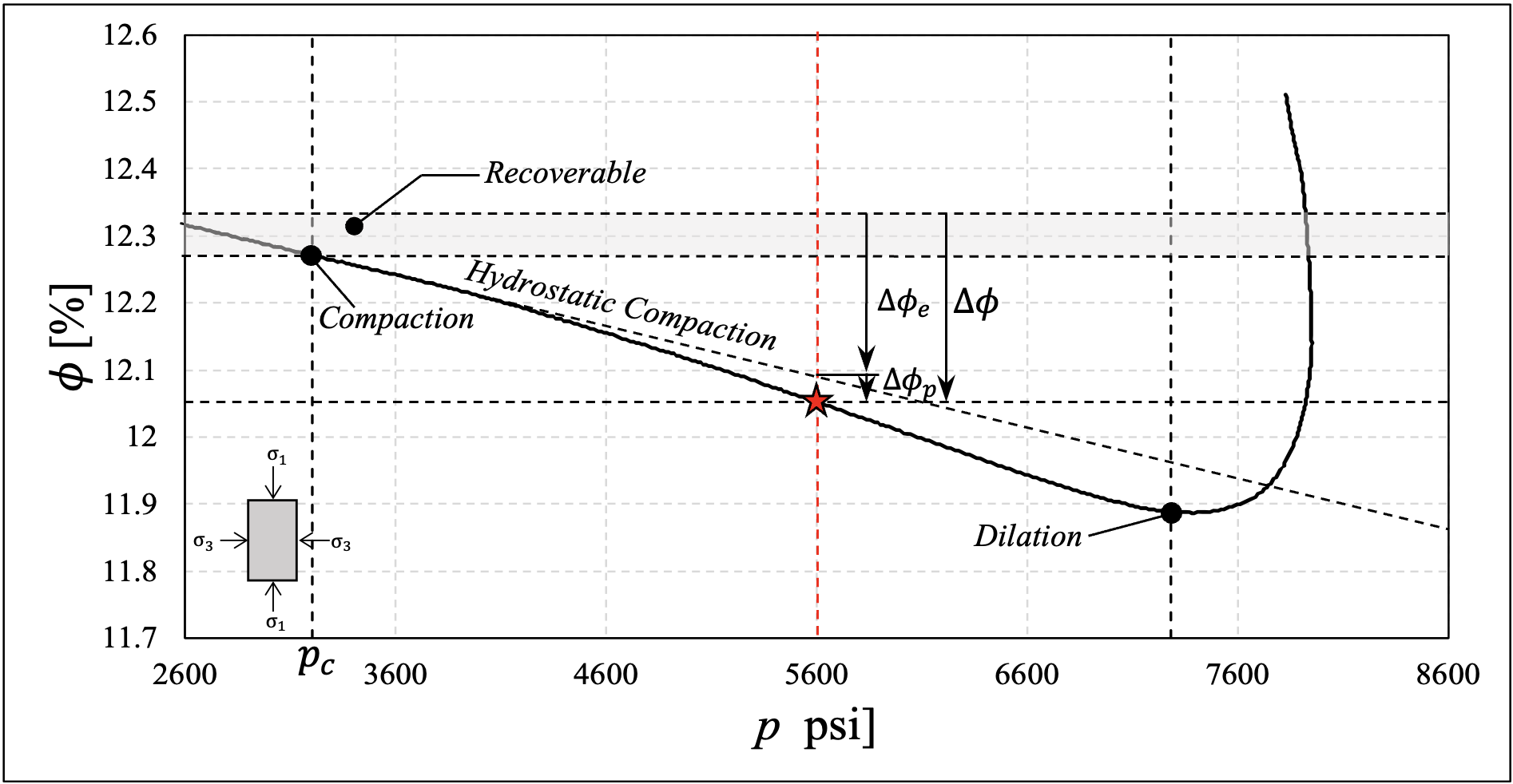}
\caption{$\phi$ vs $p$.}
\end{subfigure}
\caption{4-41v Triaxial Test Results.}
\label{fig:4_41v_triaxial_results}
\end{figure}

A triaxial test was conducted after the hydrostatic and uniaxial stages until failure. Each sample was subjected to a confinement pressure ($\upsigma_c$) of 2500 psi during the test. Figures~\ref{fig:2_44v_triaxial_results} to~\ref{fig:4_41v_triaxial_results} show Vaca Muerta mudstone material response during the triaxial test. These figures show a linear elastic response until 0.05 \% of axial deformation. Within the linear elastic range, we estimate Young's Modulus and Poisson's Ratio; we summarize these elastic constants in Table~\ref{tab:vm_elastic_constants}.
\begin{table}[h!]
\centering
\begin{tabular}{| c | c | c | c | c |}
 \hline
 \textbf{Parameter} & \textbf{2-44v} & \textbf{3-14-2v} & \textbf{4-31-1v} & \textbf{4-41v}\\
 \hline
 $E [\text{Mpsi}]$  & 2.328 & 2.245 & 3.040 & 2.370\\
 $\nu$ & 0.186 & 0.168 & 0.196 & 0.162\\
 \hline
\end{tabular}
 \caption{Elastic Constants for each Vaca Muerta sample.}
 \label{tab:vm_elastic_constants}
\end{table}
From 0.05 \%  to 1-1.2 \% of axial strain, the material exhibits unrecoverable compaction. Finally, the material dilates until localized shear failure. Nonlinear compaction is evident when we plot the hydrostatic stress $p$ against $\upvarepsilon_v$ where the graph departs from the theoretical hydrostatic linear trend (point $p_c$ in $p \, \text{vs}\, \upvarepsilon_v$ charts). Thus,  plastic volumetric strains develop due to compaction until a limit point where the material dilates at constant hydrostatic pressure. We observe a similar response when representing the $\phi$ evolution, when considering Assumption~\ref{assmp:matrix_incompr} with increasing hydrostatic pressure. Initially, the volumetric plastic deformation induces a monotonic porosity reduction until a critical point where the material dilates and, ultimately, fails in shear.

The state equation~\eqref{eq:porosity_state_equation} describes the porosity evolution in rate form. The porosity degradation parameter $\psi$ is conventionally derived from plotting $\phi$ against $\upvarepsilon_v$.
\begin{figure}[h!]
\centering
\includegraphics[scale=0.20]{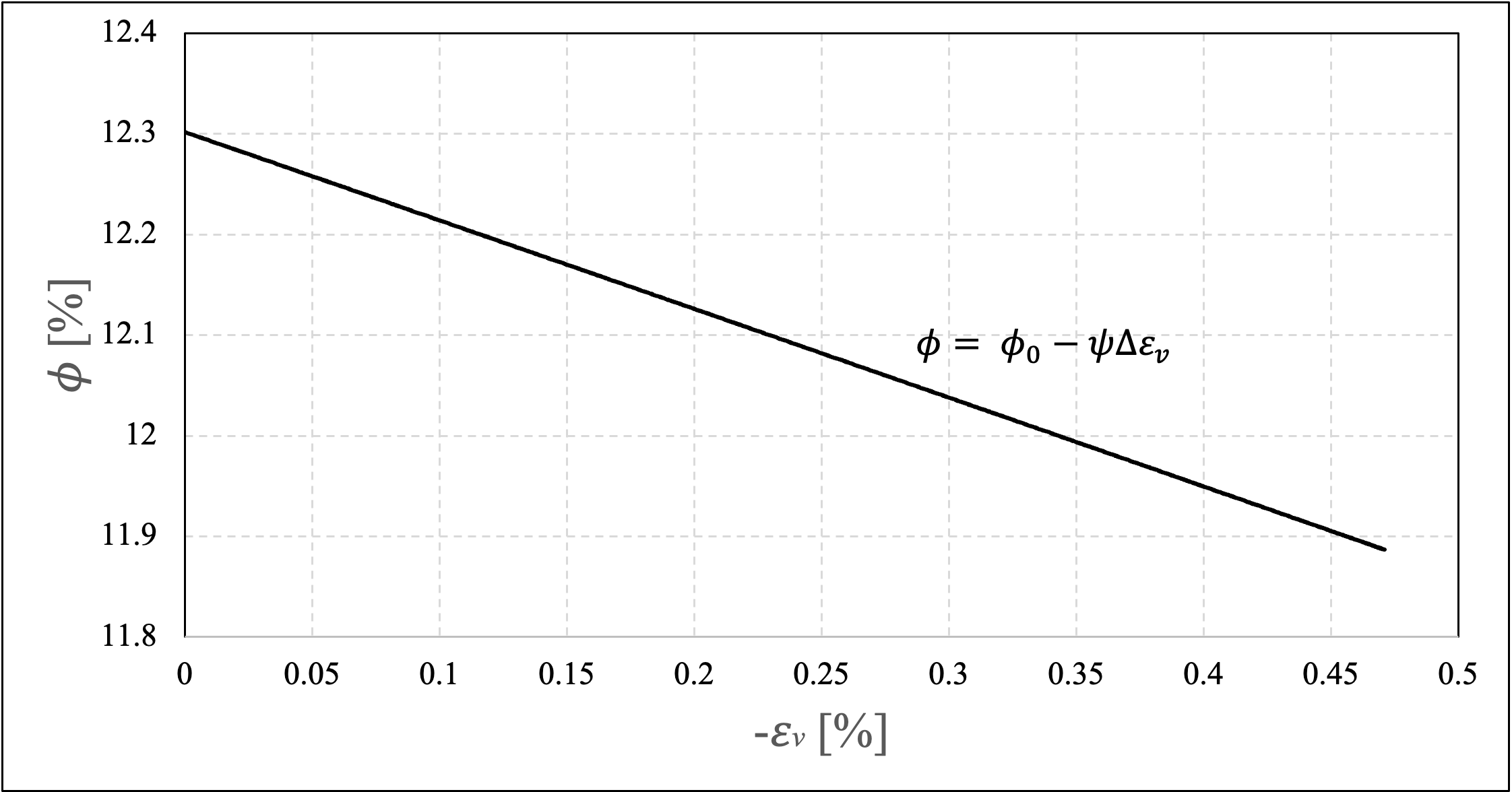}
\caption{Typical Porosity degradation profile at increasing volumetric strain.}
\label{fig:porosity_def_profile}
\end{figure}
Figure~\ref{fig:porosity_def_profile} shows a typical porosity degradation profile of our laboratory data in conjunction with Assumption~\ref{assmp:matrix_incompr}. The porosity degradation material parameter $\psi$ ranges from 0.0085 to 0.0088 in the analyzed samples; thus, we adopt hereafter $\psi = 0.0088$.
\begin{figure}[h!] 
\centering
\begin{subfigure}[b]{0.45\textwidth}
\centering
\includegraphics[scale=0.21]{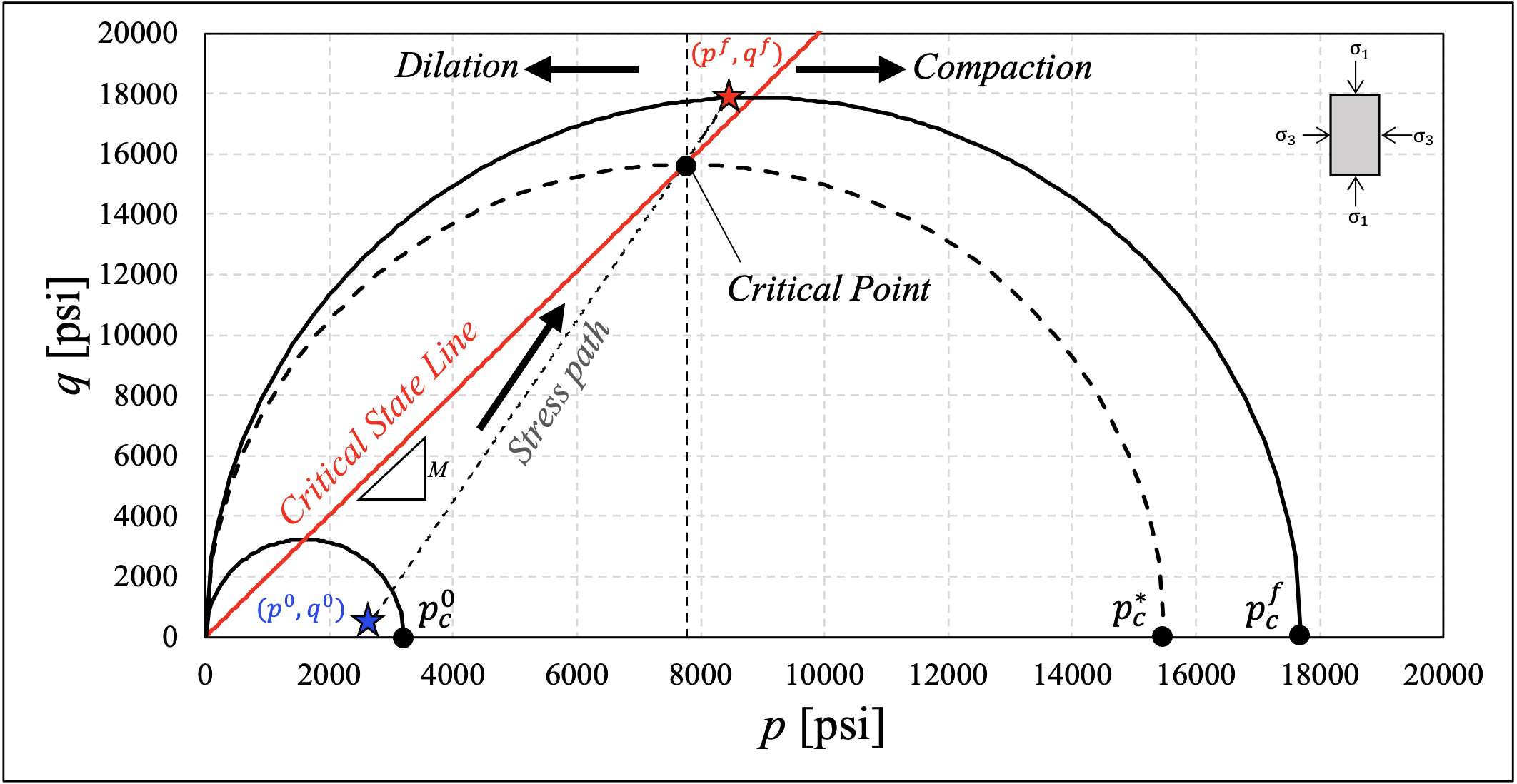}
\caption{Sample: 2-44v.}
\end{subfigure}
\hspace{1.2cm}
\begin{subfigure}[b]{0.45\textwidth} 	 
\centering
\includegraphics[scale=0.21]{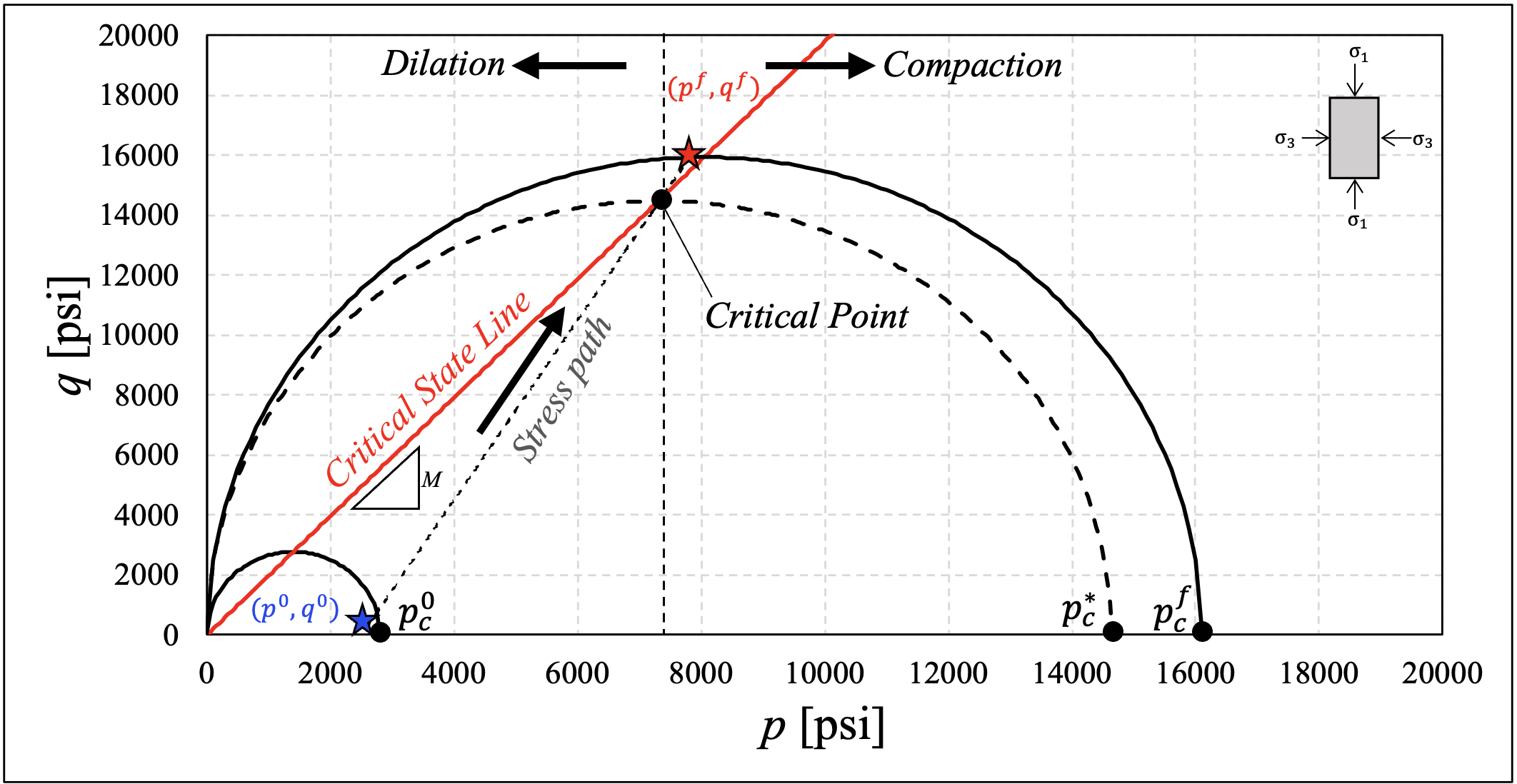}
\caption{Sample: 3-14-2v.}
\end{subfigure}
\begin{subfigure}[b]{0.45\textwidth} 	 
\bigskip
\centering
\includegraphics[scale=0.21]{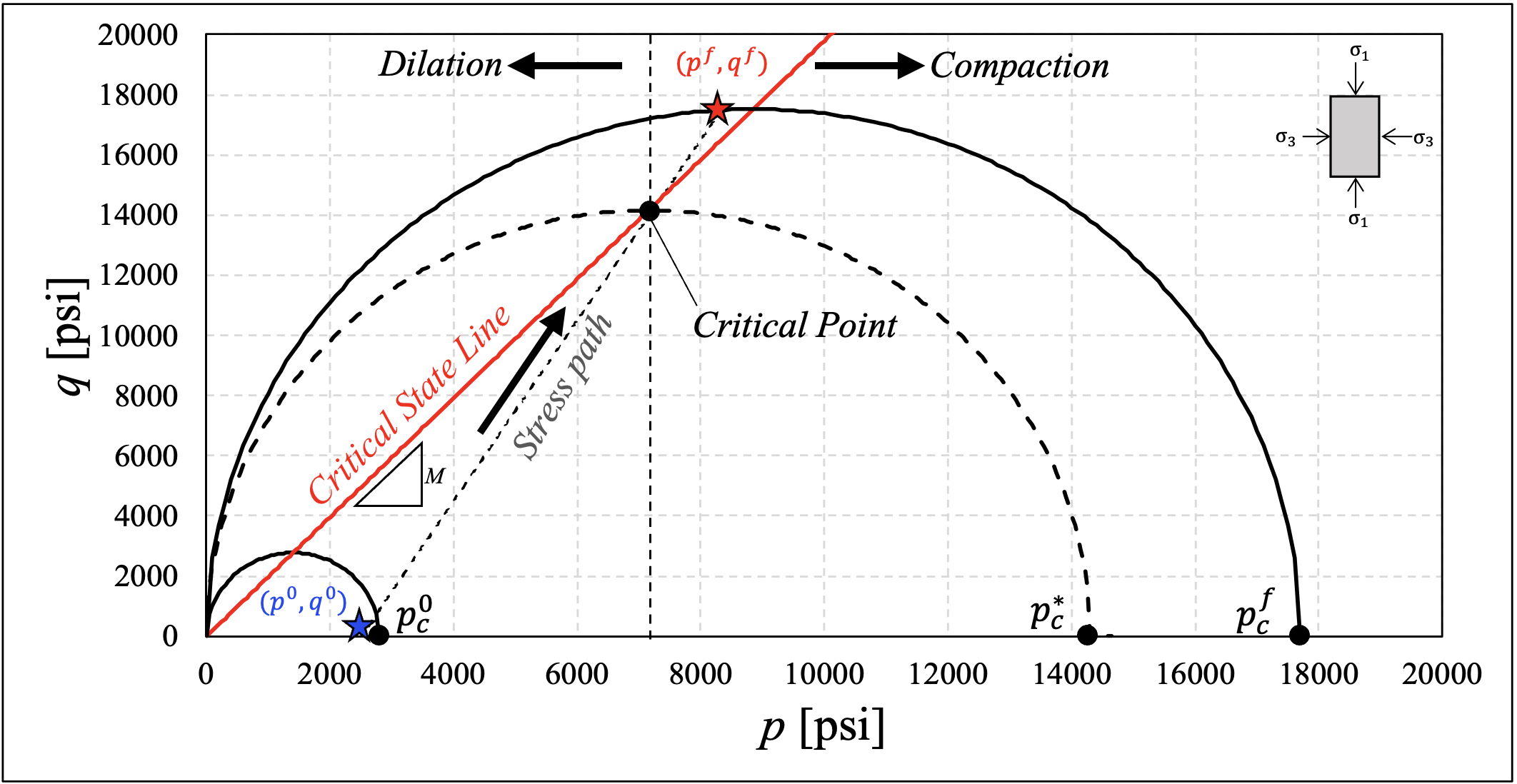}
\caption{Sample: 4-31-1v.}
\end{subfigure}
\hspace{1.2cm}
\begin{subfigure}[b]{0.45\textwidth} 	 
\bigskip
\centering
\includegraphics[scale=0.21]{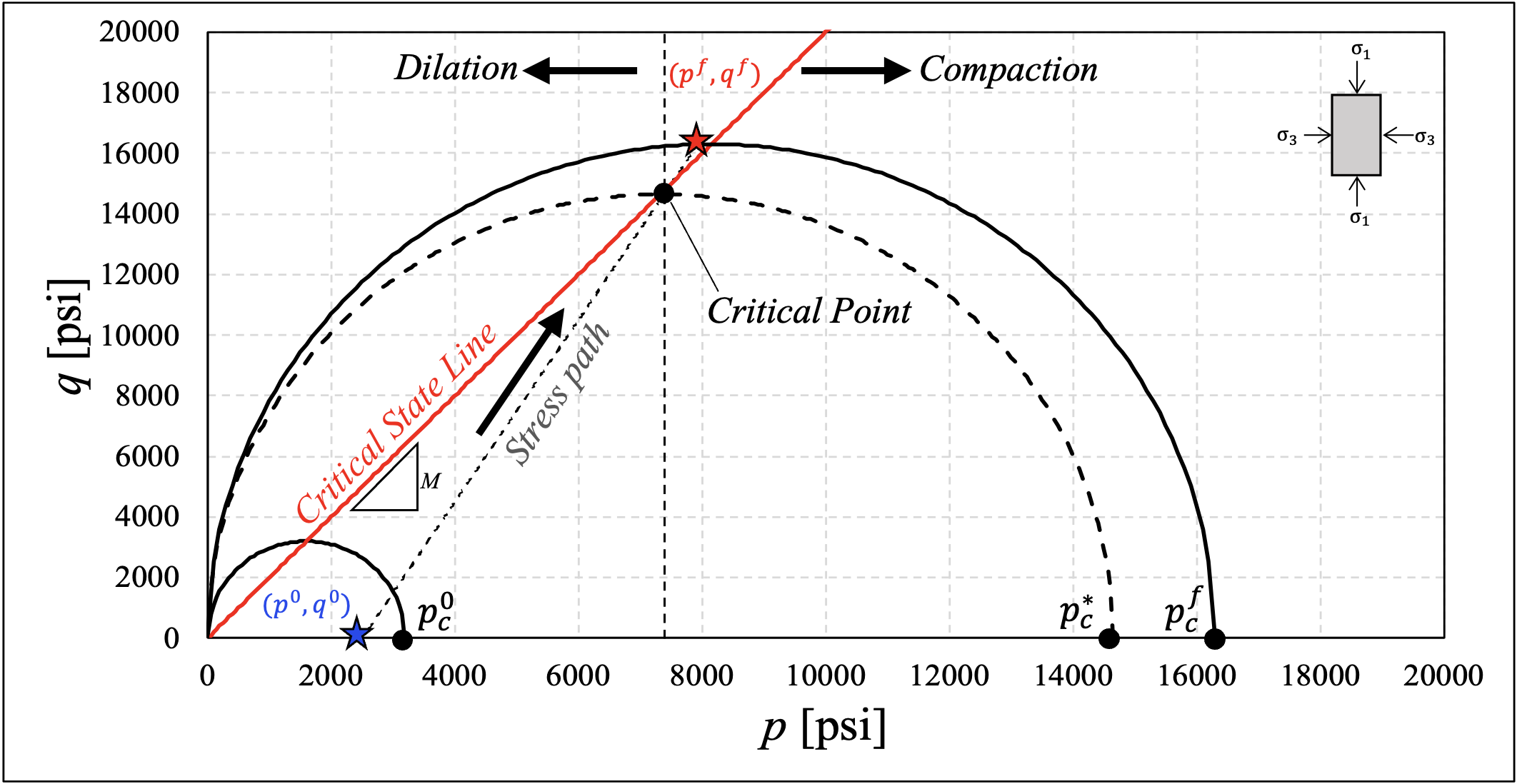}
\caption{Sample: 4-41v.}
\end{subfigure}
\caption{Cam-Clay Model Calibration to Vaca Muerta Experimental Data.}
\label{fig:VM_cam_clay_calibration}
\end{figure}
\begin{table}[h!]
\centering
\begin{tabular}{| c | c | c | c | c |}
 \hline
 \textbf{Parameter} & \textbf{2-44v} & \textbf{3-14-2v} & \textbf{4-31-1v} & \textbf{4-41v}\\
 \hline
 $M$  & 2.02 & 1.98 & 1.98 & 2.00\\
 \hline
\end{tabular}
 \caption{Critical-State-Line (CSL) slope for each Vaca Muerta sample.}
 \label{tab:vm_constitutive_parameters}
\end{table}
Figure~\ref{fig:VM_cam_clay_calibration} shows the Modified Cam-Clay  evolution through the experimental stress path for the Vaca Muerta mudstone samples. The slope of the stress path is 3, the theoretical slope for triaxial conditions. The samples are initially loaded to an initial state $\left(p^0,\, q^0\right)$ inside the initial yield locus defined by the major axis $p^0_c$ (the initial hardening parameter captured by hydrostatic cycles). The sample response is elastic as Figures~\ref{fig:2_44v_triaxial_results} to~\ref{fig:4_41v_triaxial_results} show. As the pair $(p,\, q)$ increase the load reaches the Critical State Line (CSL), where the yield locus is defined by the critical hardening parameter $p^*_c$. The intersection between the critical Cam-Clay ellipse and the CSL coincides with the initiation of  dilation that Figures~\ref{fig:2_44v_triaxial_results} to~\ref{fig:4_41v_triaxial_results} display. Once the dilatant response starts, samples fail in shear, and the final Cam-Clay ellipse is defined by the final hardening parameter $p^f_c$.

The slope of the CSL ($M$) is defined by the intersection between the triaxial test stress path and the hydrostatic pressure  $p$ at which the sample starts to increase volumetric strain at constant pressure looking at the $\upvarepsilon_v\, \text{vs} \, p$ chart. Thus, the CSL is a straight line from the origin to the latter intersection on the $q-p$ projection. Table~\ref{tab:vm_constitutive_parameters} summarizes the MCC constitutive parameters for the Vaca Muerta mudstone samples we tested.
%

\section{Numerical Formulation}
\subsection{Closest Point Projection Mapping for the Modified Cam-Clay Model}
Following~\cite{ Borja1990, Borja1991, Simo1998}, let $\Omega \in \R^d, \, d=2, \,3$. Consider the discretization $\Omega_h \subset \Omega$. Let us take an arbitrary Gauss point on a finite element $e\in\Omega_h$. Let $t \in [0,\, T], \, T >0$ be a pseudo-time anf $n,\, k \in \mathbb{N}^+$ be the pseudo-time increment and iteration counters, respectively. The incremental strain tensor $\Delta \strain^{k}_{n+1}$ is,
\begin{equation}\label{eq:strain_increment}
\Delta \strain^{k}_{n+1} := \strain^k_{n+1} - \strain_n.
\end{equation}
Additionally, consider the trial state defined by freezing all the internal variables as:
\begin{align}
\hat{\stress}_{n+1} &:= \stress_n + \C^e:\Delta\strain^k_{n+1} \label{eq:trial_stress},\\
\hat{p} &:= \dfrac{1}{3} \hat{\stress}_{n+1}:\1 \label{eq:trial_p},\\
\hat{\s}_{n+1} &:= \hat{\stress}_{n+1} - \hat{p}_{n+1}\label{eq:trial_s} \1,\\
\hat{q}_{n+1} &:= \sqrt{\dfrac{3}{2}} \|\hat{\s}_{n+1}\|\label{eq:trial_q}.
\end{align}
where $\stress_n$ and $\strain_n$ are the converged effective stress and strain tensors of the previous pseudo-time step $n$. The return mapping tensor equations in their general form are:
\begin{equation}\label{eq:stress_update}
\stress^k_{n+1} = \hat{\stress}_{n+1} - \C^e : \Delta \strain^p.
\end{equation}
Integrating~\eqref{eq:generalized_flow_rule} within $[t_n,\, t_{n+1}]$  leads to the following discrete plastic-strain increment $\Delta \strain^p$,
\begin{equation}\label{eq:incremental_plastic_deformation}
\Delta \strain^p := \Delta \lambda \dfrac{\partial F_f}{\partial \stress},
\end{equation}
where $\Delta\lambda$ is the discrete consistency parameter. Consider the volumetric part of $\stress^k_{n+1}$:
\begin{align}\label{eq:p_converged}
p^k_{n+1} &= \dfrac{1}{3} \stress^k_{n+1} : \1 \nonumber\\
	             &= \dfrac{1}{3} \hat{\stress}_{n+1} : \1 - \dfrac{1}{3} \C^e : \Delta\strain^p : \1 \nonumber\\
				  &= \hat{p}_{n+1} -K_{n+1}\Delta\upvarepsilon^p_v,
\end{align}
where $\Delta\upvarepsilon^p_v = \Delta \strain^p : \1= \Delta\lambda \left(2p^k_{n+1} - p_c\right)$. Additionally, consider the deviatoric measure of $\stress^k_{n+1}$ (see~\ref{sec:discrete_deviatoric_measure}):
\begin{equation}\label{eq:updated_q_4}
q^k_{n+1} (\Delta\lambda) = \dfrac{\hat{q}_{n+1}}{\left(1 + 6G_{n+1}\, \dfrac{\Delta\lambda}{M^2}\right)}.
\end{equation}
Combining~\eqref{eq:p_converged} and~\eqref{eq:updated_q_4}, results in the following system of scalar equations on $\Delta\lambda$:
\begin{equation}\label{eq:return_map_system}
\begin{aligned}
p^k_{n+1} (\Delta\lambda) &= \hat{p}_n+1 - K\Delta\lambda\,(2\, p^k_{n+1} - p_c), \\
q^k_{n+1} (\Delta\lambda) &= \dfrac{\hat{q}_{n+1}}{\left(1 + 6G\, \dfrac{\Delta\lambda}{M^2}\right)}. 
\end{aligned}
\end{equation}
\begin{figure}[h!]
\centering
\includegraphics[scale=0.35]{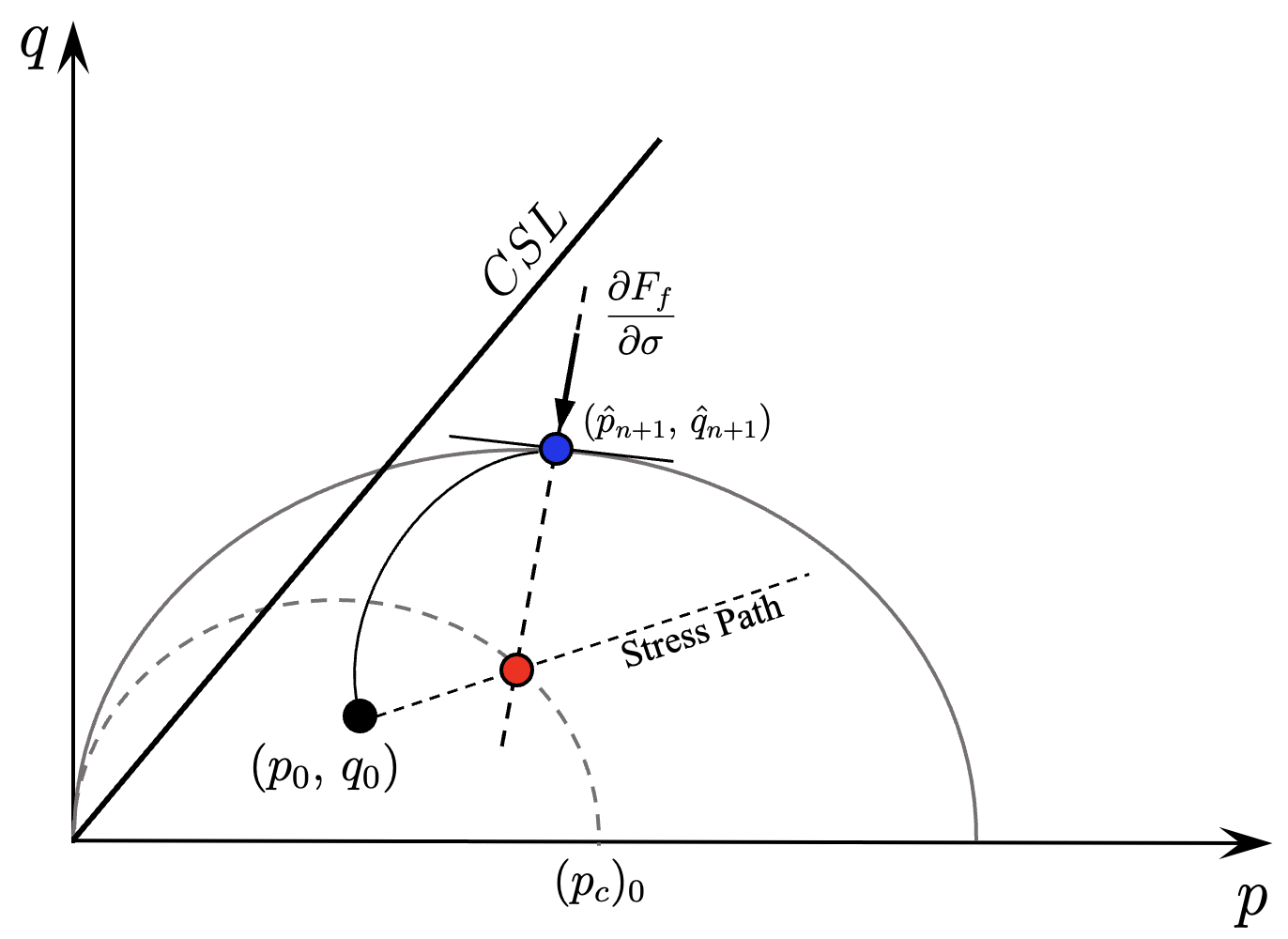}
\caption{Internal variables update by the Closest Point Projection Algorithm}
\label{fig:closest_point_proj}
\end{figure}
\begin{algorithm}[h!]
\caption{Closest Point Projection}
\label{algo:CPP}
\begin{algorithmic}
\State \textbf{1.} Given $\{\stress_n,\, \strain^p_n, \, \Delta\strain, \, (p_c)_n\}$\\
\State \textbf{2.} Calculate the trial state: \\
$\qquad \strain^k_{n+1} = \strain_n + \Delta\strain^k_{n+1}$,\\
$\qquad \hat{\stress}_{n+1} = \stress_n + \C^e:\Delta\strain^k_{n+1}$,\\
$\qquad \hat{p} = \dfrac{1}{3} \hat{\stress}_{n+1}:\1 $,\\
$\qquad \hat{\s}_{n+1} = \hat{\stress}_{n+1} - \hat{p}_{n+1}$,\\
$\qquad \hat{q}_{n+1} = \sqrt{\dfrac{3}{2}} \|\hat{\s}_{n+1}\|$.\\
\State \textbf{3.} Evaluate the Cam-Clay Yield function at the trial state: $F_f\left[ \hat{\stress}_{n+1}, (p_c)_n\right]$\\
\State \textbf{4.} \textbf{IF} $F_f\left[\hat{\stress}_{n+1}, (p_c)_n\right] < 0$: \Comment{Elastic Step} \\
$\qquad \stress^k_{n+1} = \hat{\stress}_{n+1}$,\\
$\qquad \strain^k_{n+1} = \strain_n + \Delta\strain^k_{n+1}$,\\
$\qquad \strain^p_{n+1} = \strain^p_n$,\\	
$\qquad  (p_c)_{n+1} =  (p_c)_n$.\\
\State \textbf{5.} \textbf{ELSE IF} $F_f\left[\hat{\stress}_{n+1}, (p_c)_n\right] \geq 0:$ \Comment{Plastic Step} \\
\hspace{1cm} i. Initialize:\\
\hspace{1.5cm} $\stress^k_{n+1} = \stress_{n}$,\\
\hspace{1.5cm}$\strain^p_{n+1} = \strain^p_n$,\\
\hspace{1.5cm}$ (p_c)_{n+1} =  (p_c)_n$.\\
\\
\hspace{1cm} ii. \textbf{WHILE} $|F_f\left[\stress^k_{n+1}, (p_c)_{n+1}\right]| > \text{FTOL}$: \Comment{Outer Newton-Raphson}\\
\\
\hspace{1.5cm} $\circ$ \textbf{WHILE} $|H_f\left[(p_c)_{n+1}\right]| > \text{HTOL}$: \Comment{Inner Newton-Raphson}\\
\hspace{2.0cm} $\star$ $(p_c)_{n+1} = (p_c)_{n+1} - H_f\left[(p_c)_{n+1}\right]\left(\dfrac{\partial H_f}{\partial p_c}\right)^{-1}$\\
\\
\hspace{1.5cm} $\circ$ $\Delta\lambda^k = \Delta\lambda^{k-1} - F_f \left[\stress^k_{n+1},\, (p_c)_{n+1}\right] \left(\dfrac{\partial F_f}{\partial \Delta\lambda}\right)^{-1}$\\
\hspace{1.5cm} $\circ$ $p^k_{n+1} = \dfrac{\hat{p}_{n+1} + \Delta\lambda^k\, K\, p_c}{1 + 2\,\Delta\lambda^k\, K}$\\
\hspace{1.5cm} $\circ$ $q^k_{n+1} = \dfrac{\hat{q}_{n+1}}{\left(1 + 6G\, \frac{\Delta\lambda^k}{M^2}\right)}$
\State \textbf{6.} \textbf{RETURN}: \\
$\qquad \Delta\strain^p_{n+1} = \Delta\lambda^k \, \dfrac{\partial F_f}{\partial \stress}$,
$\qquad \stress^k_{n+1} = \hat{\stress}_{n+1} - \C^e : \Delta\strain^p_{n+1}$,
$\qquad \strain^p_{n+1} = \strain^p_n + \Delta\strain^p_{n+1}$
\end{algorithmic}
\end{algorithm}
Exact integration of the hardening law~\eqref{eq:hardening_law}, gives:
\begin{align}\label{eq:updated_hardening}
p_c(\Delta\lambda) &= \left(p_c\right)_n \exp\left(\chi\,\Delta\lambda\dfrac{\partial F_f}{\partial p^k_{n+1}}\right) \nonumber \\
&= \left(p_c\right)_n \exp\left[\chi\,\Delta\lambda \left(2\,p^k_{n+1} - p_c \right)\right].
\end{align}
The consistency parameter $\Delta\lambda$ in~\eqref{eq:return_map_system})-\eqref{eq:updated_hardening} is calculated by imposing the consistency condition on $F_f(\Delta\lambda)$:
\begin{equation}\label{eq:consistency_condition_MCC}
F_f(\Delta\lambda) = \left(\dfrac{q^k_{n+1}}{M}\right)^2 + p^k_{n+1}\,\left(p^k_{n+1} - p_c\right)=0.
\end{equation}
Since~\eqref{eq:updated_hardening} couples the variables $p^k_{n+1}$ and $p_c$, $F_f(\Delta\lambda)$ cannot be evaluated explicitly. Therefore, $p^k_{n+1}$ and $p_c$ are updated iteratively by a local Newton-Raphson iteration. Thus, we rewrite the first equation in~\eqref{eq:return_map_system} as,
\begin{equation}\label{eq:update_p_2}
p^k_{n+1} = \dfrac{\hat{p}_{n+1} + \Delta\lambda\, K_{n+1}\, p_c}{1 + 2\,\Delta\lambda\, K_{n+1}},
\end{equation}
and substitute~\eqref{eq:update_p_2} into the second equation in~\eqref{eq:return_map_system} we have the following scalar equation on $p_c$:
\begin{equation}\label{eq:hardening_NR}
H_f(p_c) = \left(p_c\right)_n \exp \left(\chi\, \Delta\lambda \dfrac{2\,\hat{p}_{n+1} + \Delta\lambda\, K_{n+1}\, p_c}{1 + 2\,\Delta\lambda\, K_{n+1}}\right)- p_c = 0.
\end{equation}
In Section~\ref{sec:continuous_formulation}, we define $K$ and $G$ as dependent on $p$ and $\phi$. Therefore, $K$ and $G$ are state variables updated at each strain increment during loading. We use an explicit integration scheme to update $\phi$, $K$ and $G$ as follows:
\begin{align}
\phi_{n+1} &= \phi_n - \psi (\Delta\upvarepsilon_v)_{n+1} \label{eq:updated_phi},\\
K_{n+1} &= \dfrac{p_{n+1}}{\kappa(1-\phi_{n+1})} \label{eq:updated_K},\\
G_{n+1} &= \dfrac{3K_{n+1}(1 - 2\nu)}{2(1+\nu)} \label{eq:updated_G}.
\end{align}
Thus, combining~\eqref{eq:strain_increment} to~\eqref{eq:updated_G} and the derivatives of $F_f(\Delta \lambda)$ and $H_f(\Delta \lambda)$ with respect to the consistency parameter $\Delta \lambda$ (see~\ref{sec:der_F_f_G_delta_lmbda}) to fully define Newton-Raphson iteration scheme, we obtain the following closest point projection algorithm (see~\cite{ Borja1990,Simo1998}) for updating $\stress^k_{n+1}$, $\strain^p_{n+1}$ and $p_c$ (see Figure~\ref{fig:closest_point_proj} for a sketch of the closest point projection algorithm detailed in Algorithm~\ref{algo:CPP}):

\begin{figure}[h!]
\centering
\includegraphics[scale=0.40]{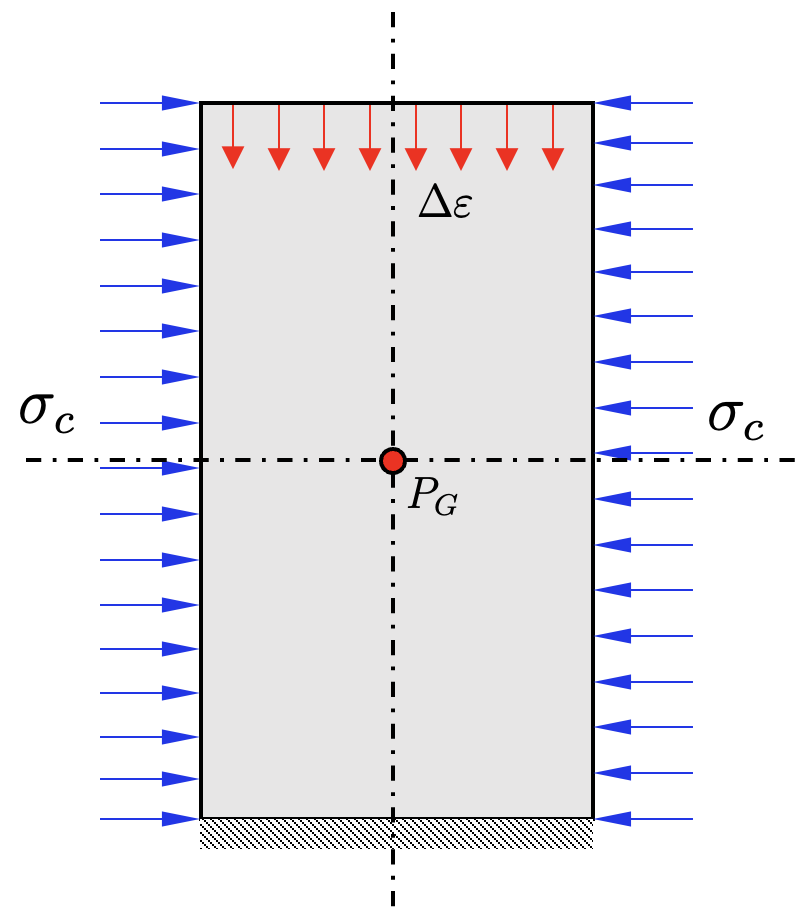}
\caption{Triaxial test general setting for numerical experiments.}
\label{fig:triaxial_test_setting}
\end{figure}

\subsection{Numerical Results: Triaxial test simulation}
Next, we evaluate the performance of the closest point projection algorithm to reproduce a triaxial test by focusing on a single Gauss point $P_G$ and inducing a triaxial stress state. We consider a confinement pressure $\upsigma_c$ applied to the sample's sides; at the sample's top,  we apply a strain increment $\Delta\strain$, which induces a deviatoric strain increment at the Gauss point. Additionally, we fix the sample's bottom during the numerical experiment. Figure~\ref{fig:triaxial_test_setting} shows the general setting for the numerical experiment.
\begin{table}[h!]
\centering
\begin{tabular}{| c | c |}
 \hline
 \textbf{Parameter} & \textbf{Values}\\
 \hline
 $\nu$  & 0.165 \\
 $\upsigma_c [\text{psi}]$ & 2500\\
 $\phi^0 [\%]$ & 12.3\\
 $\gamma$ & 2.43E-03\\
 $\kappa$ & 1.48E-03\\
 $\chi$ & 0.0088\\ 
 $p^0_c [\text{psi}]$ & 3200\\
 $M$ & 2.0\\
 $\delta\epsilon$ & 8E-05\\
 $FTOL$ & 1E-06\\
 $HTOL$ & 1E-6\\
 \hline
\end{tabular}
\caption{Numerical Experiment. Constitutive and Material Parameters}
\label{tab:numerical_experiment_param}
\end{table}

We consider an initial hydrostatic stress state at $P_G$ of the form,
\begin{equation}\label{eq:initial_stress_triaxial}
\stress_0 = 
\begin{bmatrix}
\upsigma_c & 0 & 0\\
0 & \upsigma_c & 0\\
0 & 0 & \upsigma_c
\end{bmatrix}
\e_i \otimes \e_j.
\end{equation}
Additionally, at $P_G$ we apply the following deviatoric strain tensor,
\begin{equation}
\Delta \strain^d_0 = \delta \upvarepsilon
\begin{bmatrix}
1  & 0 & 0\\
0 & 0 & 0\\
0 & 0 & 0
\end{bmatrix}
\e_i \otimes \e_j
- \dfrac{\delta\upvarepsilon}{3}
\begin{bmatrix}
1  & 0 & 0\\
0 & 1 & 0\\
0 & 0 & 1
\end{bmatrix}
\e_i \otimes \e_j,
\end{equation}
where $\stress^k_{n+1}$ is the converged stress tensor in principal components (i.e., $\upsigma_1$, $\upsigma_2$, $\upsigma_3$ where $\upsigma_1\ge\upsigma_2\ge\upsigma_3$). During the triaxial test, the stress components at $P_G$ must satisfy $\upsigma_2=\upsigma_3 = \upsigma_c$. Therefore, after each strain step, we iteratively enforce this condition by increasing the volumetric component of the deviatoric strain tensor increment $\Delta \strain^d_0$.\\

\begin{center}
\begin{figure}[h!]
\centering
\begin{subfigure}[b]{0.75\textwidth}
\includegraphics[width=\textwidth]{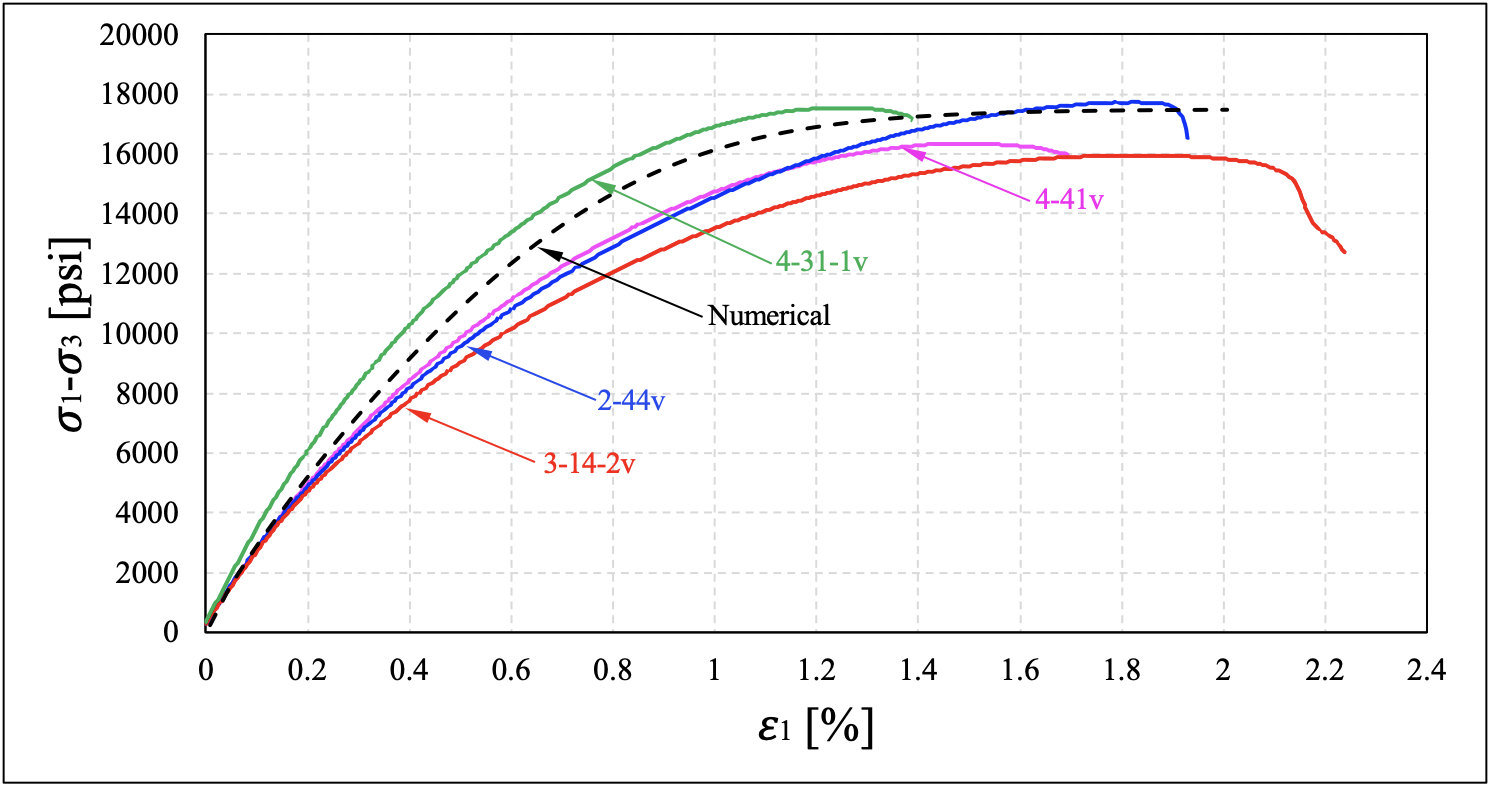}
\caption{Stress-Strain response.}
\end{subfigure}
\hspace{1.2cm}
\centering
\begin{subfigure}[b]{0.75\textwidth} 	 
\includegraphics[width=\textwidth]{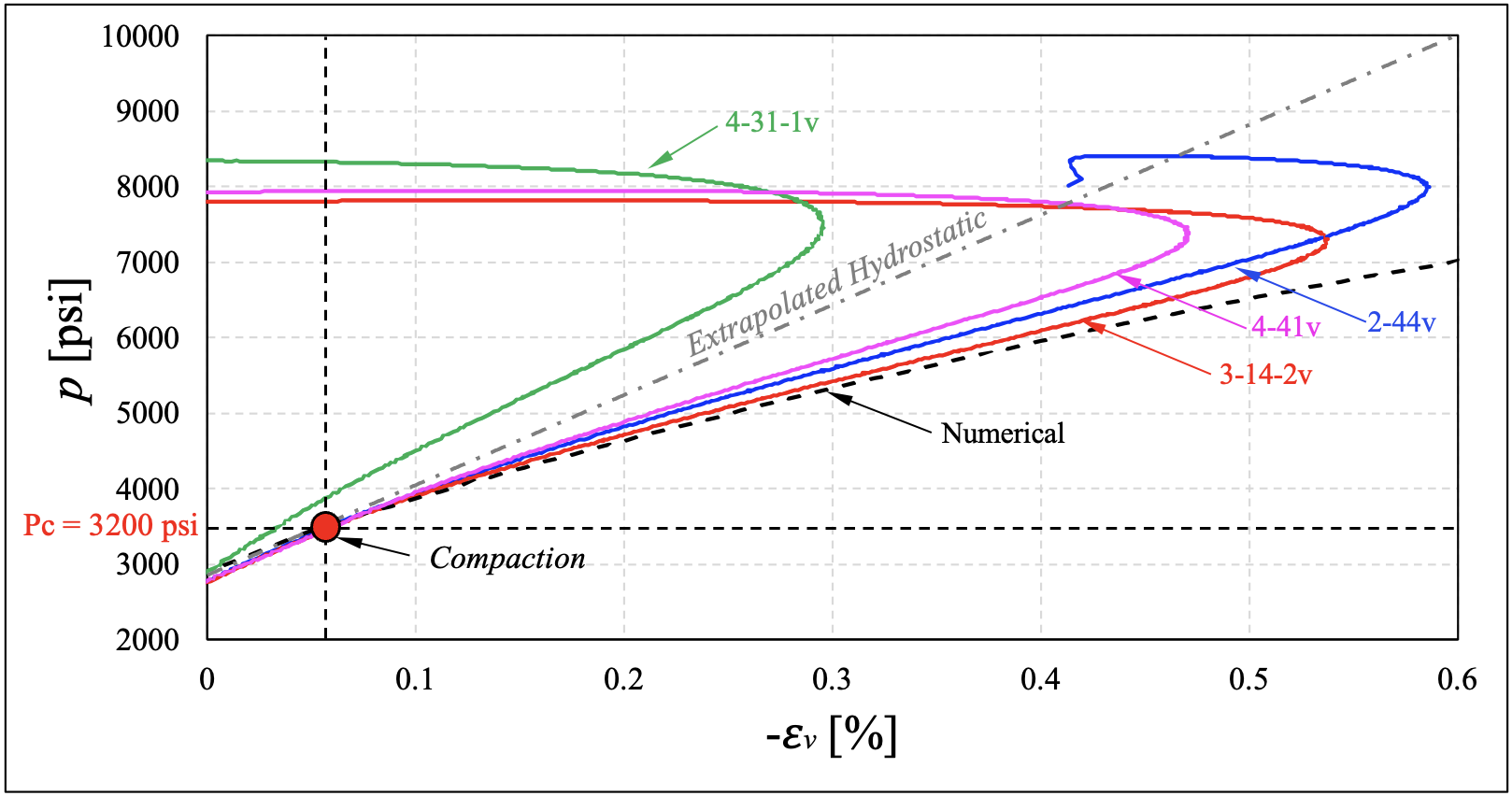}
\caption{Volumetric response.}
\end{subfigure}
\caption{Vaca Muerta mudstone triaxial test simulation.}
\label{fig:vm_triaxial_test_sim}
\end{figure}
\end{center}

We consider material parameters within the range we obtain for the Vaca Muerta mudstone samples. Table~\ref{tab:numerical_experiment_param}  summarizes the input data we use to simulate the triaxial test stress response. Figure~\ref{fig:vm_triaxial_test_sim} shows the triaxial test simulation using the MCC Yield function and Closest Point Projection algorithm to update state variables at the Gauss point $P_G$ and the triaxial response for Vaca Muerta mudstone samples. The calibrated constitutive model reproduces the main features we observe in the stress-strain material response in the lab tests. This model properly captures plasticity due to shear-enhanced compaction despite using an associative flow rule for updating state variables. However, the model parametrization cannot represent the dilation response, as our state equation we introduce does not capture this phenomenon since the hardening law induces compaction as the yield function evolves. Moreover, this model predicts perfect plasticity instead of dilation after sample reaches its critical deformation. Even though some samples showed dilation and softening, these phenomena happens at the end of the triaxial test just before the shear failure of the sample. Thus, our model properly captures the main plastic response due to compaction.

\section{Conclusions and future work}

We calibrate a constitutive model capable of capturing the volumetric plasticity of the Vaca Muerta mudstone using a standard laboratory testing program. The integration algorithm inspired by~\cite{ Borja1990} can properly update the state variables $\stress^k_{n+1}$ and $\strain^p_{n+1}$; this algorithm uses a closest point projection strategy proposed primarily by~\cite{ Simo1998} with an associative flow rule and a simple hardening law. Although we adopt an associative flow rule, the model is capable of capturing compaction without over-predicting volumetric strains. Our mathematical model does not seek to describe the dilatant behavior of this rock as the plastic load evolves and the algorithm updates the state variables. Therefore, the material behaves as perfectly plastic after a critical compaction value. The model captures the main plastic dissipation mechanism well due to compaction for Vaca Muerta mudstone samples, even though dilatancy was not included in the constitutive model formulation.
The calibrated MCC model and the integration algorithm we describe can be implemented in standard Finite Element routines to properly capture the compaction of Vaca Muerte mudstone in more complex stress states and geomechanics applications such as wellbore stability problems and hydraulic fracture propagation problems. Our future work will incorporate dilatancy to the hardening law to properly capture this phenomenon .
\section*{Acknowledgements}
This publication was made possible in part by the support of Vista Energy Company which provided Vaca Muerta samples, and W. D Von Gonten Engineering that provided their laboratory facilities and technical team to conduct all the laboratory tests. 

\newpage
\appendix
\section{Complementary Calculations}
\subsection{Derivatives of the Modified Cam-Clay Yield Function}\label{sec:modified_cam_clay_derivatives}

We use a Modified Cam-Clay Yield function with the following form,
$$
F_f (\stress) = \dfrac{q^2}{M^2} + p(p-p_c),
$$
where $q = \sqrt{\frac{3}{2}}\, \|\s\|$ measures the deviatoric effective stress, $p$ is the hydrostatic stress, $M$ and $p_c$ are material parameters. We calculate the derivative of $F_f$ with respect to $\stress$, which by the chain rule results in,
\begin{equation}\label{eq:der_Ff_sigma}
\dfrac{\partial F_f(\stress)}{\partial \stress} = \dfrac{\partial F_f(\stress)}{\partial p}\,\dfrac{\partial p}{\partial \stress} + \dfrac{\partial F_f(\stress)}{\partial q} \, \dfrac{\partial q}{\partial \stress}.
\end{equation}

We obtain the derivatives of $F_f$ with respect to $p$, $q$  and $p_c$ directly if we reformulate $F_f$ as
$$
F_f (\stress) = \dfrac{q^2}{M^2} + p^2-p\,p_c.
$$
Thus,
\begin{align}
\dfrac{\partial F_f(\stress)}{\partial p} &= 2p-p_c \label{eq:der_Ff_p}, \\
\dfrac{\partial F_f(\stress)}{\partial q} &= \dfrac{2q}{M^2},  \label{eq:der_Ff_q}\\
\dfrac{\partial F_f(\stress)}{\partial p_c} &= -p \label{eq:der_Ff_pc}.
\end{align}
\vs
The derivative of $p$ with respect to the effective stress tensor is collected in the following complementary  results:
\begin{lem} \label{lem:derivative_of_tensor}
Let $\stress \in \R^{3 \times 3}$ be a second-order tensor and $\upsigma_{ij},\, i,\,j = 1,\,2,\,3$ be its components, the following result holds,
\begin{equation*}
\dfrac{\partial\upsigma_{ij}}{\partial\upsigma_{kl}} = \delta_{ik} \delta_{jl}, \quad i,\,j,\,k,\,l = 1,\,2,\,3.
\end{equation*}
\begin{proof}
The result follows from index inspection,
$$
\dfrac{\partial \upsigma_{ij}}{\partial \upsigma_{kl}} =
\begin{cases}
1,& \text{for } \, i=k \, \text{ and } \, j=l.   \\
0,& \text{otherwise.}
\end{cases}
$$
Additionally,
$$
\delta_{ik}\delta_{jl} =
\begin{cases}
1,& \text{for } \, i=k \, \text{ and } \, j=l.   \\
0,& \text{otherwise.}
\end{cases}
$$
\end{proof} 
\end{lem}
\begin{prpsn}\label{prpsn:der_p_sigma}
Let $p$ be the trace of Cauchy's Effective stress tensor; thus, the following holds,
\begin{equation}
\dfrac{\partial p}{\partial \stress} = \dfrac{1}{3} \delta_{ij} \,\e_i \otimes \e_j = \dfrac{1}{3} \mathrm{\boldsymbol{1}}.
\end{equation}
\begin{proof}
Since $p = \dfrac{1}{3} \stress: \1$ and using index notation, the derivative with respect to the effective stress tensor is,
\begin{align*}
\dfrac{\partial p}{\partial \stress} &= \dfrac{1}{3} \dfrac{\partial}{\partial \stress} \left( \stress : \1 \right)\\
&= \dfrac{1}{3} \dfrac{\partial}{\partial \upsigma_{kl}} \left(\upsigma_{ij}\,\delta_{ij} \right)\e_k \otimes \e_l\\
&= \dfrac{1}{3} \dfrac{\partial \upsigma_{ij}}{\partial \upsigma_{kl}}\, \delta_{ij} \, \e_k \otimes \e_l\\
&= \dfrac{1}{3} \delta_{ik}\,\delta_{jl} \, \delta_{ij} \e_k \otimes \e_l \quad \text{(by Lemma~\ref{lem:derivative_of_tensor})}\\
&= \dfrac{1}{3} \delta_{ij} \,\e_i \otimes \e_j. 
\end{align*}
\end{proof}
\end{prpsn}
\vs
The derivative of $q$ with respect to the effective stress tensor is collected in the following results:
\begin{lem}\label{lem:derivative_dev_tensor}
Let $\stress,\, \s \in \R^{3 \times 3}$ be a second-order tensor and its deviatoric part. Let $\upsigma_{ij}$ and $\textrm{s}_{ij}$, $i,\,j =1,\,2,\,3$ the components of $\stress$ and $\s$ respectively. The following holds,
$$
\dfrac{\partial \textrm{s}_{ij}}{\partial \upsigma_{kl}} = \delta_{ik}\,\delta_{lj} - \dfrac{1}{3} \delta_{ij} \,\delta_{kl}.
$$
\begin{proof}
Recall that $\textrm{s}_{ij} = \upsigma_{ij} - \dfrac{1}{3} p\delta_{ij}$; thus,
$$
\dfrac{\partial \,\textrm{s}_{ij}}{\partial \upsigma_{kl}} = \dfrac{\partial}{\partial \upsigma_{kl}} \left(\upsigma_{ij} - \dfrac{1}{3}\,p\delta_{ij}\right).
$$ 
By Lemma~\ref{lem:derivative_of_tensor} and Proposition~\ref{prpsn:der_p_sigma} and expanding the left-hand side of the previous equality,
\begin{align*}
\dfrac{\partial \,\textrm{s}_{ij}}{\partial\,\upsigma_{kl}} &= \dfrac{\partial \upsigma_{ij}}{\partial \upsigma_{kl}} - \dfrac{1}{3}\,\dfrac{\partial\,p}{\partial \upsigma_{ij}}\,\delta_{kl}\\
&=\delta_{ik}\delta_{lj} - \dfrac{1}{3}\,\delta_{ij}\delta_{kl}.
\end{align*}
\end{proof}
\end{lem}
\vs
\begin{prpsn} \label{prpsn:der_q_sigma}
Let $q = \sqrt{\dfrac{3}{2}}\,\| \s \| = \sqrt{\dfrac{3}{2} \, \s:\s}$ be a measure for the deviatoric stress tensor; thus, the following holds,
$$
\dfrac{\partial q}{\partial \stress} = \sqrt{\dfrac{3}{2}}\,  \dfrac{\s}{\|\s\|}.
$$
\begin{proof}
We expand the left-hand side of the equality using index notation, the definition of the norm, and double contraction for second-order tensors, 
\begin{align*}
\dfrac{\partial q}{\partial \stress} &= \dfrac{\partial q}{\partial \upsigma_{ij}} \, \e_i \otimes \e_j\\
\\
&= \underbrace{\dfrac{\partial}{\partial \upsigma_{ij}} \left(\sqrt{\dfrac{3}{2} \textrm{s}_{kl}\,\textrm{s}_{kl}} \right)}_{\textrm{(A)}}\, \e_i \otimes \e_j.
\end{align*}

Applying the chain rule for derivatives in (A), we obtain,
\begin{align*}
\dfrac{\partial}{\partial \upsigma_{ij}} \left(\sqrt{\dfrac{3}{2} \textrm{s}_{kl}\,\textrm{s}_{kl}} \right) &= \sqrt{\dfrac{3}{2}}\, \dfrac{\partial}{\partial \upsigma_{ij}} \big(\textrm{s}_{kl} \, \textrm{s}_{kl} \big)^{\frac{1}{2}}\\
&= \sqrt{\dfrac{3}{2}} \left[\dfrac{1}{2} \big(\textrm{s}_{pq} \, \textrm{s}_{qp} \big)^{-\frac{1}{2}}\, \dfrac{\partial}{\partial\,\upsigma_{ij}} \big(\textrm{s}_{kl}\,\textrm{s}_{kl} \big) \right] \\
&= \sqrt{\dfrac{3}{2}} \left\{\dfrac{1}{2} \big(\textrm{s}_{pq}\,\textrm{s}_{pq}\big)^{-\frac{1}{2}}\, \left[ \dfrac{\partial\,\textrm{s}_{kl}}{\partial \upsigma_{ij}}\, \textrm{s}_{kl} + \textrm{s}_{kl}\,\dfrac{\partial\,\textrm{s}_{kl}}{\partial \textrm{s}_{kl}} \right] \right\}\\
&= \sqrt{\dfrac{3}{2}} \big(\textrm{s}_{pq}\,\textrm{s}_{pq}\big)^{-\frac{1}{2}} \left( \delta_{ik}\,\delta_{lj} - \dfrac{1}{3}\, \delta_{ij}\,\delta_{kl} \right) \, \textrm{s}_{kl} \qquad \quad \textrm{(by Lemma~\ref{lem:derivative_dev_tensor})}\\
&= \sqrt{\dfrac{3}{2}} \big(\textrm{s}_{pq}\,\textrm{s}_{pq}\big)^{-\frac{1}{2}} \big( \upsigma_{ij} - \dfrac{1}{3} \upsigma_{ll} \, \delta_{ij} \big)\\
&= \sqrt{\dfrac{3}{2}} \big(\textrm{s}_{pq}\,\textrm{s}_{pq}\big)^{-\frac{1}{2}} \big( \upsigma_{ij} - p\, \delta_{ij} \big) \qquad \qquad \left(\textrm{since } p = \dfrac{1}{3} \upsigma_{ll} = \dfrac{1}{3} \textrm{tr}(\stress) \right). \\
\end{align*}
Replacing (A) and contracting indexes, we get the desired expression.
\begin{align*}
\dfrac{\partial\,q}{\partial \stress} &= \sqrt{\dfrac{3}{2}} \big(\textrm{s}_{pq}\,\textrm{s}_{pq}\big)^{-\frac{1}{2}} \big( \upsigma_{ij} - p\, \delta_{ij} \big)\, \e_i \otimes \e_j\\
&= \sqrt{\dfrac{3}{2}}\, \dfrac{\textrm{s}_{ij}}{\|\s\|} \, \e_i \otimes \e_j\\
&= \sqrt{\dfrac{3}{2}}\,  \dfrac{\s}{\|\s\|}.
\end{align*}
\end{proof}
\end{prpsn}
Therefore, we collect the derivatives of the modified Cam-Clay yield function is collected in the following,
\begin{prpsn}[Cam-Clay Yield Function Derivative] Let $F_f(\stress) : \R^{3 \times 3} \mapsto \R$ be the Cam-Clay yield function then,
$$
F_f(\stress) := \dfrac{q^2}{M^2}+p\,(p-p_c), \, \forall M,\, p_c >0.
$$
Thus, the derivative of $F_f$ with respect to the effective stress tensor adopts the following expression,
$$
\dfrac{\partial F_f(\stress)}{\partial \stress} = \dfrac{1}{3}\, (2p - p_c)\, \1 + \sqrt{\dfrac{3}{2}} \, \dfrac{2q}{M^2}\, \dfrac{\s}{\| \s \|}.
$$
\begin{proof}
The result follows by applying the chain rule for derivation,
$$
\dfrac{\partial\, F_f(\stress)}{\partial \stress} = \dfrac{\partial\,F_f(\stress)}{\partial\,p}\,\dfrac{\partial\, p}{\partial \stress} + \dfrac{\partial\,F_f(\stress)}{\partial\,q}\, \dfrac{\partial\, q}{\partial\, \stress},
$$
and replacing the corresponding terms with~\eqref{eq:der_Ff_p},~\eqref{eq:der_Ff_q}, and the results from Proposition~\ref{prpsn:der_p_sigma} and Proposition~\ref{prpsn:der_q_sigma}.
\end{proof}
\end{prpsn}
\subsection{Discrete deviatoric measure}
\label{sec:discrete_deviatoric_measure}

Let $\stress^k_{n+1}$ the updated effective stress update of~\eqref{eq:stress_update}. The updated deviatoric effective stress tensor is,
\begin{equation}\label{eq:updated_s}
\s^k_{n+1} = \stress^k_{n+1} - p^k_{n+1}\,\1.
\end{equation}
Replacing~\eqref{eq:stress_update} and~\eqref{eq:p_converged} in~\eqref{eq:updated_s} the following holds,
\begin{align}\label{eq:updated_s_2}
\s^k_{n+1} &= \hat{\stress}_{n+1} - \C^e:\Delta\strain^p - \hat{p}_{n+1} \1 + K\,\Delta \upvarepsilon^p_v \1 \nonumber \\
&= \hat{\s}_{n+1} - \underbrace{\C^e : \Delta\strain^p}_{(A)} + K \Delta \upvarepsilon^p_v \1.
\end{align}
$(A)$ can be written as
\begin{align*}
\qquad \qquad \C^e : \Delta\strain^p &= \left[K\1 \otimes \1 + 2G \left(\I - \dfrac{1}{3} \, \1 \otimes \1\right)\right] : \Delta \strain^p\\
&= K\underbrace{\1\otimes\1 : \Delta\strain^p}_{(B)} + \underbrace{2G\, \I:\Delta\strain^p}_{(C)} - \dfrac{2}{3} G \1\otimes\1:\Delta\strain^p.
\end{align*}
Now, expanding $(B)$,
\begin{align*}
\Delta\strain^p : \1\otimes\1 &= \Delta \upvarepsilon^p_{ij}\, \e_i \otimes \e_j : \delta_{kl}\, \e_k \otimes \e_l \otimes \delta_{pq}\, \e_p \otimes \e_q\\
&= \Delta\upvarepsilon^p_{ij}\, \delta_{ik}\delta_{jl}\, \delta_{pq} \e_p \otimes \e_q\\
&= \Delta\upvarepsilon^p_{kk}\, \delta_{pq} \e_p \otimes \e_q\\
&= \Delta\upvarepsilon^p_v\, \1.
\end{align*}
Considering the definition of $\I$ in~\eqref{eq:fourth_order_unit_tensor}, $(C)$ admits the following expansion,
\begin{align*}
2G\, \I : \Delta\upvarepsilon^p &= 2G\, \left[\dfrac{1}{2}\left(\delta_{ik}\,\delta_{jl} + \delta_{il}\,\delta_{jk}\right) \Delta\upvarepsilon^p_{ij} \right]\\
&= 2G\, \left[ \dfrac{1}{2} \left(\Delta\upvarepsilon^p_{kl} + \Delta\upvarepsilon^p_{lk}\right)\right]\\
&= 2G\, \Delta\strain^p\qquad \qquad \text{(thanks to the symmetry of } \Delta\strain^p \text{)}.
\end{align*}
Replacing $(B)$ and $(C)$ in $(A)$ and then replacing in~\eqref{eq:updated_s_2} we obtain the following expression:
\begin{align}
\s^k_{n+1} &= \hat{\s}_{n+1} -\cancel{K\, \Delta\upvarepsilon_v\, \1} - 2G\, \Delta\strain^p + \dfrac{2}{3} G\Delta \upvarepsilon^p_v \1 + \cancel{K\Delta\upvarepsilon^p_v \1} \nonumber \\
&= \hat{\s}_{n+1} - 2G\,\left(\Delta\strain^p - \dfrac{1}{3} \Delta\upvarepsilon^p_v \1\right) \nonumber \\
&= \hat{\s}_{n+1}  - 2G\,\Delta\strain^p_d,
\end{align}
where $\Delta\strain^p_d = \Delta\strain^p - \dfrac{1}{3} \Delta \upvarepsilon^p_v\, \1$. By the definition of the updated deviatoric measure $q^k_{n+1}$ the following holds,
\begin{equation}\label{eq:updated_q_2}
q^k_{n+1} = \sqrt{\dfrac{3}{2}} \| \hat{\s}_{n+1} - 2G\, \Delta\strain^p_d\|.
\end{equation}
From the discrete flow rule~\eqref{eq:incremental_plastic_deformation} and the derivative of $F_f(\stress)$ with respect to $\stress$ we deduce that $\hat{\s}_{n+1}$, $\s^k_{n+1}$ and $\Delta\strain^p_d$ are colinear, thus 
\begin{align}\label{eq:updated_q_3}
q^k_{n+1} &= \sqrt{\dfrac{3}{2}} \|\hat{s}_{n+1}\| - 2G\,\sqrt{\dfrac{3}{2}}\|\Delta\strain^p\| \nonumber \\
&= \hat{q}_{n+1} - 3G \Delta\strain^p,
\end{align}
where $\|\Delta\strain^p\| = \Delta\lambda \sqrt{\dfrac{3}{2}} \, \dfrac{2 q^k_{n+1}}{M^2}$.

\subsection{Derivative of Modified Cam-Clay Yield Function and Hardening Function respect to $\Delta\lambda$}
\label{sec:der_F_f_G_delta_lmbda}
We fully define the iterative scheme for determining the discrete consistency parameter $\Delta\lambda$ using the following:
\begin{prpsn}[Derivative of $F_f$ respect to $\Delta\lambda$]
Let $F_f(\Delta\lambda)$ be given by~\eqref{eq:consistency_condition_MCC} and, consider~\eqref{eq:update_p_2},~\eqref{eq:updated_q_4},~\eqref{eq:updated_hardening},~\eqref{eq:der_Ff_p},~\eqref{eq:der_Ff_q} and~\eqref{eq:der_Ff_pc}. Then the following result holds,
\begin{equation}\label{eq:der_F_f_delta_lmbda}
\dfrac{\partial F_f}{\partial \Delta\lambda} = -K \dfrac{(2\,p -p_c)^2}{1 + (2K + \chi\,p_c)\Delta\lambda} - \dfrac{2q}{M^2} \dfrac{q}{\Delta\lambda +\frac{M^2}{6G}} -\chi \, p \, p_c \dfrac{(2p - p_c)}{1 + (2K + \chi\,p_c)\Delta\lambda}.
\end{equation}
\begin{proof}
By the chain rule, we obtain,
\begin{equation}\label{eq:chain_rule_Ff_lambda}
\dfrac{\partial F_f}{\partial \Delta\lambda} = \dfrac{\partial F_f}{\partial p}\dfrac{\partial p}{\partial \Delta\lambda} + \dfrac{\partial F}{\partial q} \dfrac{\partial q}{\partial \Delta\lambda} + \dfrac{\partial p_c}{\partial\Delta\lambda}.
\end{equation}
The derivatives of $F_f$ respect to $p$, $q$ and $p_c$ are given by~\eqref{eq:der_Ff_p},~\eqref{eq:der_Ff_q} and~\eqref{eq:der_Ff_pc}. Thus we only need to calculate $\dfrac{\partial p}{\partial \Delta\lambda}$, $\dfrac{\partial q}{\partial \Delta\lambda}$ and $\dfrac{\partial p_c}{\partial\Delta\lambda}$. Considering~\eqref{eq:update_p_2},~\eqref{eq:updated_q_4},~\eqref{eq:updated_hardening},  we have
\begin{align}
\dfrac{\partial p}{\partial \Delta\lambda} &= -K\dfrac{(2p - p_c)}{1 + (2K+ \chi\,p_c)\Delta\lambda} \label{eq:der_p_Dlmda}, \\
\dfrac{\partial q}{\partial \Delta\lambda} &= - \dfrac{q}{\Delta\lambda +\frac{M^2}{6G}} \label{eq:der_q_Dlmda}, \\
\dfrac{\partial p_c}{\partial \Delta\lambda} &= \chi p_c \dfrac{(2p-p_c)}{1+(2K+\chi p_c)\Delta\lambda} \label{eq:der_pc_Dlmda}. 
\end{align}
The result follows after substituting~\eqref{eq:der_Ff_p},~\eqref{eq:der_Ff_q},~\eqref{eq:der_Ff_pc},~\eqref{eq:der_p_Dlmda},~\eqref{eq:der_q_Dlmda} and~\eqref{eq:der_pc_Dlmda} into~\eqref{eq:chain_rule_Ff_lambda}.
\end{proof}
\end{prpsn}
\begin{prpsn}[Derivative of $H_f$ respect to $\Delta\lambda$]
Let $H_f(\Delta\lambda)$ be given by~\eqref{eq:hardening_NR}. Then, the following result holds,
\begin{equation}\label{eq:der_Hf_lmbda}
\dfrac{\partial H_f}{\partial \Delta\lambda} = -\dfrac{\left(p_c\right)_n\,\chi\,\Delta\lambda}{1+2\,\Delta\lambda\,K}\, \exp\left[\dfrac{\chi \, \left(2\,\hat{p}_{n+1} - p_c \right)}{1+2\,\Delta\lambda\,K}\right] -1.
\end{equation}
\begin{proof}
Get~\eqref{eq:hardening_NR} and perform chain rule for derivation respect to $\Delta\lambda$.
\end{proof}
\end{prpsn}

\bibliography{mybibfile}

\begin{thebibliography}{32}
\expandafter\ifx\csname natexlab\endcsname\relax\def\natexlab#1{#1}\fi
\providecommand{\url}[1]{\texttt{#1}}
\providecommand{\href}[2]{#2}
\providecommand{\path}[1]{#1}
\providecommand{\DOIprefix}{doi:}
\providecommand{\ArXivprefix}{arXiv:}
\providecommand{\URLprefix}{URL: }
\providecommand{\Pubmedprefix}{pmid:}
\providecommand{\doi}[1]{\href{http://dx.doi.org/#1}{\path{#1}}}
\providecommand{\Pubmed}[1]{\href{pmid:#1}{\path{#1}}}
\providecommand{\bibinfo}[2]{#2}
\ifx\xfnm\relax \def\xfnm[#1]{\unskip,\space#1}\fi
\bibitem[{Borja(1991)}]{Borja1991}
\bibinfo{author}{Borja, R.~I.} (\bibinfo{year}{1991}).
\newblock \bibinfo{title}{Cam-clay plasticity, part ii: Implicit integration of
  constitutive equation based on a nonlinear elastic stress predictor}.
\newblock {\it \bibinfo{journal}{Computer Methods in Applied Mechanics and
  Engineering}\/},  {\it \bibinfo{volume}{88}\/}, \bibinfo{pages}{225--240}.
  \URLprefix \url{http://doi.org/10.1016/0045-7825%2891%2990256-6}.
  \DOIprefix\doi{10.1016/0045-7825(91)90256-6}.
\bibitem[{Byerlee(1968)}]{Byerlee1968}
\bibinfo{author}{Byerlee, J.~D.} (\bibinfo{year}{1968}).
\newblock \bibinfo{title}{Brittle-ductile transition in rocks}.
\newblock {\it \bibinfo{journal}{Journal of Geophysical Research}\/},  {\it
  \bibinfo{volume}{73}\/}, \bibinfo{pages}{4741--4750}. \URLprefix
  \url{http://doi.org/10.1029/jb073i014p04741}.
  \DOIprefix\doi{10.1029/jb073i014p04741}.
\bibitem[{Carman(1997)}]{Carman1997}
\bibinfo{author}{Carman, P.} (\bibinfo{year}{1997}).
\newblock \bibinfo{title}{Fluid flow through granular beds}.
\newblock {\it \bibinfo{journal}{Chemical Engineering Research and Design}\/},
  {\it \bibinfo{volume}{75}\/}. \URLprefix
  \url{http://doi.org/10.1016/s0263-8762%2897%2980003-2}.
  \DOIprefix\doi{10.1016/s0263-8762(97)80003-2}.
\bibitem[{Cier et~al.(2022)Cier, Labanda \& Calo}]{Cier2022}
\bibinfo{author}{Cier, R.~J.}, \bibinfo{author}{Labanda, N.~A.}, \&
  \bibinfo{author}{Calo, V.~M.} (\bibinfo{year}{2022}).
\newblock \bibinfo{title}{Compaction band localization in geomaterials: a
  mechanically consistent failure criterion}.
\newblock \href{http://arxiv.org/abs/2202.03849}{\tt arXiv:2202.03849}.
\bibitem[{Coussy(2004)}]{Coussy2004}
\bibinfo{author}{Coussy, O.} (\bibinfo{year}{2004}).
\newblock {\it \bibinfo{title}{Poromechanics}\/}.
\newblock (\bibinfo{edition}{2nd} ed.).
\newblock \bibinfo{publisher}{Wiley}.
\bibitem[{Curran \& Carroll(1979)}]{Curran1979}
\bibinfo{author}{Curran, J.~H.}, \& \bibinfo{author}{Carroll, M.~M.}
  (\bibinfo{year}{1979}).
\newblock \bibinfo{title}{Shear stress enhancement of void compaction}.
\newblock {\it \bibinfo{journal}{Journal of Geophysical Research}\/},  {\it
  \bibinfo{volume}{84}\/}, \bibinfo{pages}{1105}. \URLprefix
  \url{http://doi.org/10.1029/jb084ib03p01105}.
  \DOIprefix\doi{10.1029/jb084ib03p01105}.
\bibitem[{Diarra et~al.(2017)Diarra, Mazel, Busignies \&
  Tchoreloff}]{Diarra2017}
\bibinfo{author}{Diarra, H.}, \bibinfo{author}{Mazel, V.},
  \bibinfo{author}{Busignies, V.}, \& \bibinfo{author}{Tchoreloff, P.}
  (\bibinfo{year}{2017}).
\newblock \bibinfo{title}{Comparative study between drucker-prager/cap and
  modified cam-clay models for the numerical simulation of die compaction of
  pharmaceutical powders}.
\newblock {\it \bibinfo{journal}{Powder Technology}\/},  (p.
  \bibinfo{pages}{S0032591017306241}). \URLprefix
  \url{http://doi.org/10.1016/j.powtec.2017.07.077}.
  \DOIprefix\doi{10.1016/j.powtec.2017.07.077}.
\bibitem[{Ewy \& Cook(1990)}]{EWY1990387}
\bibinfo{author}{Ewy, R.}, \& \bibinfo{author}{Cook, N.}
  (\bibinfo{year}{1990}).
\newblock \bibinfo{title}{Deformation and fracture around cylindrical openings
  in rock-i. observations and analysis of deformations}.
\newblock {\it \bibinfo{journal}{International Journal of Rock Mechanics and
  Mining Sciences and Geomechanics}\/},  {\it \bibinfo{volume}{27}\/},
  \bibinfo{pages}{387--407}.
  \DOIprefix\doi{https://doi.org/10.1016/0148-9062(90)92713-O}.
\bibitem[{Geertsma(1973)}]{Geertsma1973}
\bibinfo{author}{Geertsma, J.} (\bibinfo{year}{1973}).
\newblock \bibinfo{title}{Land subsidence above compacting oil and gas
  reservoirs}.
\newblock {\it \bibinfo{journal}{Journal of Petroleum Technology}\/},  {\it
  \bibinfo{volume}{25}\/}, \bibinfo{pages}{734--744}. \URLprefix
  \url{http://doi.org/10.2118/3730-pa}. \DOIprefix\doi{10.2118/3730-pa}.
\bibitem[{G.Kirsch(1898)}]{Kirsch1898}
\bibinfo{author}{G.Kirsch} (\bibinfo{year}{1898}).
\newblock \bibinfo{title}{Die theorie der elstizitat und die bedurfnisse der
  fetigkeitslehre}.
\newblock {\it \bibinfo{journal}{Antralblatt Verlin Deutscher Ingenieure}\/},
  (pp. \bibinfo{pages}{797--807}).
\bibitem[{Griffith(1921)}]{Griffith1921}
\bibinfo{author}{Griffith, A.~A.} (\bibinfo{year}{1921}).
\newblock \bibinfo{title}{The phenomena of rupture and flow in solids}.
\newblock {\it \bibinfo{journal}{Philosophical Transactions Mathematical
  Physical and Engineering Sciences}\/},  {\it \bibinfo{volume}{221}\/},
  \bibinfo{pages}{163--198}. \URLprefix
  \url{http://doi.org/10.1098/rsta.1921.0006}.
  \DOIprefix\doi{10.1098/rsta.1921.0006}.
\bibitem[{Hasbani \& Hryb(2018)}]{Hasbani2018}
\bibinfo{author}{Hasbani, J.~H.}, \& \bibinfo{author}{Hryb, D.~E.}
  (\bibinfo{year}{2018}).
\newblock \bibinfo{title}{On the characterization of the viscoelastic response
  of the vaca muerta formation.}
\newblock {\it \bibinfo{journal}{U.S. Rock Mechanics/Geomechanics
  Symposium}\/}, .
\newblock \bibinfo{note}{ARMA-2018-1214}.
\bibitem[{Horii \& Nemat-Nasser(1986)}]{Horii1986}
\bibinfo{author}{Horii, H.}, \& \bibinfo{author}{Nemat-Nasser, S.}
  (\bibinfo{year}{1986}).
\newblock \bibinfo{title}{Brittle failure in compression: Splitting, faulting
  and brittle-ductile transition}.
\newblock {\it \bibinfo{journal}{Philosophical Transactions Mathematical
  Physical and Engineering Sciences}\/},  {\it \bibinfo{volume}{319}\/},
  \bibinfo{pages}{337--374}. \URLprefix
  \url{http://doi.org/10.1098/rsta.1986.0101}.
  \DOIprefix\doi{10.1098/rsta.1986.0101}.
\bibitem[{J.~C.~Simo(1998)}]{Simo1998}
\bibinfo{author}{J.~C.~Simo, T. J. R. H.~a.} (\bibinfo{year}{1998}).
\newblock {\it \bibinfo{title}{Computational Inelasticity}\/}.
\newblock Interdisciplinary Applied Mathematics 7 (\bibinfo{edition}{1st} ed.).
\newblock \bibinfo{publisher}{Springer-Verlag New York}.
\newblock \URLprefix
  \url{http://gen.lib.rus.ec/book/index.php?md5=8229c50d9cddeb87352a3e944dcf1dcb}.
\bibitem[{Kias et~al.(2015)Kias, Maharidge \& Hurt}]{Kias2015}
\bibinfo{author}{Kias, E.}, \bibinfo{author}{Maharidge, R.}, \&
  \bibinfo{author}{Hurt, R.} (\bibinfo{year}{2015}).
\newblock \bibinfo{title}{Mechanical versus mineralogical brittleness indices
  across various shale plays}.
\newblock {\it \bibinfo{journal}{SPE Annual Technical Conference and
  Exhibition}\/}, . \DOIprefix\doi{10.2118/174781-MS}.
\bibitem[{Lecampion et~al.(2017)Lecampion, Bunger \& Zhang}]{Lecampion2017}
\bibinfo{author}{Lecampion, B.}, \bibinfo{author}{Bunger, A.}, \&
  \bibinfo{author}{Zhang, X.} (\bibinfo{year}{2017}).
\newblock \bibinfo{title}{Numerical methods for hydraulic fracture propagation:
  A review of recent trends}.
\newblock {\it \bibinfo{journal}{Journal of Natural Gas Science and
  Engineering}\/},  (p. \bibinfo{pages}{S1875510017304006}). \URLprefix
  \url{http://doi.org/10.1016/j.jngse.2017.10.012}.
  \DOIprefix\doi{10.1016/j.jngse.2017.10.012}.
\bibitem[{Lee(1990)}]{Borja1990}
\bibinfo{author}{Lee, R. I. B. L.~S.} (\bibinfo{year}{1990}).
\newblock \bibinfo{title}{Cam-clay plasticity, part 1: Implicit integration of
  elasto-plastic constitutive relations}.
\newblock {\it \bibinfo{journal}{Computer Methods in Applied Mechanics and
  Engineering}\/},  {\it \bibinfo{volume}{78}\/}, \bibinfo{pages}{49--72}.
  \URLprefix \url{http://doi.org/10.1016/0045-7825%2890%2990152-c}.
  \DOIprefix\doi{10.1016/0045-7825(90)90152-c}.
\bibitem[{Lubiner(1985)}]{Lubiner1985}
\bibinfo{author}{Lubiner, J.} (\bibinfo{year}{1985}).
\newblock \bibinfo{title}{Thermomechanics of deformable bodies}.
\newblock \bibinfo{address}{Berckley}.
\newblock \bibinfo{note}{University of California}.
\bibitem[{Nygård et~al.(2006)Nygård, Gutierrez, Bratli \&
  Høeg}]{Nygård2006}
\bibinfo{author}{Nygård, R.}, \bibinfo{author}{Gutierrez, M.},
  \bibinfo{author}{Bratli, R.~K.}, \& \bibinfo{author}{Høeg, K.}
  (\bibinfo{year}{2006}).
\newblock \bibinfo{title}{Brittle–ductile transition, shear failure and
  leakage in shales and mudrocks}.
\newblock {\it \bibinfo{journal}{Marine and Petroleum Geology}\/},  {\it
  \bibinfo{volume}{23}\/}, \bibinfo{pages}{0--212}. \URLprefix
  \url{http://doi.org/10.1016/j.marpetgeo.2005.10.001}.
  \DOIprefix\doi{10.1016/j.marpetgeo.2005.10.001}.
\bibitem[{Papanastasiou(1997)}]{papanastasiou1997}
\bibinfo{author}{Papanastasiou, P.} (\bibinfo{year}{1997}).
\newblock \bibinfo{title}{The influence of plasticity in hydraulic fracturing}.
\newblock {\it \bibinfo{journal}{International Journal of Fracture}\/},  {\it
  \bibinfo{volume}{84}\/}, \bibinfo{pages}{61--79}.
\bibitem[{Roscoe \& Burland()}]{Roscoe1968}
\bibinfo{author}{Roscoe, K.~H.}, \& \bibinfo{author}{Burland, J.~B.} ().
\newblock \bibinfo{title}{On the generalized stress-strain behavior of
  “wet” clay}.
\newblock {\it \bibinfo{journal}{Engineering Plasticity}\/},  (p.
  \bibinfo{pages}{535–609}). \DOIprefix\doi{10.1016/0022-4898(70)90160-6}.
\bibitem[{Roscoe et~al.(1963)Roscoe, Schofield \& Thurairajah}]{Roscoe1963}
\bibinfo{author}{Roscoe, K.~H.}, \bibinfo{author}{Schofield, A.}, \&
  \bibinfo{author}{Thurairajah, A.} (\bibinfo{year}{1963}).
\newblock \bibinfo{title}{Yielding of clays in states wetter than critical.
  géotechnique,}.
\newblock {\it \bibinfo{journal}{Geotechnique}\/},  {\it
  \bibinfo{volume}{13}\/}, \bibinfo{pages}{211--240}.
\bibitem[{Sagasti et~al.(2014)Sagasti, Ortiz, Hryb, Foster \&
  Lazzari}]{Sagasti2014}
\bibinfo{author}{Sagasti, G.}, \bibinfo{author}{Ortiz, A.},
  \bibinfo{author}{Hryb, D.}, \bibinfo{author}{Foster, M.}, \&
  \bibinfo{author}{Lazzari, V.} (\bibinfo{year}{2014}).
\newblock \bibinfo{title}{Understanding geological heterogeneity to customize
  field development. an example from the vaca muerta unconventional play,
  argentina}.
\newblock {\it \bibinfo{journal}{SPE/AAPG/SEG Unconventional Resources
  Technology Conference}\/}, . \DOIprefix\doi{10.15530/URTEC-2014-1923357}.
\newblock \bibinfo{note}{URTEC-1923357-MS}.
\bibitem[{EA~de Souza~Neto(2009)}]{Neto2019}
\bibinfo{author}{EA~de Souza~Neto, P. D.~O., Prof. D~Periæ}
  (\bibinfo{year}{2009}).
\newblock {\it \bibinfo{title}{Computational Methods for Plasticity Theory and
  Applications}\/}.
\newblock \bibinfo{publisher}{Wiley}.
\newblock \URLprefix
  \url{http://gen.lib.rus.ec/book/index.php?md5=6d9787869c93dc6ee5b4dedbbf0aa0ee}.
\bibitem[{Suarez-Rivera et~al.(2023)Suarez-Rivera, Kias, Degenhardt, Alvarez,
  Mesdour, Eichmann \& Gupta}]{SuarezRivera2023}
\bibinfo{author}{Suarez-Rivera, R.}, \bibinfo{author}{Kias, E.},
  \bibinfo{author}{Degenhardt, J.}, \bibinfo{author}{Alvarez, A.~R.},
  \bibinfo{author}{Mesdour, R.}, \bibinfo{author}{Eichmann, S.}, \&
  \bibinfo{author}{Gupta, A.} (\bibinfo{year}{2023}).
\newblock \bibinfo{title}{Compaction in unconventional carbonate reservoir
  rocks and its effect on well completions and hydraulic fracturing}.
\newblock {\it \bibinfo{journal}{SPE/AAPG/SEG Unconventional Resources
  Technology Conference}\/},  {\it \bibinfo{volume}{Day 3 Thu, June 15,
  2023}\/}. \URLprefix \url{https://doi.org/10.15530/urtec-2023-3864914}.
  \DOIprefix\doi{10.15530/urtec-2023-3864914}.
\newblock \bibinfo{note}{D031S059R003}.
\bibitem[{Sundaram(1996)}]{Sundaram1996}
\bibinfo{author}{Sundaram, R.~K.} (\bibinfo{year}{1996}).
\newblock {\it \bibinfo{title}{A First Course in Optimization Theory}\/}.
\newblock \bibinfo{publisher}{Cambridge University Press}.
\newblock \URLprefix
  \url{http://gen.lib.rus.ec/book/index.php?md5=8eca9be5fddf21716bebfcb8e6ebb8b4}.
\bibitem[{Varela \& Hasbani(2017)}]{Varela2017}
\bibinfo{author}{Varela, R.~A.}, \& \bibinfo{author}{Hasbani, J.~G.}
  (\bibinfo{year}{2017}).
\newblock \bibinfo{title}{A rock mechanics laboratory characterization of vaca
  muerta formation}.
\newblock {\it \bibinfo{journal}{U.S. Rock Mechanics/Geomechanics
  Symposium}\/}, .
\newblock \bibinfo{note}{ARMA-2017-0167}.
\bibitem[{Wang et~al.(2019)Wang, Ge, Wang, Shen, Liu, Zhang \& Meng}]{Wang2019}
\bibinfo{author}{Wang, J.}, \bibinfo{author}{Ge, H.}, \bibinfo{author}{Wang,
  X.}, \bibinfo{author}{Shen, Y.}, \bibinfo{author}{Liu, T.},
  \bibinfo{author}{Zhang, Y.}, \& \bibinfo{author}{Meng, F.}
  (\bibinfo{year}{2019}).
\newblock \bibinfo{title}{Effect of clay and organic matter content on the
  shear slip properties of shale}.
\newblock {\it \bibinfo{journal}{Journal of Geophysical Research: Solid
  Earth}\/},  (p. \bibinfo{pages}{2018JB016830}). \URLprefix
  \url{http://doi.org/10.1029/2018JB016830}.
  \DOIprefix\doi{10.1029/2018JB016830}.
\bibitem[{Wang et~al.(1994)Wang, Scott \& Dusseault}]{wang1994}
\bibinfo{author}{Wang, Y.}, \bibinfo{author}{Scott, J.~D.}, \&
  \bibinfo{author}{Dusseault, M.~B.} (\bibinfo{year}{1994}).
\newblock \bibinfo{title}{Borehole rupture from plastic yield to hydraulic
  fracture—a nonlinear model including elastoplasticity}.
\newblock {\it \bibinfo{journal}{Journal of Petroleum Science and
  Engineering}\/},  {\it \bibinfo{volume}{12}\/}, \bibinfo{pages}{97--111}.
\bibitem[{Wong(1990)}]{Wong1990}
\bibinfo{author}{Wong, T.~F.} (\bibinfo{year}{1990}).
\newblock \bibinfo{title}{Mechanical compaction and the brittle-ductile
  transition in porous sandstones}.
\newblock {\it \bibinfo{journal}{Geological Society London Special
  Publications}\/},  {\it \bibinfo{volume}{54}\/}, \bibinfo{pages}{111--122}.
  \URLprefix \url{http://doi.org/10.1144/GSL.SP.1990.054.01.12}.
  \DOIprefix\doi{10.1144/GSL.SP.1990.054.01.12}.
\bibitem[{Wong \& Baud(1999)}]{Wong1999}
\bibinfo{author}{Wong, T.~F.}, \& \bibinfo{author}{Baud, P.}
  (\bibinfo{year}{1999}).
\newblock \bibinfo{title}{Mechanical compaction of porous sandstone}.
\newblock {\it \bibinfo{journal}{Oil and Gas Science and Technology – Revue
  d'IFP Energies nouvelles}\/},  {\it \bibinfo{volume}{54}\/},
  \bibinfo{pages}{715--727}. \URLprefix
  \url{http://doi.org/10.2516/ogst%3A1999061}.
  \DOIprefix\doi{10.2516/ogst:1999061}.
\bibitem[{Wong et~al.(1997)Wong, David \& Zhu}]{Wong1997}
\bibinfo{author}{Wong, T.~F.}, \bibinfo{author}{David, C.}, \&
  \bibinfo{author}{Zhu, W.} (\bibinfo{year}{1997}).
\newblock \bibinfo{title}{The transition from brittle faulting to cataclastic
  flow in porous sandstones: Mechanical deformation}.
\newblock {\it \bibinfo{journal}{Journal of Geophysical Research}\/},  {\it
  \bibinfo{volume}{102}\/}, \bibinfo{pages}{3009}. \URLprefix
  \url{http://doi.org/10.1029/96jb03281}. \DOIprefix\doi{10.1029/96jb03281}.

\end{thebibliography}
\end{document}